\documentclass[prb,aps,floats,floatfix,superscriptaddress]{revtex4}
\usepackage{epsfig}
\usepackage{colordvi}
\usepackage{graphicx}
\usepackage{amssymb}
\usepackage{color}

\begin{document}
\title{Nonperturbative quantum theory of multiplasmonic electron emission from surfaces:\\
Gauge-specific cumulant expansions vs. Volkov ansatz over plasmonic coherent states}
\author{Branko Gumhalter\footnote{Corresponding author. Email: branko@ifs.hr}}
\affiliation{Institute of Physics, HR 10000 Zagreb, Croatia}

\begin{abstract}
Energetic electromagnetic fields produce a variety of elementary excitations in solids that can strongly modify their primary photoemission spectra. Such is the plasmon excitation or pumping mechanism which, although indirect, is very efficient and hence may give rise to formation of plasmonic coherent states. In turn, these states may act as a source or sink of energy and momentum for escaping electrons. Starting from the model Hamiltonian approach we show that prepumped plasmonic bath of coherent states gives rise to ponderomotive potentials and Floquet electronic band structure that support multiple plasmon-induced electron emission or plasmoemission from metals. Theoretical description of multiple plasmoemission requires a nonperturbative approch which is here formulated by applying cumulant expansion and Volkov ansatz to the calculations of electron wavefunctions and emission rates. The calculations are performed in the standard length gauge as well as in the Pauli-transformed velocity gauge for electron-plasmon interaction. The applicability of two nonperturbative approaches to calculation of excitation amplitudes are examined in each gauge. They smoothly interpolate between the fully quantal first order Born approximation and semiclassical multiplasmon-induced electron excitation limit. This is illustrated on the example of plasmoemission from Floquet surface bands on Ag(111) from which this channel of electron yield has been detected. Our calculations indicate that even subsingle mode occupations of plasmonic coherent states can support multiplasmon electron emission from surface bands. A way of calibration of plasmonic coherent states is proposed.
\end{abstract}

\date{\today}
\maketitle

\newcommand{\bq}{\begin{equation}}
\newcommand{\eq}{\end{equation}}

\newcommand{\barr}{\begin{eqnarray}}
\newcommand{\earr}{\end{eqnarray}}

\renewcommand{\bibnumfmt}[1]{[#1]}
\renewcommand{\citenumfont}[1]{#1}

\def\bcalP{{\hbox{\boldmath${\cal P}$}}}
\def\balpha{{\hbox{\boldmath$\alpha$}}}
\def\brho{{\hbox{\boldmath$\rho$}}}
\def\bvarepsilon{{\hbox{\boldmath$\varepsilon$}}}
\def\bepsilon{{\hbox{\boldmath$\epsilon$}}}
\def\bnu{{\hbox{\boldmath$\nu$}}}
\def\bxi{{\hbox{\boldmath$\xi$}}}
\def\bcalH{{\hbox{\boldmath$\cal{H}$}}}
\def\bcalL{{\hbox{\boldmath$\cal{L}$}}}
\def\bcalW{{\hbox{\boldmath$\cal{W}$}}}
\def\bcalP{{\hbox{\boldmath$\cal{P}$}}}
\def\bcalR{{\hbox{\boldmath$\cal{R}$}}}
\def\bpi{{\hbox{\boldmath$\pi$}}}
\def\bcalA{{\hbox{\boldmath$\cal{A}$}}}
\def\bcalE{{\hbox{\boldmath$\cal{E}$}}}

\newpage

\begin{widetext}
\tableofcontents
\end{widetext}


\section{ Motivation for studying plasmonically assisted electron emission from surfaces}
\label{sec:introduction}

\begin{figure}[tb]
\rotatebox{0}{\epsfxsize=8cm \epsffile{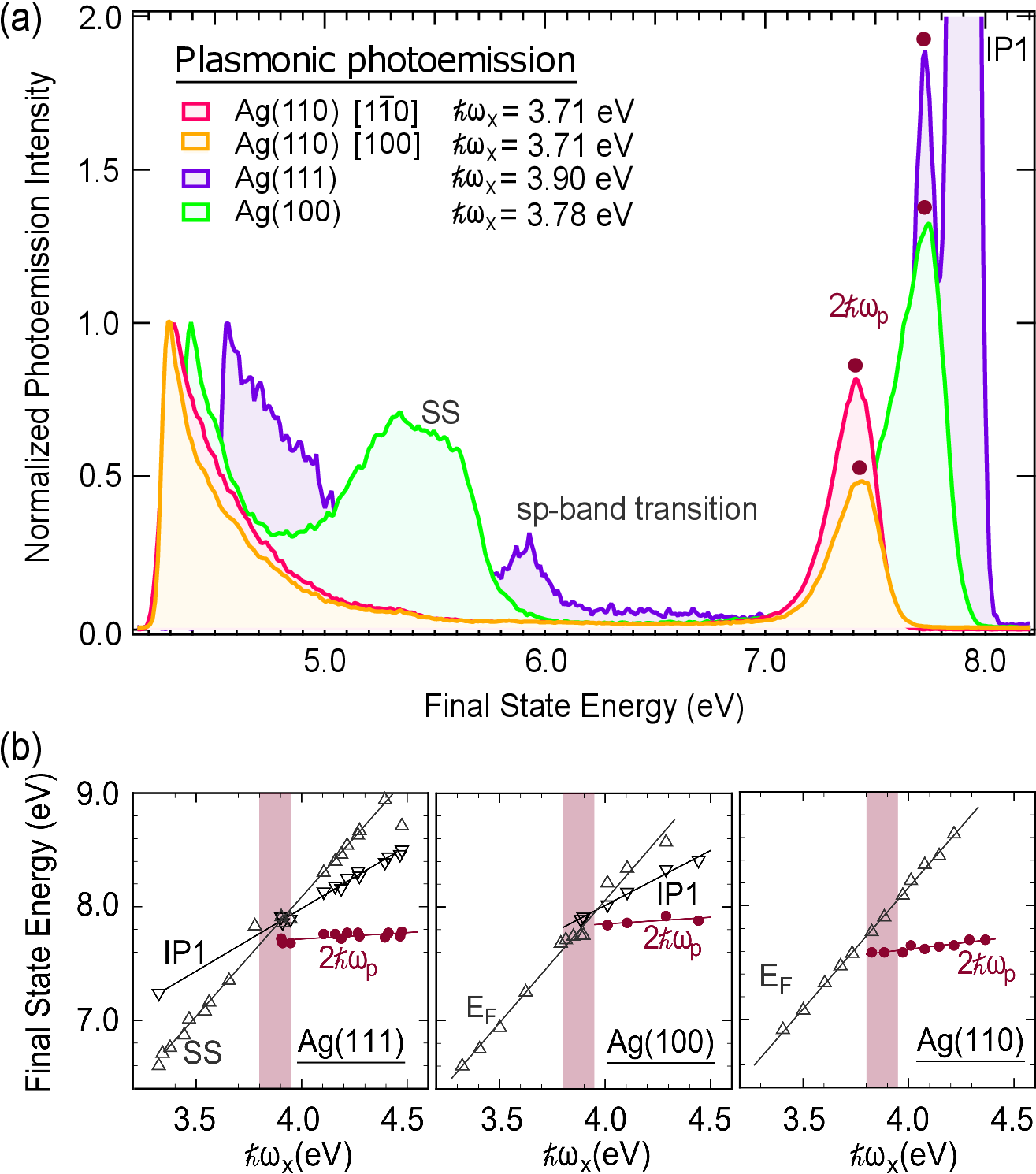}} 
\caption{Panel (a): Selection of 2PP spectra from three low index surfaces of Ag for photon field excitation energies $\hbar\omega_x=3.7-3.9$ eV in the epsilon near zero (ENZ) region where the real part of Ag  dielectric function $\varepsilon(\omega)$ is near zero.[\onlinecite{ACSPhotonics}]  The latter puts  bulk plasmon energy in the interval $\hbar\omega_{p}=3.7-3.9$ eV. [\onlinecite{ACSPhotonics}] The manifolds of all the dots as shown above the four representative plasmonic peak maxima make the $2\hbar\omega_p$ scaling curves in panel (b) below.
Panel (b): Dependence of measured final state energies of 2PP electron yields from Ag on the variation of energy $\hbar\omega_x$. The data shown by triangles were recorded in the constant inital state mode (CIS) involving the surface state (SS) and the first image potential state (IP1) on Ag(111), the IP1 and the Fermi level $E_F$ on Ag(100), and $E_F$ on Ag(110). They exhibit the standard scaling of 2PP yield energy with $2\hbar\omega_x$ in emission from SS and $E_F$, and with $\hbar\omega_x$ in emission from IP1. In contrast to this, there also appear yields with energy $\sim 2\hbar\omega_p$ above $E_F$ (dots) which do not scale with the multiples of the radiation field energy $m\hbar\omega_x$. This $2\hbar\omega_p$ emission extends far beyond the ENZ region denoted by vertical shading. Similar processes involving surface plasmons may also be envisaged. (After Ref. [\onlinecite{PSS2023}])}
\label{AllAgFloquet}
\end{figure}

Interaction of electrons in solids with external electromagnetic (EM) fields gives rise to photoionization and photoeffect whose quantum formulation was first proposed by Einstein in his seminal paper published in 1905.[\onlinecite{Einstein}]. However, it can be envisaged that strong electron interactions with other bosonic fields with similar range of frequencies can also give rise to electron emission from solids. We shall designate such non-EM-field induced emissions as non-Einsteinian. 
Recently we have proposed a scenario for plasmonically induced electron emission or plasmoemission from metal surfaces[\onlinecite{plasPE,plasFloquet,PSS2023}] with the aim to explain the non-Einsteinian emission channels detected in a series of earlier multi-photon photoemission ($m$PP) experiments.[\onlinecite{ReutzelPRX,MarcelPRL,AndiNJP,ACSPhotonics,AndiPRB22,PetekPlasmonicsPSS}]
 A peculiar characteristics of these channels is the absence of linear scaling of the emitted electron excitation energy $\epsilon_{f}-\epsilon_{b}$ with the multiples $m$ of the absorbed photon energy $\hbar\omega_{x}$ that are connected through the generalized Einstein's relation  
\bq
\epsilon_{f}-\epsilon_{b}=m\hbar\omega_{x}.
\label{eq:Einstein}
\eq
Here $\epsilon_{f}$ and $\epsilon_{b}$ denote the electron final state and the relaxed initial state electron binding energy, respectively.[\onlinecite{plasPE,UebaGumhalter}]
The energetics of such nonlinear non-Einsteinian and nearly monochromatic yields starting from the bulk plasmon onset energy observed in two-photon photoemission (2PP) from (111), (100) and (110) surfaces of silver is illustrated in Fig. \ref{AllAgFloquet}. 
 Similar observation regarding one-plasmon mediated electron emission was previously reported albeit differently interpreted.[\onlinecite{Horn}]  The non-Einsteinian character of these spectra was interpreted in Ref. [\onlinecite{plasPE}] in terms of one and two bulk plasmon (BP) induced electron emission using perturbative approach. In this model the plasmons required for electron excitations are supplied from the plasmonic coherent state(s) generated following the primary interactions of the external pump EM field with the system. 

Completely analogous framework can also be established for electron emission from quasi-two dimensional (Q2D) surface electronic bands on metals when transitions are induced by the fields of surface electron density fluctuations or surface plasmons (SP). In metals these fields are localized within few atomic radii across (perpendicular to) the outermost crystal plane but propagate in the lateral (parallel to the surface) direction in the form of plane waves characterized by the two-dimensional lateral wavevector ${\bf Q}$. A variety of SP modes have been predicted and detected at metal surfaces like monopole,  multipole and acoustic SPs.[\onlinecite{Rocca,Liebsch,accousticSP}].  Each of them contributes with specific oscillator strength and dispersion to the spectrum of plasmonic excitations. 
   The pumping of coherent states constituted of surface plasmons may be realized in the same fashion as for bulk plasmons, i.e. in the primary interactions of external charged probes or EM fields with electrons in the system. This mechanism was described in Secs. II and III.A of Ref. [\onlinecite{plasPE}]. However, since the simultaneous treatment of plasmoemission induced by all the above mentioned components of SP spectrum may prove very tedious we shall restrict our attention to the one with strongest oscillator strength, viz. the surface electronic charge density fluctuation arising in response to the longitudinal Coulomb field. The corresponding model Hamiltonians with electron-surface plasmon (e-SP) coupling  and SP dispersions have been described in detail in Refs. [\onlinecite{SunjicLucas,Lucas}].

Quite generally, as required by the unitarity of the full scattering matrix the higher order multiexcitation processes induced by strong electron-plasmon coupling may strongly affect also the low order ones, as is also evident from semiclassical descriptions of electron dynamics.[\onlinecite{SunjicLucas,Lucas}] 
However, due to the simple structure of Q2D surface electronic bands and simple form of e-SP coupling, one may also invoke fully quantal nonperturbative methods to describe and analyze the SP-induced electron emission from surface bands. This program is followed in the present work and is carried out intermittently in the length and velocity gauge formulations of e-SP interaction.  In either gauge the electron spectral properties, inelastic energy transfer, as well as transition amplitudes and emission rates can be based on cumulant expansion and Volkov operator ansatz in averaging over plasmonic coherent states. Here we elaborate and examine comparatively these approaches which represent generalizations of earlier multiexcitation theories [\onlinecite{Faisal,Reiss1980,4cumul}] to momentum conserving electron interactions with quantized boson fields. The effects arising from the other SP modes with different oscillator strengths and dispersions would largely
appear additive in the final plasmoemission spectra.

In Sec. \ref{sec:System} we introduce the standard model for e-SP field  interaction in the length gauge and define plasmoemission amplitudes and rates. In Sec. \ref{sec:EPdynlg} we employ nonperturbative method of truncated cumulant expansion to calculate correlation functions from which we obtain the electron inelastic spectral properties and energy transfer obeying temporal boundary conditions appropriate to  {\it (i)} sudden or nonadiabatic switching of e-SP interaction, and {\it (ii)} adiabatic limit or scattering boundary conditions (SBC).  In Sec. \ref{sec:velocitygauge} we introduce the velocity gauge representation of e-SP interaction and the corresponding electron transition amplitudes. The aspects of e-SP interaction dynamics equivalent to those presented in Sec. \ref{sec:EPdynlg} for the length gauge are studied again in the velocity gauge in Sec. \ref{sec:EPdynvg} using the same nonperturbative methods. In Sec. \ref{sec:Tlength} and \ref{sec:TwithcalV} we introduce nonperturbative $T$-matrix formulation of surface plasmoemission based on the Volkov ansatz for electron wavefunctions in the length and velocity gauge, respectively. This ansatz is based on the equivalent truncation criterion for higher order correlated scattering processes described by the evolution operator as employed in cumulant expansion and hence is expected to yield equivalent results within the same boundary conditions. This is used in Sec. \ref{sec:emissionFloquet} to calculate and discuss plasmoemission currents from plasmonically induced Floquet bands on Ag(111) surface both in the high and low plasmon frequency regime. Summary section pinpoints the domains of applicability and prospects of  application of the presented theory to the interpretation of multiplasmon electron emission spectra. Tedious formal mathematical procedures have been deferred to Appendices.

%
\section{Model description of surface plasmon-induced electron excitations at metal surfaces}
\label{sec:System}

\begin{figure}[tb] 
\rotatebox{0}{\epsfxsize=8.5cm \epsffile{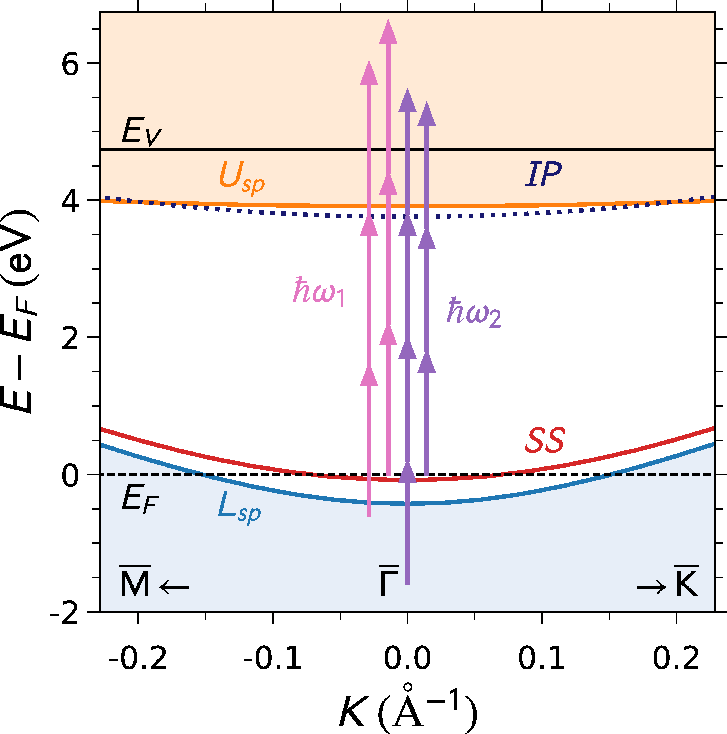}} 
\caption{Electronic band structure of the Ag(111) surface along the high symmetry lines of the 2D surface Brillouin zone. The surface projected band structure of Ag(111) has a band gap extending between -0.4 and 3.9 eV from the lower to the upper sp-band ($L_{sp}$ and $U_{sp}$, respectively). Within the band gap, the quasi-two-dimensional Shockley surface state band, SS (red line), and $n=1$ IP state band (blue dotted line) with minima at -0.063 and 3.79 eV, respectively, form quantum wells at the metal-vacuum interface. The SS-band is occupied up to ${\bf K}=0.07$ \AA$^{-1}$ where it intersects the Fermi level ($E_{F}$). The energy of $n=1$ IP quantum well state is the first member of a Rydberg-like series converging to the vacuum level, $E_{V}$. Vertical arrows indicate independent excitation pathways for 3PP and 4PP via the initial SS- or the penultimate IP-state for photon energies $\hbar\omega_{x}=2.20$ and $\hbar\omega_{x}=1.81$ eV employed in the $m$PP investigations of these bands.[\onlinecite{TE}] }
\label{SSband}
\end{figure}

Low index surfaces of some metals may exhibit surface projected band gaps. This means that there are no electrons with energies within the gap that can propagate normal to the surface. The effective one-electron potential at such surfaces can support  the set of Q2D surface state (SS) and image potential state (IP) bands. Localization of SS and IP electrons in the direction perpendicular to the surface is only few atomic radii over the image potential well and centered outside the physical surface[\onlinecite{FeibelmanPSS}], as exemplified in Fig. 1 of Ref. [\onlinecite{ultrafastPRB}]. Their Q2D  Bloch state dynamics in the lateral direction parallel to the surface is well described in the effective mass approximation.[\onlinecite{FausterSteinmann,Chulkov,TE}] The energetics of the SS-band and the first IP-band on the Ag(111) surface is illustrated in Fig. \ref{SSband}. 
Hence, in the context of plasmoemission the Q2D surface localized bands and the corresponding electron wavefunctions represent an ideal platform for application of nonperturbative approach to electron emission by surface plasmons.

\subsection{Length gauge description of electron-surface plasmon interaction and dynamics}
\label{sec:lengthgauge}

To study nonperturbative plasmoemission from surface bands we start from quantum description of the system of electrons and nonretarded surface plasmons based on a standard model Hamiltonian $H$ expressed in the length gauge regarding the e-SP interaction.[\onlinecite{SunjicLucas,Lucas}] Here $H$ comprises the component that describes the unperturbed quantized electron and SP system and the coupling of electron charge to  SP field[\onlinecite{PlasGauge}]
\bq
H=H_{0}^{e}+H_{0}^{pl}+V=H_{0}^{syst}+V.
\label{eq:HsystSP}
\eq
$H_{0}^{e}$ describes an electron in a crystal band, $H_{0}^{pl}$ is the Hamiltonian of  unperturbed surface plasmon field, and $V$ describes their interaction. The electron part reads
\bq
H_{0}^{e}=\frac{{\bf p}^2}{2m}+v({\bf r}),
\label{eq:H0el}
\eq
where ${\bf p}=({\bf P},p_z)$ and ${\bf r}=(\brho,z)$ are the electron momentum and radius vector expressed in cylindrical coordinates, respectively, with $z$  measured perpendicular to the physical surface for later convenience taken at $z=0$.[\onlinecite{FeibelmanPSS}]  $m$ is the bare electron mass and $v({\bf r})$ is the effective one-electron crystal potential.  ${\bf r}$ and ${\bf p}$ are the conjugate noncommuting operators satisfying $[{\bf r},{\bf p}]=i\hbar$. 
In the following formulations of electron transition rates in the length and velocity gauge it will prove convenient to describe the unperturbed electron and plasmon dynamics by the same wavefunctions. This is achieved by adding to (\ref{eq:H0el}) the local electron image potential $\Phi({\bf r})$ arising from the polarization of plasmon ground state (see derivation in Sec. \ref{sec:canonical})  and then subtract it from $V$, viz.  
\barr
H_{0}^{e}&\rightarrow& \frac{{\bf p}^2}{2m}+v({\bf r})+\Phi({\bf r})=\frac{{\bf p}^2}{2m}+v_{ps}({\bf r}), 
\label{eq:H0}
\\
V&\rightarrow& V-\Phi({\bf r}).
\label{eq:Vred}
\earr
This leaves the full $H$ defined in (\ref{eq:HsystSP}) unchanged. Since the same $H_{0}^{e}$ (\ref{eq:H0}) with the pseudopotential $v_{ps}({\bf r})=v({\bf r})+\Phi({\bf r})$ results from the transformation of $H$ to the velocity gauge (see Eq. (\ref{eq:H'}) below), this will enable definitions of the transition amplitudes in the same basis of eigenstates of $H_{0}^{e}$ in both gauges.

Employing  second quantization to represent the boson field of surface plasmons characterized by their two-dimensional (2D) wavevector ${\bf Q}$ and dispersion $\omega_{\bf Q}$ we have for the Hamiltonian of unperturbed plasmons 
\bq
H_{0}^{pl}=\sum_{\bf Q} \hbar\omega_{\bf Q}\hat{a}_{\bf Q}^{\dag}\hat{a}_{\bf Q}=\sum_{\bf Q} \hbar\omega_{\bf Q}\hat{n}_{\bf Q}. 
\label{eq:H0pl}
\eq
 Here the plasmon creation and annihilation operators are denoted by $\hat{a}_{\bf Q}^{\dag}$ and $\hat{a}_{\bf Q}$, respectively, and they satisfy commutation relations  $[\hat{a}_{\bf Q},\hat{a}_{\bf Q'}^{\dag}]=\delta_{\bf Q,Q'}$. The SP number operator is $\hat{n}_{\bf Q}=\hat{a}_{\bf Q}^{\dag}\hat{a}_{\bf Q}$. The summation over the ${\bf Q}$-quasicontinuum is performed according to $\sum_{\bf Q}\rightarrow (L^2/(2\pi)^2)\int d^2{\bf Q}$ where $L$ is the SP-field quantization length in the $(x,y)$ plane.

%

We shall consider electronic transitions from quasi-two dimensional surface bands located outside the physical surface taken at $z=0$.[\onlinecite{FeibelmanPSS}]. Then the e-SP coupling outside the surface may be described in the length gauge by the anisotropic scalar potential[\onlinecite{SunjicLucas,Lucas}]
\bq
V=\sum_{\bf Q}V_{\bf Q}e^{i{\bf Q}\brho-Qz}(\hat{a}_{\bf Q}+\hat{a}_{\bf -Q}^{\dag})=
\sum_{\bf Q}V({\bf Q},{\bf r})(\hat{a}_{\bf Q}+\hat{a}_{\bf -Q}^{\dag})
=
\sum_{\bf Q}\left[V({\bf Q},{\bf r})\hat{a}_{\bf Q}+V^{\dag}({\bf Q},{\bf r})\hat{a}_{\bf Q}^{\dag}\right],
\label{eq:V}
\eq
with $Q=|{\bf Q}|$, $\omega_{\bf Q}=\omega_{\bf -Q}$, and
\barr
& & V({\bf Q,r})=V_{\bf Q}(\brho,z)=V_{\bf Q}e^{i{\bf Q}\brho-Qz}=e^{i{\bf Q}\brho}V_{\bf Q}(z),
\label{eq:VQr}\\
& & V({\bf Q,r})^{\dag}=V_{\bf Q}(\brho,z)^{\dag}=V({\bf -Q,r})=e^{-i{\bf Q}\brho}V_{\bf Q}(z),
\label{eq:VdagV}\\ 
& & V_{\bf Q}(z)=V_{\bf Q}e^{-Qz},
\label{eq:VQz}\\
& & V_{\bf Q}=\left(\frac{\pi e^{2}\hbar\omega_{\bf Q}}{Q L^2}\right)^{1/2}=\left(\frac{e^{2}\hbar\omega_{\bf Q}}{4\pi Q a_{B}^2}\right)^{1/2}{\cal N}_{1D}(Q),
\label{eq:V_Q}
\earr
where $e$ is the electron charge. For later convenience we have in (\ref{eq:V_Q}) introduced the  atomic unit of length $a_{B}=\hbar^{2}/me^{2}$ (Bohr radius) to factorize the density of ${\bf Q}$-states in one dimension expressed in atomic units, viz.
\bq
{\cal N}_{1D}(Q)=\frac{2\pi}{L/a_{B}}.
\label{eq:N1DQ}
\eq
Expression (\ref{eq:V_Q}) is valid in the long wavelegth limit in which surface plasmons are well defined stable excitations. Generically, it describes the Coulomb coupling of the point electron charge with surface electronic charge density fluctuations.[\onlinecite{Lucas,Badro}] The length gauge formulation of the interaction (\ref{eq:V}) is particularly convenient for retrieval of the surface image potential in an accurate one-step procedure [cf. Eq. (\ref{eq:Phi})] which is prerequisite for introduction of image potential states. By contrast, it does not lead to ponderomotive effects in a simple fashion because these arise in nonpertrbative treatments of the velocity gauge formulation of the e-SP interaction (cf. Sec. \ref{sec:Elspecvel}.1). For the clarity of presentation we have left out from the present model the weaker coupled multipole[\onlinecite{Liebsch}] and accoustic surface plasmons[\onlinecite{accousticSP}] that may also occur at surfaces, and whose role in multiexcitation processes was addressed in Ref. [\onlinecite{Rivera}]. Due to the different phase space of these modes their effects in the final spectra would generally be additive relative to the ones induced by the primary interaction (\ref{eq:V}).
Lastly we note that linear coupling of the general form (\ref{eq:V}) dominates the quasiparticle-induced multiple excitations of bosonic fields.[\onlinecite{nonlinscat}]  

The one-electron Hamiltonian $H$ defined in (\ref{eq:HsystSP}), with the constituting terms (\ref{eq:H0el})-(\ref{eq:V}), is not explicitly time dependent. The electron-plasmon interaction $V$ (\ref{eq:V}) is active as long as the electron occupies the position ${\bf r}$. However, the history of bringing the electron into that position or state may follow various temporal paths of which the adiabatic and sudden switching on of the interaction $V$ are the most common examples. The former is standardly associated with the scattering boundary conditions in which the scattered particle is brought from the remote distance and past into the region of interaction and this is the common approach in collision theory.[\onlinecite{GW}] The latter is appropriate to the situation in which a particle is injected at a specified time instant $t_{0}$ into a state subject to the interaction $V$. Here the act of particle injection into (or removal from) a particular state determines the instant of the switching on (or off) of the interaction. This is most clearly exploited in the formulations of X-ray photoemission from core levels.[\onlinecite{Langreth,PSS84}] In the present work we shall investigate the behaviour of the system described by (\ref{eq:HsystSP}) in both of these experimentally relevant limits.   

To facilitate tracing the time evolution of the processes induced by $V$  defined in (\ref{eq:V}) we inspect its interaction representation generated by $H_{0}^{syst}=H_{0}^{e}+H_{0}^{pl}$ and find
%
\bq
V_{I}(t)=e^{iH_{0}^{syst}t/\hbar}Ve^{-iH_{0}^{syst}t/\hbar}
=
e^{iH_{0}^{e}t/\hbar}\sum_{\bf Q}\left[V({\bf Q},{\bf r})\hat{a}_{\bf Q}e^{i\omega_{\bf Q}t}
+
V^{\dag}({\bf Q},{\bf r})\hat{a}_{\bf Q}^{\dag}e^{-i\omega_{\bf Q}t}\right]e^{-iH_{0}^{e}t/\hbar},
\label{eq:lgVI} 
\eq
%
where the last line follows because 
\bq
[H_{0}^{e},H_{0}^{pl}]=0.
\label{eq:HeHpl}
\eq
We also observe that outside the surface the interaction (\ref{eq:V}) exhibits the property
\bq
\nabla^{2}V({\bf r})=0,
\label{eq:nabla2V}
\eq
which will be exploited in the derivation of transition rates in the length gauge.


\subsection{Generation of SP coherent states in electron systems perturbed by external EM fields}
\label{sec:SPgeneration} 

Unlike the charged particles (electrons, ions), external EM fields can not directly excite SP or surface plasmon polaritons (SPP) at flat metal surfaces due to the mismatch of polarizations and/or momenta of the two fields (see Refs. [\onlinecite{Ritchie1973,Economou,Devaux,Pitarke,KuboPetek1,KuboPetek2,KuboPetek3}] and references therein). Therefore, we shall consider the external EM field that drives photoemission and the SP field from (\ref{eq:HsystSP}) uncoupled and described with commutative respective canonical coordinates. However, as described in Secs. II.B and III of Ref. [\onlinecite{plasPE}], there exists an efficient indirect and nonresonant EM field-plasmon excitation mechanism mediated by the metal electrons that is also applicable to the present situation. In this picture the  electron interaction $W(t)$ with an external EM field first excites energetic electrons which can subsequently excite or pump SP coherent state clouds via the e-SP interaction (\ref{eq:V}). These states incorporate excited SP modes described  by the quantum numbers ${\bf Q}$, frequency $\omega_{\bf Q}$ and mode occupation amplitude $\lambda_{\bf Q}$.[\onlinecite{Glauber,GlauberPR,SunkoBG}] They are represented by[\onlinecite{PSS2023}] 
\bq
|\mbox{coh}\rangle=
e^{\sum_{\bf Q}(\lambda_{\bf Q}\hat{a}_{\bf Q}^{\dag}-\lambda_{\bf Q}^{*}\hat{a}_{\bf Q})}|0\rangle
=
e^{-i\hat{\cal P}_{\rm{\small pu}}}|0\rangle.  
\label{eq:cohstate}
\eq
To calculate the averages of $\hat{a}_{\bf Q}$ and $\hat{a}_{\bf Q}^{\dag}$ over the coherent state (\ref{eq:cohstate}) we use the commutator expansion $e^{iA}Be^{-iA}=B+[iA,B]+\frac{1}{2!}[iA,[iA,B]]+\dots$ and find
\barr
\langle\mbox{coh}|\hat{a}_{\bf Q}|\mbox{coh}\rangle&=&\lambda_{\bf Q}, 
\label{eq:lambdaSP}\\
\langle\mbox{coh}|\hat{a}_{\bf Q}^{\dag}|\mbox{coh}\rangle&=&\lambda_{\bf Q}^{*}
\label{eq:lambdaSPdag}
\earr
and other implications arising thereof. According to the findings of Sec. III.A in Ref. [\onlinecite{plasPE}], in such excitation processes the SP coherent state parameters $\lambda_{\bf Q}$ are real and positive. Hence in the following we assume $\lambda_{\bf Q}^{*}=\lambda_{\bf Q}\geq 0$.

To calculate the coherent state averages of bilinear combinations of plasmon operators we make use of (\ref{eq:cohstate}), (\ref{eq:lambdaSP}), (\ref{eq:lambdaSPdag}), and the following relations
\barr
 \langle \mbox{coh}|\hat{a}_{\bf Q}\hat{a}_{\bf Q'}|\mbox{coh}\rangle&=&\lambda_{\bf Q}\lambda_{\bf Q'},
\label{eq:cohaverage1}
\\
 \langle \mbox{coh}|\hat{a}_{\bf Q}^{\dag}\hat{a}_{\bf Q'}|\mbox{coh}\rangle&=&\lambda_{\bf Q}\lambda_{\bf Q'},\label{eq:cohaverage2}
\\
 \langle \mbox{coh}|\hat{a}_{\bf Q}\hat{a}_{\bf Q'}^{\dag}|\mbox{coh}\rangle&=&\lambda_{\bf Q}\lambda_{\bf Q'}+\delta_{\bf Q,Q'},\label{eq:cohaverage3}
\\
 \langle \mbox{coh}|\hat{a}_{\bf Q}^{\dag}\hat{a}_{\bf Q'}^{\dag}|\mbox{coh}\rangle&=&\lambda_{\bf Q}\lambda_{\bf Q'}.
\label{eq:cohaverage4}
\earr
Using this we obtain (with $\lambda_{\bf Q}=\lambda_{\bf -Q}$ implied) for the energy of the coherent state  
\bq
\epsilon_{coh}=\langle \mbox{coh}|H_{0}^{pl}|\mbox{coh}\rangle=\sum_{\bf Q}\lambda_{\bf Q}^{2}\hbar\omega_{\bf Q},
\label{eq:epscoh}
\eq
and with (\ref{eq:V}) the semiclassical form of e-SP interaction
\bq
V_{sc}=\langle\mbox{coh}|V|\mbox{coh}\rangle=\sum_{\bf Q}\lambda_{\bf Q}\left[V({\bf Q},{\bf r})+V^{\dag}({\bf Q},{\bf r})\right]
\label{eq:Vsc}
\eq
Analogously we find the averages
\barr
\langle \mbox{coh}|\hat{a}_{\bf Q}^{\dag}H_{0}^{pl}\hat{a}_{\bf Q}|\mbox{coh}\rangle&=&\lambda_{\bf Q}^{2}\epsilon_{coh}
\label{eq:epscoh+-}\\
\langle \mbox{coh}|\hat{a}_{\bf Q}H_{0}^{pl}\hat{a}_{\bf Q}^{\dag}|\mbox{coh}\rangle&=&\epsilon_{coh}+\hbar\omega_{\bf Q}(1+2\lambda_{\bf Q}^{2}).
\label{eq:epscoh-+}\\
\langle \mbox{coh}|\hat{a}_{\bf -Q}^{\dag}\hat{a}_{\bf Q}^{\dag}H_{0}^{pl}\hat{a}_{\bf Q}\hat{a}_{\bf -Q}|\mbox{coh}\rangle&=&\lambda_{\bf Q}^{4}\epsilon_{coh}, \hspace{5mm}\mbox{etc.}
\label{eq:epscoh++--}
\earr
which will all prove useful in the calculations to follow. 
Note that expressions (\ref{eq:epscoh}) and (\ref{eq:epscoh+-})-(\ref{eq:epscoh++--}) are plasmon momentum-diagonal, i.e. the momentum transferred from $|\mbox{coh}\rangle$ to $\langle \mbox{coh}|$ is zero, and so is the phase. 
\subsection{Semiclassical limit of e-SP interaction}
\label{sec:calV}

The semiclassical limit in which the electron system is perturbed by the interaction potential $V$ defined in (\ref{eq:V}) and arising from a highly excited semiclassical plasmonic field is most conveniently established from the interaction picture of the interaction (\ref{eq:lgVI}).  Assuming that the pumped plasmonic coherent state represents a predetermined complete set fixed by the experimental conditions we may apply in all interaction operators  the coherent state averaging of (\ref{eq:lgVI}) by using (\ref{eq:lambdaSP}) and the Baker-Hausdorff-Campbell formula to obtain 
\barr  
{\cal V}_{I}({\bf r},t)&=&
\langle\mbox{coh}|V_{I}({\bf r},t)|\mbox{coh}\rangle
=
\langle\mbox{coh}|e^{iH_{0}t}V({\bf r})e^{-iH_{0}t}|\mbox{coh}\rangle
=
e^{iH_{0}^{el}t}\langle 0|e^{i\hat{\cal P}_{\rm{\small pu}}}e^{iH_{0}^{pl}t}V({\bf r})e^{-iH_{0}^{pl}t}e^{-i\hat{\cal P}_{\rm{\small pu}}}|0\rangle e^{-iH_{0}^{el}t}\nonumber\\
&=&
e^{iH_{0}^{e}t}{\cal V}_{S}({\bf r},t)e^{-iH_{0}^{e}t}.
\label{eq:Vcoh}
\earr
Here the desired semiclassical form of the length gauge interaction in the Schr\"{o}dinger picture reads
\bq
{\cal V}_{S}({\bf r},t)=\sum_{\bf Q}\lambda_{\bf Q}\left[V({\bf Q,r})e^{-i\omega_{\bf Q}t}+V^{\dag}({\bf Q,r})e^{i\omega_{\bf Q}t}\right]
=
2\sum_{\bf Q}\lambda_{\bf Q}V_{\bf Q}e^{i({\bf Q},iQ){\bf r}}\cos(\omega_{\bf Q}t).
\label{eq:calV}
\eq
where 
\bq
({\bf Q},iQ)={\bf Q}+i{\bf\hat{e}}_{\perp}Q
\label{eq:e_Q}
\eq
is a complex vector with lateral and perpendicular to the surface components ${\bf Q}$ and $iQ$, respectively. 
This enables introducing the semiclassical counterpart of $H$
\bq
{\cal H}=H_{0}^{e}+{\cal V}_{S}({\bf r},t).
\label{eq:calH}
\eq
Using (\ref{eq:VQr})-(\ref{eq:V_Q}) we can also bring ${\cal V}_{S}({\bf r},t)$ to the form
\bq
{\cal V}_{S}({\bf r},t)=2\sum_{\bf Q}\lambda_{\bf Q}V({\bf Q},z)\cos({\bf Q}\brho-\omega_{\bf Q}t).
\label{eq:calVcos}
\eq
The representation (\ref{eq:calVcos}) is convenient for introducing the dipole approximation in which ${\bf Q}\brho$ can be neglected relative to $\omega_{\bf Q}t$. However, implementation of this approximation eliminates the momentum selection rules in the intermediate states between consecutive electron-plasmon interaction vertices.  

Lastly, using (\ref{eq:calV}) we can define the plasmonic coherent state-induced electric field as  
\bq
\bcalE({\bf r},t)=-\nabla{\cal V}_{S}({\bf r},t)
=
-2i\sum_{\bf Q}\lambda_{\bf Q}({\bf Q},iQ)V_{\bf Q}e^{i({\bf Q},iQ){\bf r}}\cos(\omega_{\bf Q}t),
\label{eq:bcalE}
\eq
which satisfies $\bcalE({\bf r},t)=\bcalE({\bf r},t)^{\dag}$.


\subsection{Surface plasmon-induced electron transition amplitudes in the length gauge}
\label{sec:TLG}

\subsubsection{$T$-matrix formulation of transition amplitudes}
\label{sec:241}

We start with the definition of the probability amplitude $T_{f,i}(t,t_{0})$ as a time-dependent coefficient of the wavefunction that has developed from the initial state of the system under the action of the perturbation $V$ in the interval $(t,t_{0})$. Here the electron initial state is taken to be a surface localized state denoted by $|\phi_{i}\rangle=|\phi_{s}\rangle$, and the final state is an outgoing wave state $|\phi_{f}\rangle$. Both are the eigenstates of (\ref{eq:H0}), viz. $H_{0}|\phi_{i,f}\rangle=E_{i,f}|\phi_{i,f}\rangle$ with the time dependence in the Schr\"{o}dinger representation $|\phi_{i,f}(t)\rangle=|\phi_{i,f}\rangle e^{-iE_{i,f}t}$.  We shall study electronic transitions that are induced by electron coupling to surface plasmons (\ref{eq:V}) prepumped by an external perturbation  into a coherent state (\ref{eq:cohstate}) in a sequence of external field interactions with the electron system (cf. Figs. 3 and 4 in Ref. [\onlinecite{plasPE}]). Such a coherent state plays the role of a plasmonic bath present both in the initial and final states of the coupled electron-plasmon system. To describe this situation we introduce the unperturbed states of the system in the Schr\"{o}dinger representation $||\phi_{i,f},\mbox{coh};(t)\rangle\rangle=||\phi_{i,f},\mbox{coh}\rangle\rangle e^{-i(E_{i,f}+\epsilon_{coh})t}$, where $||\phi_{f},\mbox{coh}\rangle\rangle$ and $||\phi_{i},\mbox{coh}\rangle\rangle$ denote Kronecker products $|\phi_{f}\rangle|\mbox{coh}\rangle$ and $|\phi_{i}\rangle|\mbox{coh}\rangle$ of final and initial electron and plasmon states, respectively. Using this notation  we obtain in the length gauge (superscript $^{(lg)}$)[\onlinecite{MessiahII,Davydov}]
\bq
 T_{f,i}^{(lg)}(t,t_{0})=\langle\langle\mbox{coh},\phi_{f},(t)||U_{S}(H,t,t_{0})||\phi_{i},\mbox{coh},(t_{0})\rangle\rangle
=
\langle\langle\mbox{coh},\phi_{f}||U_{I}(H,t,t_{0})||\phi_{i},\mbox{coh}\rangle\rangle,
\label{eq:Tlength}
\eq
where $U_{S}(H,t,t_{0})$ and $U_{I}(H,t,t_{0})$ are the evolution operators of the system governed by H and expressed in the Schr\"{o}dinger and interaction picture, respectively. We assume that the coincidence time $\tau$ of the wavefunctions in the two representations is at $\tau=0$ which generates $U_{I}(t,t_{0})=e^{iH_{0}t}U_{S}(t,t_{0})e^{-iH_{0}t_{0}}$ (see Appendix \ref{sec:Gross}). The chosen coincidence time leads to $|\phi_{i}^{I}(t_{0})\rangle=|\phi_{i}\rangle$ and $|\phi_{f}^{I}(t)\rangle=|\phi_{f}\rangle$ which enables the use of compact notation  on the RHS of (\ref{eq:Tlength}). The same notation will be retained throughout the evaluations of the transition amplitudes.  

\subsubsection{Low order perturbative treatment of plasmonically induced electronic transitions}
\label{sec:242}

Calculations of electron emission amplitudes and rates based on (\ref{eq:Tlength}) can be carried out either perturbatively or nonperturbatively with respect to the plasmon field that induces electron transitions. In Ref. [\onlinecite{plasPE}] and in Sec. 3 of Ref. [\onlinecite{PSS2023}] we started from (\ref{eq:Tlength}) in the length gauge to derive the one- and two-bulk plasmon induced electron yield from metals in the quadratic and quartic response.  Thus, in perturbative approach the lowest, i.e. first order expression for  the final state energy resolved and initial state integrated one plasmon-induced electron excitation probability is obtained in the form
\begin{widetext}
\bq
P^{(1)}(\epsilon)=\sum_{\bf K',K}\left|\sum_{\bf Q} 2\pi \lambda_{\bf Q}V_{\bf K',K}^{\bf Q}n_{\bf K}(1-n_{\bf K'}) \delta(E_{\bf K'}-E_{\bf K}-\hbar\omega_{\bf Q})\right|^{2}\delta(\epsilon-E_{\bf K'})
\label{eq:1plasPE}
\eq
\end{widetext}
where $E_{\bf K}$ and $E_{\bf K'}$ (including the band indices) are the electron energies in the initial and final state, respectively, and $V_{\bf K',K}^{\bf Q}=\langle{\bf K'}|V({\bf Q,r})|{\bf K}\rangle$. 
The initial state $|{\bf K}\rangle$ can be any occupied electron state in the solid whereas the final state $|{\bf K'}\rangle$ should be taken in the form of asymptotic outgoing scattered wave solution of $H_{0}^{el}$ as explained long ago by Breit and Bethe[\onlinecite{BreitBethe}] and implemented to surface photoeffect by Makinson[\onlinecite{Makinson}] and Adawi[\onlinecite{Adawi}](for more details see next Sec. \ref{sec:SSenergetics}). The selection rules obeyed by the electron and plasmon quasimomenta are then contained in these matrix elements. Energy conservation expressed through the $\delta$-function appearing in the absolute square on the RHS of (\ref{eq:1plasPE}) applies between the energies of the initial state(s) $|{\bf K}\rangle$ and the final asymptotic scattering state $|{\bf K'}\rangle$.

Expression (\ref{eq:1plasPE}) is proportional to the duration of electron-plasmon interaction and hence the squares of energy conserving $\delta$-functions can be manipulated using the standard procedures of scattering theory to produce electron transition rates per unit time.[\onlinecite{MessiahII,Davydov}] 

In a completely analogous fashion one obtains from second order perturbation theory the energy resolved two-plasmon mediated transition probability[\onlinecite{UebaGumhalter,PSS2023}] 
\begin{widetext}
\bq
P^{(2)}(\epsilon)=\sum_{\bf K',K}\left|2\pi\sum_{\bf K'',Q',Q}\frac{\lambda_{\bf Q'}\lambda_{\bf Q}V_{\bf K',K''}^{\bf Q'}V_{\bf K'',K}^{\bf Q}n_{\bf K}(1-n_{\bf K'})(1-n_{\bf K''})}{(E_{\bf K''}-E_{\bf K}-\hbar\omega_{\bf Q}+i\delta)}\delta(E_{\bf K'}-E_{\bf K}-\hbar\omega_{\bf Q}-\hbar\omega_{\bf Q'})\right|^{2}\delta(\epsilon-E_{\bf K'}),
\label{eq:2plasPE}
\eq
\end{widetext}
with the same notation as in expression (\ref{eq:1plasPE}), and with $|{\bf K''}\rangle$ and $E_{\bf K''}$ standing for the intermediate states and their energy, respectively. The same remark concerning the energy conservation between the initial and final electron states applies here as well.

 Characteristic features of both formulas are perturbative expressions for transition amplitudes appearing under the  sign of absolute squares. In (\ref{eq:1plasPE}) this is simply first order Fermi's golden rule expression for the plasmonically driven electron transitions. In (\ref{eq:2plasPE}) this role is played by the second order transition amplitude or generalized Fermi's golden rule in which, likewise in Eq. (\ref{eq:1plasPE}), the total energy conservation expressed through the $\delta$-function on the RHS of (\ref{eq:2plasPE}) involves only initial and final electron state energies. This property also holds true over all higher multiexcitation processes discussed in Secs. \ref{sec:Tlength}-\ref{sec:emissionFloquet}. Specifically, expression (\ref{eq:2plasPE}) incorporates the possibility of electron propagation over the intermediate states ${\bf K'}$ which are not directly subject to energy conservation, i.e. those for which the denominator in (\ref{eq:2plasPE}) does not vanish. In this case the propagation proceeds via virtual states.[\onlinecite{MGoeppert}] If, on the other hand, the denominator vanishes for energies of certain intermediate states, the corresponding transitions are resonant, and they strongly enhance the total transition rates.    

One of the clear advantages of low order perturbative approach is the possibility of rigouros account of the electronic band structure of the system in single and multiple transitions. However, all separate contributions from truncated expansions of the unitary evolution operator like expressions (\ref{eq:1plasPE}) and (\ref{eq:2plasPE}) neither satisfy unitarity nor exhibit mutual interference. Moreover, they also diverge with increasing $\lambda_{\bf Q}^{2}$ and $\lambda_{\bf Q}^{4}$ and that may not always be acceptable in the semiclassical limit. A prerequisite for estimating the interplay of multiplasmon processes on electron emission from surfaces is a nonperturbative assessment of electron-plasmon dynamics. In Secs. \ref{sec:EPdynlg}   and \ref{sec:EPdynvg} we shall outline the approaches which enable such estimates and use the acquired information in the construction of the off-diagonal scattering matrix elements and ensuing plasmoemission yields in Secs. \ref{sec:Tlength}-\ref{sec:emissionFloquet}.

%


\section{Electron-surface plasmon interaction dynamics in the length gauge: Application to excitation from surface bands}
\label{sec:EPdynlg}

\subsection{Model wavefunctions and energetics of Ag(111) surface electron bands}
\label{sec:SSenergetics}

Electron emission currents from surfaces are associated with the asymptotic form of outgoing wave components of wavefunctions satisfying the scattering boundary conditions.[\onlinecite{Makinson,Adawi}] Plasmoemission from Q2D surface bands will be described using the model Hamiltonian developed in the preceding section. Within this framework the electron wavefunctions appropriate to the description of SP induced electronic transitions belong to the spectrum of bound and itinerant states. Typical examples of wavefunctions localized (i.e. bound) at surfaces are those of the SS- and IP-states with energies below the vacuum level.[\onlinecite{FausterSteinmann,Chulkov}] Together with the itinerat states belonging to the continuous spectrum of $H^{el}$ (\ref{eq:H0}) they constitute the complete and normalizable set of states for description of electron emission from surfaces. 

We first elaborate the one-electron wave functions describing electron motion in the $s$-th Q2D surface band in the absence of the electron-plasmon coupling embodied in the interaction $V$.  They are written as[\onlinecite{Chulkov}]
\bq
\langle{\bf r}|\phi_{i}(t)\rangle=\phi_{{\bf K},s}(\brho,z,t)
=
\frac{e^{i{\bf K}_{s}\brho}}{\sqrt{L^{2}}}u_{s}(z)e^{-i(\hbar^{2}{\bf K}_{s}^{2}/2m^{*}+E_{s})t/\hbar},
\label{eq:phi_s}
\eq
where $s$ is the surface band index and ${\bf K}_{s}$ is an eigenvalue of the electron 2D lateral momentum operator ${\bf P}$. The corresponding quantization length $L$ is for convenience taken to be the same as for the plasmon wavevector ${\bf Q}$.  The component of electron wavefunction $u_{s}(z)$ describes its localization in the $s$-state at the surface, and $E_{s}$ is the electron energy at the $s$-band bottom. The effective electron masses for motion in the lateral and perpendicular directions in the $s$-th band are denoted by $m^{*}$ and $m_{s}$, respectively. Here we shall treat them as fit parameters taken from earlier calculations (cf. Table I in [\onlinecite{SGLD}]. The wavefunctions (\ref{eq:phi_s}) satisfy box normalization $\langle \phi_{{\bf K'},s'}|\phi_{{\bf K},s}\rangle=\delta_{\bf K',K}\delta_{s',s}$. This implies that the dimension of $u_{s}(z)$ is inverse square-root of length. To facilitate further calculations we introduce $\tilde{u}_{s}(k_{z})$ as the FT of $u_{s}(z)$ in the $k_{z}$-space by
\barr
u_{s}(z)&=&\frac{1}{2\pi}\int dk_{z}e^{ik_{z}z}\tilde{u}_{s}(k_{z}), 
\label{eq:u(ks)}\\
\tilde{u}_{s}(k_{z})&=&\int dz e^{-ik_{z}z}u_{s}(z).
\earr
According to normalization adopted in (\ref{eq:phi_s}) the dimension of $\tilde{u}_{s}(k_{z})$ is square-root of length.

In the present context of plasmoemission of particular interest are the (111) surfaces of Ag and Cu with well defined SS- and IP-bands, and whose plasmonic response has also been well explored.[\onlinecite{SGLD,KrasovskiiSilkin}] However, since the two-plasmon data from Fig. \ref{AllAgFloquet} do not indicate any plasmonically mediated resonant $\mbox{SS}\rightarrow\mbox{IP}$ transitions we shall exclude from our further considerations the role of IP-states in  plasmoemission from Ag(111) surface. Therefore, we shall consider only the SS-band states as  possible initial states, i.e. $u_{s}(z)=u_{SS}(z)$. The band structure associated with these states is shown in Fig. \ref{SSband}. In our calculations we shall refer to $u_{s}(z)$ computed using the DFT methods described in Refs. [\onlinecite{plasPE,ACSPhotonics,AndiPRB22}] which yielded $E_{s}=-0.081 \mbox{eV}=-0.003\mbox{H}$ relative to the Fermi level (1 H=27.3 eV is the atomic unit of energy). Previous estimates have given the nonretarded longwavelength limit of surface plasmon energy $\hbar\omega_{sp}\simeq 3.7 \mbox{eV}=0.136 \mbox{H}$, and the location of maximum of $u_{s}(z)$ relative to the position of the image plane is at $z_{s}=1.1-1.2 a_{B}$ ($a_{B}$ is the Bohr radius or atomic unit of length).[\onlinecite{Liebsch,SGLD}] The effective SS-state electron mass for perpendicular motion is $m_{s}\simeq m$ and for lateral motion $m^{*}=0.28 m$[\onlinecite{TE}].

In a general approach the final outgoing electron states $|\phi_{f}\rangle$ in plasmoemission should be taken, likewise in photoemission,[\onlinecite{Pendry,Krasovskii}] in the asymptotic form of "time reversed" or "inverse LEED" states which are the ingoing wave scattering solutions first introduced by Breit and Bethe.[\onlinecite{BreitBethe}] However, in the present context we shall for calculational convenience approximate them by the leading order term from the here applicable solution of Eq. (2.12) from Ref. [\onlinecite{Adawi}(a)], viz. 
\bq
\langle{\bf r}|\phi_{f}\rangle\rightarrow\frac{e^{i{\bf K}_{f}\brho}}{\sqrt{L^{2}}} \frac{e^{ik_{z,f}}}{\sqrt{L_{z}}},
\label{eq:phifin}
\eq
where $k_{z,f}>0$ is the outgoing electron momentum in the positive $z$-direction and $L_{z}$ is the associated quantization length.

\subsection{Electron energy spectrum in the length gauge}
\label{sec:EspectrumLG}

\begin{figure}[tb]
\rotatebox{0}{\epsfxsize=8.5cm \epsffile{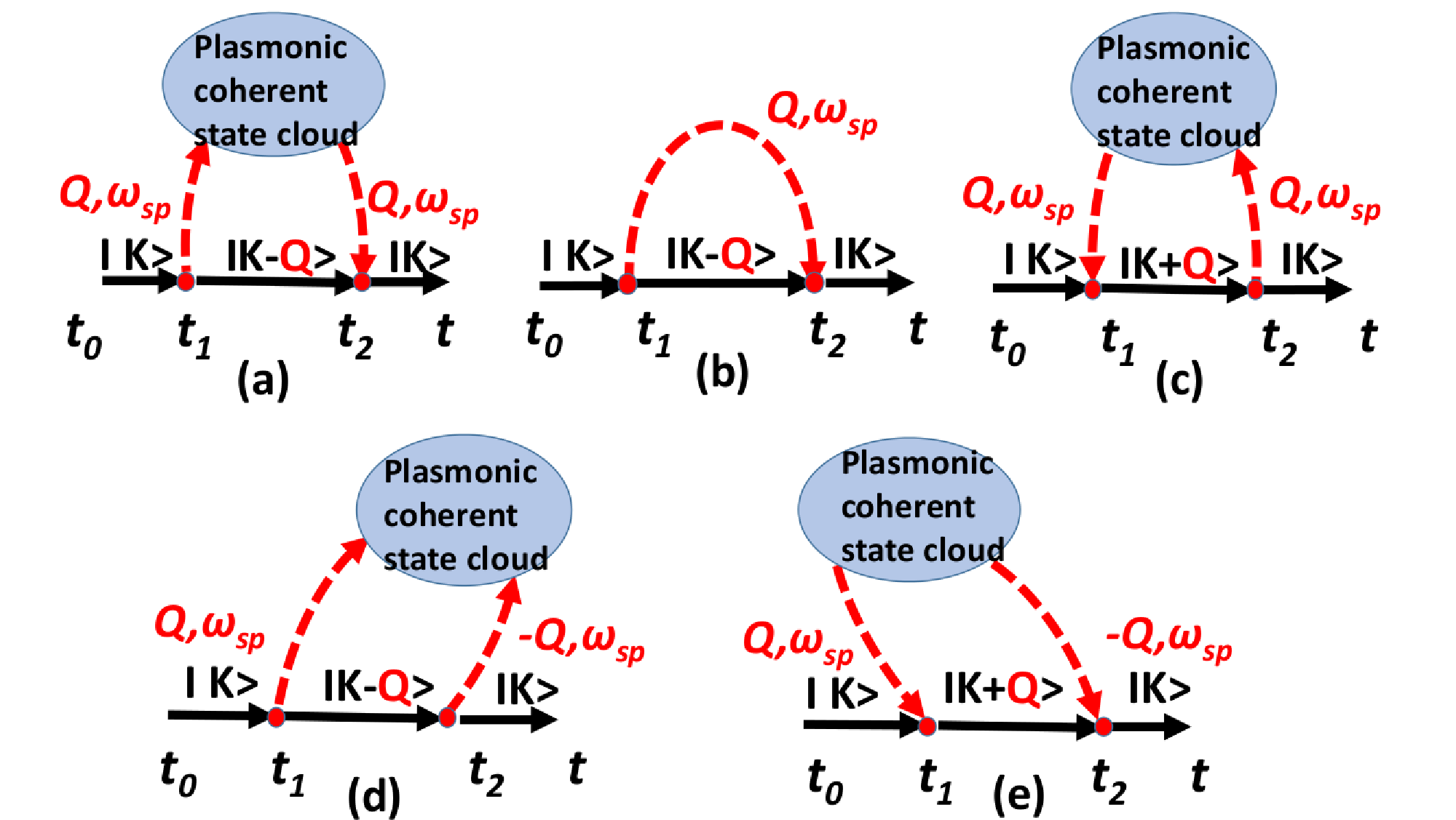}} 
\caption{Diagrams of second order cumulants generated by $V_{I}(t')$ over the state $||{\bf K},s;\mbox{coh}\rangle\rangle$. Full arrows denote electron and dashed red arrows plasmon propagators. Plasmon momentum transfer can be either outgoing, ${\bf Q}$, or incoming, ${\bf -Q}$, with respect to the same vertex. Lateral momentum is conserved in all interaction vertices denoted by red dots. Panels (a) and (b) denote plasmon emission and reabsorption process via the coherent state and directly by closing onto themselves, respectively. Quantum probability of these processes is described by expression (\ref{eq:Ms+}) where the factor $1$ in $(\lambda_{\bf Q}^{2}+1)$ accounts for process (b). Panel (c) describes the reverse process of plasmon absorption by electron from and its consecutive emission back to the coherent state, as described by expression (\ref{eq:Ms-}). Panels (d) and (e) denote two-plasmon exchange (emission and absorption) with opposite ${\bf Q}$. All diagrams are diagonal regarding lateral momentum transfer. Diagrams (d) and (e) are offdiagonal regarding the energy transfer.}
\label{C2lg}
\end{figure}

Standard method for studying the quasiparticle properties is the construction of the corresponding Green's functions.  Their spectral representations describe the dynamics of quasiparticles in interaction with classial and quantized fields and thereby also contain information on electron excitation from the occupied band structure.[\onlinecite{Adawi,Bunkin,Mahan,Ashcroft,Caroli,TimmBennemann}]. In the limit of high excitation energies, also called the sudden approximation, the electron emission yield becomes proportional to the spectrum of initially occcupied quasiparticle states obtained from the imaginary part of the corresponding Green's function.[\onlinecite{PSS84}] This is manifest in both expressions (\ref{eq:1plasPE}) and (\ref{eq:2plasPE}) where the energy conserving $\delta$-functions involve the initial state energies $E_{\bf K}$. In Secs. \ref{sec:EPdynlg} and \ref{sec:EPdynvg} we shall extract the relevant characteristics of plasmoemission yield from the corresponding quasiparticle propagators calculated in the length and velocity gauge for electron-plasmon interactions, respectively.  
    
Information on the spectrum of a single electron in interaction with a plasmonic coherent state cloud as described by the Hamiltonian (\ref{eq:HsystSP}) is obtained from the imaginary part of {\it diagonal} single-electron propagator or retarded Green's function[\onlinecite{Doniach}]  
\barr
G_{{\bf K},s}^{(lg)}(t-t_{0})&=&-i\langle\mbox{coh}|\langle 0|c_{{\bf K},s}(t)c_{{\bf K},s}^{\dag}(t_{0})|0\rangle |\mbox{coh}\rangle\Theta(t-t_{0})\nonumber\\
&=&
-i\Theta(t-t_{0})e^{-iE_{{\bf K},s}(t-t_{0})/\hbar}
\langle\langle\mbox{coh},\phi_{{\bf K},s}||U_{I}(H,t,t_{0})||\phi_{{\bf K},s},\mbox{coh}\rangle\rangle,
\label{eq:GKslength}
\earr
where $(lg)$ denotes the length gauge, $|0\rangle$ is the electron vacuum, $c_{{\bf K},s}^{\dag}(t)$ and $c_{{\bf K},s}(t)$ are the electron creation and annihilation operators for the state $|\phi_{{\bf K},s}\rangle=|{\bf K},s\rangle$ in the Heisenberg picture, respectively. Here ${\bf K}$ is a 2-dimensional lateral (parallel to the surface) electron wavevector, and $\{s\}$ are the quantum numbers describing electron motion perpendicular to the surface (e.g. surface band indices).  The time dependence of $c_{{\bf K},s}^{\dag}(t)$ and $c_{{\bf K},s}(t)$ is driven by the full Hamiltonian $H$, in contrast to $V_{I}(t)$ which is driven by $H_{0}$.  
The spectrum of (\ref{eq:GKslength}) contains information on the complete manifold of electron excitation processes generated by the interaction $V$ into all eigenstates of $H_{0}^{el}$. 
The diagonal matrix element on the RHS of (\ref{eq:GKslength}) is in contrast to expression (\ref{eq:Tlength}) for the $T$-matrix which is obtained from {\it non-diagonal} matrix elements of the evolution operator in either picture.

The average of the evolution operator $\langle\langle\mbox{coh},\phi_{{\bf K},s}||U_{I}(t,t_{0})||\phi_{{\bf K},s},\mbox{coh}\rangle\rangle$ on the RHS of (\ref{eq:GKslength}) can be calculated using cumulant expansion which is a convenient method for the calculations of averages of generalized exponential operators.[\onlinecite{Kubo}] Here it generates specially arranged exponentiated averages of ascending powers of the integrals over the repeated interactions $V_{I}(t_{1})\dots V_{I}(t_{n})$.[\onlinecite{GW+C,Dunn,Mahanbook,4cumul}] The requirement is that the averaging procedure (here implemented through $\langle\langle\mbox{coh},{\bf K},s||\dots||{\bf K},s,\mbox{coh}\rangle\rangle$) preserves unity, i.e. the thus taken average of unity is unity.[\onlinecite{Kubo}] 
This nonperturbative method is fast converging and systematic, and  amenable to diagrammatic approach.[\onlinecite{Dunn,Mahanbook,4cumul}]. Thus, we obtain
\bq
G_{{\bf K},s}^{(lg)}(t-t_{0})=-i\Theta(t-t_{0})e^{-iE_{{\bf K},s}(t-t_{0})/\hbar}\exp[C_{{\bf K},s}^{(lg)}(t-t_{0})].
\label{eq:Gcumlength}
\eq
Here  $E_{{\bf K},s}$ is the quasiparticle energy in the state $|{\bf K},s\rangle$ (and likewise for all bands $H_{0}^{el}|{\bf K},i\rangle=E_{{\bf K},i}|{\bf K},i\rangle$), and
\bq
C_{{\bf K},s}^{(lg)}(t-t_{0})=\sum_{n=1}^{\infty} C_{{\bf K},s}^{(lg,n)}(t-t_{0})
\label{eq:Ccumlength}
\eq
is given by the sum of all order cumulants generated on the state $|{\bf K},s\rangle$ by the powers of electron-plasmon interaction $V_{I}$. Thus, in the propagator formulation (\ref{eq:Gcumlength}) the sums over the matrix elements of the interaction appear in the exponent, which is in contrast to the $T$-matrix formalism outlined in Sec. \ref{sec:TLG}.  However, besides being an expansion in powers of the coupling constant, each $n$-th term in the series (\ref{eq:Ccumlength}) is additionally characterized by another potentially small quantity. This arises from weak momentum and energy correlations or statistical independence in the differences of contributions to each $C_{{\bf K},s}^{(lg,n)}(t-t_{0})$. If these differences are small than a small error is incurred if the series (\ref{eq:Ccumlength}) is truncated beyond the second order term $n=2$.
In one-band models (i.e. $|f\rangle=|s\rangle$) this property holds among the higher order cumulants exactly if there exists translational invariance in the lateral momentum space of both the quasiparticle energies and interaction matrix elements,[\onlinecite{Dunn,Mahanbook,4cumul}] viz.   
\barr
E_{{\bf K+P+Q},3}-E_{{\bf K+P},2}&\leftrightarrow& E_{{\bf K+Q},2}-E_{{\bf K},1},
\label{eq:trans_E}\\
V_{{\bf K+P+Q},3;{\bf K+P},2}&\leftrightarrow& V_{{\bf K+Q},2;{\bf K},1},
\label{eq:trans_V} 
\earr
where the second subscripts $_{1,2,3}$ denote the quantum numbers for motion in the $z$-direction.
This is trivially satisfied for quasiparticles with infinite mass (e.g. core electrons or holes) for which the second order cumulant provides an exact solution, see Appendix \ref{sec:Isoenergetic}.

In the following we shall assume the weakly correlated scattering regime in which the lowest order cumulants $C_{1}^{(lg)}({\bf K},s,t-t_{0})$ and $C_{2}^{(lg)}({\bf K},s,t-t_{0})$ that are linear and quadratic in $\hat{a}_{\bf Q}$ or $\hat{a}_{\bf Q'}^{\dag}$, respectively, give the major contribution to the cumulant sum (\ref{eq:Ccumlength}) generated by the length gauge interaction $V_{I}(t')$. Taking into account only these contributions brings (\ref{eq:Gcumlength}) to the form 
\bq
G_{{\bf K},s}^{(lg)}(t-t_{0})=
-i\Theta(t-t_{0}) e^{-\frac{i}{\hbar}E_{{\bf K},s}(t-t_{0})}
\exp[C_{1}^{(lg)}({\bf K},s,t-t_{0})+C_{2}^{(lg)}({\bf K},s,t-t_{0})]. 
\label{eq:GKslengthC}
\eq

For the sake of simplicity we shall in the forthcoming calculations use the nonretarded longwavelength limit of surface plasmon frequency $\omega_{\bf Q}\simeq \omega_{sp}$ unless stated otherwise.
Owing to the symmetry of the considered problem that is reflected in the wavefunctions (\ref{eq:phi_s}) and (\ref{eq:phifin}) the lateral momentum is conserved in the interaction vertices generated by $V$. This implies the form of cumulants comprising emission and absorption of plasmons of the same wavevector, or double emission or double absorption of plasmons of opposite wavevector (see Fig. \ref{C2lg}). In the present case this yields 
\bq
C_{1}^{(lg)}({\bf K},s,t-t_{0})=0.
\label{eq:C1=0}
\eq
The sum of second order cumulants 
\bq
C_{2}^{(lg)}({\bf K},s,t-t_{0})=C_{2}^{(lg)}(1\omega_{sp};{\bf K},s,t-t_{0})+ C_{2}^{(lg)}(2\omega_{sp};{\bf K},s,t-t_{0}),
\label{eq:C2}
\eq
that describes one-plasmon $(1\omega_{sp})$ and two-plasmon $(2\omega_{sp})$ exchange processes is calculated following the rules outlined in Ref. [\onlinecite{4cumul}]. After performing $\int d^{2}\brho$ integration which yields lateral momentum conservation arising from the matrix elements (\ref{eq:VQr}) and (\ref{eq:VdagV}) we obtain
\begin{widetext}
\barr
C_{2}^{(lg)}(1\omega_{sp};{\bf K},s,t-t_{0})
&=&
-\frac{1}{\hbar^{2}}\sum_{{\bf Q},f}|\langle f|V_{\bf Q}(z)|s\rangle|^{2}\int_{t_{0}}^{t}dt_{2}\int_{t_{0}}^{t_{2}}dt_{1}\nonumber\\
&\times&
\left\{(1+\lambda_{\bf Q}^{2})\exp[i(E_{{\bf K},s}-E_{{\bf K-Q},f}-\hbar\omega_{sp})t_{2}/\hbar]
\exp[-i(E_{{\bf K},s}-E_{{\bf K-Q},f}-\hbar\omega_{sp})t_{1}/\hbar]\right.\nonumber\\
&+& 
\lambda_{\bf Q}^{2}\left.\exp[i(E_{{\bf K},s}-E_{{\bf K-Q},f}+\hbar\omega_{sp})t_{2}/\hbar]
\exp[-i(E_{{\bf K},s}-E_{{\bf K-Q},f}+\hbar\omega_{sp})t_{1}/\hbar]\right\},
\label{eq:C1omegas},
\earr
\end{widetext} 
where $s$ and $f$ stand for the initial and final state quantum numbers describing electron motion in the $z$-direction. The exponentials in upper and lower line describe one-plasmon emission and reabsorption (Figs. \ref{C2lg}(a) and (b)) and one plasmon absorption and re-emission processes (Fig. \ref{C2lg}(c)), respectively. Analogously 
\begin{widetext}
\barr
C_{2}^{(lg)}(2\omega_{sp};{\bf K},s,t-t_{0})
&=&
-\frac{1}{\hbar^{2}}\sum_{{\bf Q},f}\lambda_{\bf Q}^{2}|\langle f|V_{\bf Q}(z)|s\rangle|^{2}\int_{t_{0}}^{t}dt_{2}\int_{t_{0}}^{t_{2}}dt_{1}\nonumber\\
&\times&
\left\{\exp[i(E_{{\bf K},s}-E_{{\bf K-Q},f}+\hbar\omega_{sp})t_{2}/\hbar]
\exp[-i(E_{{\bf K},s}-E_{{\bf K-Q},f}-\hbar\omega_{sp})t_{1}/\hbar]\right.\nonumber\\
&+&
\left.\exp[i(E_{{\bf K},s}-E_{{\bf K-Q},f}-\hbar\omega_{sp})t_{2}/\hbar]
\exp[-i(E_{{\bf K},s}-E_{{\bf K-Q},f}+\hbar\omega_{sp})t_{1}/\hbar]
\right\},
\label{eq:C2omegas}
\earr
\end{widetext} 
where the exponentials in upper and lower line correspond to two-plasmon emission and absorption processes depicted in Fig. \ref{C2lg}(d) and (e), respectively. 

Owing to the zero net energy transfer past the interaction interval, the processes represented by the diagrams (a)-(c) of Fig. \ref{C2lg} and expression (\ref{eq:C1omegas}) can be assigned to the self-energy class. Conversely, expression (\ref{eq:C2omegas}) arising from the diagrams (d)-(e) of Fig. \ref{C2lg}  and inducing $\pm 2\omega_{sp}$ net energy transfer, can be assigned to the vertex correction class. 

Lastly, we reiterate that lateral momentum conservation is treated exactly in the cumulants shown in Fig. \ref{C2lg} that describe lowest order scattering amplitudes. Since in the truncated cumulant form of the propagator (\ref{eq:GKslengthC}) the  amplitudes of higher order processes are approximated by exponentiation of lowest order cumulants, the lateral momenta between successive higher order scattering events are uncorrelated or independent.

In the following we shall investigate two opposite cases of switching on of the e-SP interaction $V$, viz. \\
{\it (i)} $V$ is switched on suddenly at the time instant $t_{0}$ with the electron injection into the system[\onlinecite{Doniach,GW+C}], and \\
{\it (ii)} $V$ is switched on adiabatically starting from $t_{0}\rightarrow-\infty$. This complies with the scattering boundary conditions.



\subsection{Sudden switching on of the interaction ${\cal V}$}
\label{sec:suddenV}

\subsubsection{Transient $1\omega_{sp}$-processes}
\label{sec:331}

We first inspect the limit of sudden or nonadiabatic switching on of the electron-plasmon potential $V$ 
at $t=t_{0}$ which occurs with electron injection into the plasmonically prepumped system.[\onlinecite{Doniach,GW+C}]  For later convenience  we shall make the notation more compact by setting  $t_{0}=0$ in (\ref{eq:GKslength}) and all ensuing expressions.

 With the electron-plasmon coupling described by (\ref{eq:V}) the first order cumulant is zero due to requirement of the lateral momentum conservation.
The second order cumulant (\ref{eq:C1omegas}) describes the processes of one-plasmon emissions(or absorptions) that is followed by their consecutive absorptions (emissions). Their diagrammatic representation is given in Fig. \ref{C2lg} (a)-(c) and they all obey total lateral momentum conservation. The one-plasmon emission and reabsorption processes  from Fig. \ref{C2lg}(a)  proceed via the coherent state cloud whereas that from Fig. \ref{C2lg}(b) proceeds directly. They give the total contribution    
\begin{widetext}
\barr
 C_{2}^{(lg)}(1\omega_{sp};{\bf K},s,t)
=
&-&
\left[{\cal M}_{{\bf K},s}^{(lg)}(\mbox{em1abs1};0)+{\cal M}_{{\bf K},s}^{(lg)}(\mbox{abs1em1};0)\right]\nonumber\\
&+&
\left[{\cal M}_{{\bf K},s}^{(lg)}(\mbox{em1abs1};t)e^{-i\omega_{sp}t}
+
{\cal M}_{{\bf K},s}^{(lg)}(\mbox{abs1em1};t)e^{i\omega_{sp}t}\right]
-
i\Omega_{{\bf K},s}^{(lg)}t,
\label{eq:C2lg1omega}
\earr 
\end{widetext}
where the symbol "$\mbox{em1abs1}$" denotes the temporal sequence of one-plasmon emission followed by one-plasmon absorption process, and the symbol "$\mbox{abs1em1}$" for the reverse order. Here the amplitudes of the one plasmon creation and subsequent absorption processes are described by 
%
\bq
{\cal M}_{{\bf K},s}^{(lg)}(\mbox{em1abs1};t)=\sum_{f,{\bf Q}}(1+\lambda_{\bf Q}^{2})\frac{|\langle f|V_{\bf Q}(z)|s\rangle|^{2}}{(E_{{\bf K},s}-E_{{\bf K-Q},f}-\hbar\omega_{sp})^{2}}
e^{i(E_{{\bf K},s}-E_{{\bf K-Q},f})t/\hbar}.
\label{eq:Ms+}
\eq
%
Analogously, the amplitude of one plasmon absorption and reemission  via the plasmonic coherent state cloud (cf. Fig. \ref{C2lg}(c)) is described by
%
\bq
{\cal M}_{{\bf K},s}^{(lg)}(\mbox{abs1em1};t)=\sum_{f,{\bf Q}}\lambda_{\bf Q}^{2}\frac{|\langle f|V_{\bf Q}(z)|s\rangle|^{2}}{(E_{{\bf K},s}-E_{{\bf K-Q},f}+\hbar\omega_{sp})^{2}}
e^{i(E_{{\bf K},s}-E_{{\bf K-Q},f})t/\hbar}.
\label{eq:Ms-}
\eq
%
These forms of  plasmon creation (\ref{eq:Ms+}) and absorption probabilities (\ref{eq:Ms-}) are typical of the regime of sudden or nonadiabatic switching on of the interaction in (\ref{eq:GKslength}) upon electron injection into the state $|{\bf K},s\rangle$ at the time $t_{0}=0$. 
The last term on the RHS of (\ref{eq:C2lg1omega}) describes the energy shift given by the Rayleigh-Schr\"{o}dinger (RS) correction to the polarization energy of the electronic $s$-level derived from the real parts of the $({\bf K},s)$-level self-energy:
\bq
\hbar\Omega_{{\bf K},s}^{(lg)}=\mbox{Re}\Sigma_{{\bf K},s}^{(lg)}(1\omega_{sp})
=
\mbox{Re}\Sigma_{{\bf K},s}^{(lg)}(\mbox{vac})+\mbox{Re}\Sigma_{{\bf K},s}^{(lg)}(\mbox{em1})+\mbox{Re}\Sigma_{{\bf K},s}^{(lg)}(\mbox{abs1})
=
\hbar\Omega_{{\bf K},s}^{(lg)}(\mbox{vac})+\hbar\Omega_{{\bf K},s}^{(lg)}(\mbox{em})+\hbar\Omega_{{\bf K},s}^{(lg)}(\mbox{abs}),
\label{eq:Omegas}
\eq
where
\barr
& &\mbox{Re}\Sigma_{{\bf K},s}^{(lg)}(\mbox{vac})=\sum_{f,{\bf Q}}\frac{|\langle f|V_{\bf Q}(z)|s\rangle |^{2}}{(E_{{\bf K},s}-E_{{\bf K-Q},f}-\hbar\omega_{sp})},
\label{eq:DeltaKs}\\
& &\mbox{Re}\Sigma_{{\bf K},s}^{(lg)}(\mbox{em1})=\sum_{f,{\bf Q}}\lambda_{\bf Q}^{2}\frac{|\langle f|V_{\bf Q}(z)|s\rangle |^{2}}{(E_{{\bf K},s}-E_{{\bf K-Q},f}-\hbar\omega_{sp})},
\label{eq:Omegas+}\\
& &\mbox{Re}\Sigma_{{\bf K},s}^{(lg)}(\mbox{abs1})=\sum_{f,{\bf Q}}\lambda_{\bf Q}^{2}\frac{|\langle f|V_{\bf Q}(z)|s\rangle |^{2}}{(E_{{\bf K},s}-E_{{\bf K-Q},f}+\hbar\omega_{sp})}.
\label{eq:Omegas-}
\earr

The corresponding imaginary parts read
\begin{widetext}
\barr
\Gamma_{{\bf K},s}^{(lg)}(\mbox{vac})&=&-\mbox{Im}\Sigma_{{\bf K},s}^{(lg)}(\mbox{vac})=\frac{\pi}{\hbar}\sum_{{\bf Q},f}|\langle f|V_{\bf Q}(z)|s\rangle|^{2}
\delta({E_{{\bf K},s}-E_{{\bf K-Q},f}-\hbar\omega_{sp}})
\label{eq:Gammavac}\\
\Gamma_{{\bf K},s}^{(lg)}(\mbox{em1})&=&-\mbox{Im}\Sigma_{{\bf K},s}^{(lg)}(\mbox{em1})=\frac{\pi}{\hbar}\sum_{{\bf Q},f}\lambda_{\bf Q}^{2}|\langle f|V_{\bf Q}(z)|s\rangle|^{2}
\delta({E_{{\bf K},s}-E_{{\bf K-Q},f}-\hbar\omega_{sp}})
\label{eq:Gamma1em}\\
\Gamma_{{\bf K},s}^{(lg)}(\mbox{abs1})&=&-\mbox{Im}\Sigma_{{\bf K},s}^{(lg)}(\mbox{abs1})=\frac{\pi}{\hbar}\sum_{{\bf Q},f}\lambda_{\bf Q}^{2}|\langle f|V_{\bf Q}(z)|s\rangle|^{2}
\delta(E_{{\bf K},s}-E_{{\bf K-Q},f}+\hbar\omega_{sp}),
\label{eq:Gamma1abs}
\earr
\end{widetext}
so that the total decay rate is
\bq
\Gamma_{{\bf K},s}^{(lg)}(1\omega_{sp})=
\Gamma_{{\bf K},s}^{(lg)}(\mbox{vac})+\Gamma_{{\bf K},s}^{(lg)}(\mbox{em1})+\Gamma_{{\bf K},s}^{(lg)}(\mbox{abs1}).
\label{eq:Gamma0+-}
\eq
The first two terms on the RHS of (\ref{eq:Gamma0+-}) describe one-plasmon emissions over the ground and coherent state, respectively, and the third term describes one plasmon absorptions over the coherent state.

The electron spectrum is given by the Fourier transform of (\ref{eq:GKslength}) obtained by using (\ref{eq:Ms+}), (\ref{eq:Ms-}), (\ref{eq:Omegas}) but with $\Theta(t-t_{0})$ removed. This amounts to calculating the time integral of an exponential function where in the exponent there are two exponential functions from the middle term in (\ref{eq:C2lg1omega}). This can be very tedious but the procedure can be formally made tractable by introducing the transformations detailed in Sec. 5.2.3 of Ref. [\onlinecite{PhysRep}]. This leads to the representation of the exponentiated second square bracket in (\ref{eq:C2lg1omega}) in the form 
\begin{widetext}
\barr
& &\exp\left[{\cal M}_{{\bf K},s}^{(lg)}(\mbox{em1abs1};t)e^{-i\omega_{sp}t}+{\cal M}_{{\bf K},s}^{(lg)}(\mbox{abs1em1};t)e^{i\omega_{sp}t}\right]\nonumber\\
&=&
\exp\left[2\sqrt{{\cal M}_{{\bf K},s}^{(lg)}(\mbox{em1abs1};t){\cal M}_{{\bf K},s}^{(lg)}(\mbox{abs1em1};t)}\cos\left(\omega_{sp}t+i\ln\sqrt{\frac{{\cal M}_{{\bf K},s}^{(lg)}(\mbox{em1abs1};t)}{{\cal M}_{{\bf K},s}^{(lg)}(\mbox{abs1em1};t)}}\right)\right].
\label{eq:AB(t)}
\earr
\end{widetext}
Introducing the generating functions for Bessel functions 
\barr
 \exp(iz\sin\varphi)&=&\sum_{n=-\infty}^{\infty}e^{in\varphi}J_{n}(z),
\label{eq:BesselJ}\\
\exp(z\cos\varphi)&=&\sum_{n=-\infty}^{\infty}e^{in\varphi}I_{n}(z),
\label{eq:BesselI}
\earr
where for integer $n$
\barr
J_{n}(iz)&=&i^{n}I_{n}(z),
\label{eq:JI}\\
J_{-n}(z)&=&(-1)^{n}J_{n}(z),
\label{eq:J-n}\\
J_{n}(-z)&=&(-1)^{n}J_{n}(z),
\label{eq:Jn-}\\
I_{-n}(z)&=&I_{n}(z),
\label{eq:I-n}\\
I_{n}(-z)&=&(-1)^{n}I_{n}(z),
\label{eq:In-}
\earr
we obtain one-plasmon generated second order cumulant contribution to the single electron propagator (\ref{eq:Gcumlength}), viz.  
\begin{widetext}
\bq
e^{C_{2}^{(lg)}(1\omega_{sp};{\bf K},s,t)}=e^{-i\Omega_{{\bf K},s}^{(lg)}t}
e^{-2v_{{\bf K},s}^{(lg)}(1\omega_{sp})}
\sum_{n=-\infty}^{\infty}e^{in\omega_{sp}t}\left(\sqrt{\frac{{\cal M}_{{\bf K},s}^{(lg)}(\mbox{abs1em1};t)}{{\cal M}_{{\bf K},s}^{(lg)}(\mbox{em1abs1};t)}}\right)^{n}I_{n}\left(2\sqrt{{\cal M}_{{\bf K},s}^{(lg)}(\mbox{em1abs1};t){\cal M}_{{\bf K},s}^{(lg)}(\mbox{abs1em1};t)}\right)
\label{eq:expAB(t)}.
\eq
\end{widetext}
 The exponent in the prefactor on the RHS of (\ref{eq:expAB(t)})
\bq
2v_{{\bf K},s}^{(lg)}(1\omega_{sp})=\left[{\cal M}_{{\bf K},s}^{(lg)}(\mbox{em1abs1};0)+{\cal M}_{{\bf K},s}^{(lg)}(\mbox{abs1em1};0)\right]  
\label{eq:2vs}
\eq
is the Debye-Waller exponent (DWE) which gives the overall normalization of the spectrum. Hence, the first term in the expansion of $I_{-1}(x)$ and $I_{1}(x)$ gives the unitarized first order Born approximation result for the processes depicted in Fig. \ref{C2lg}(a)-(c).

The Fourier transform of $e^{-iE_{{\bf K},s}t/\hbar}\exp\left[C_{2}^{(lg)}(1\omega_{sp};{\bf K},s,t)\right]$ gives the electron spectrum within the approximation of uncorrelated multiplasmon exchange embodied in (\ref{eq:expAB(t)}). It is characterized by the Floquet or multiplasmon structure generated by the terms $e^{in\omega_{sp}t}$ on the RHS of (\ref{eq:expAB(t)}). The other factors determine the positions and weights 
of the peaks in this structure. Their time dependences are affected by the electron recoil energies $E_{{\bf K},s}-E_{{\bf K-Q},f}$. For localized electrons $E_{{\bf K-Q},f}=E_{{\bf K},s}$ and hence  ${\cal M}_{{\bf K},s}^{(\pm1\omega_{sp})}$ become time independent. In this limit the spectrum consists of discrete equidistant $\delta$-peaks weighted by the static limit of the last two factors in the sum on the RHS of (\ref{eq:expAB(t)}). 

\vskip 3mm

\subsubsection{Transient $2\omega_{sp}$-processes}
\label{sec:332}

\vskip 2mm

In addition to the processes described by (\ref{eq:expAB(t)}), the momentum-correlated consecutive emission or absorption of two plasmons of opposite wavevector (see panels (d) and (e) in Fig. \ref{C2lg}) give rise to extra highly nontrivial second order cumulants generated in the expansion of (\ref{eq:Gcumlength}). These are important contributions that may lead to ponderomotive effects encounterred in strong field theories.
Such effects ought to be considered in the context of cumulant averaging over coherent states intended to describe the classical field limit. Their seeming off-diagonality in plasmon exchange and energy transfer is compensated by the coherent state averaging, likewise in (\ref{eq:cohaverage1})-(\ref{eq:epscoh++--}).  Taking into account (\ref{eq:VdagV}) and $|\langle {\bf K-Q},f|V({\bf -Q,r})|{\bf K},s\rangle|^{2}\leftrightarrow |\langle {\bf K-Q},f|V^{\dag}({\bf Q,r})|{\bf K},s\rangle|^{2}$, expression (\ref{eq:C2omegas}) arising from the diagrams in Fig. \ref{C2lg}(d) and \ref{C2lg}(e) for $t_{0}=0$ produce  
\begin{widetext}
\barr
C_{2}^{(lg)}(2\omega_{sp};{\bf K},s,t)&=&\sum_{{\bf Q},f}\lambda_{\bf Q}^{2}|\langle f|V_{\bf Q}(z)|s\rangle|^{2}\nonumber\\
&\times&
\left[\left(\frac{e^{2i\omega_{sp}t}-1}{(-2\hbar\omega_{sp})(E_{{\bf K},s}-E_{{\bf K-Q},f}-\hbar\omega_{sp})}
+
\frac{e^{i(E_{{\bf K},s}-E_{{\bf K-Q},f}+\hbar\omega_{sp})t/\hbar}-1}{(E_{{\bf K},s}-E_{{\bf K-Q},f})^{2}-(\hbar\omega_{sp})^{2}}  \right)_{\mbox{em}}
\right.                              
\label{eq:C2omegas+}\\
&+&
\left.
\left(\frac{e^{-2i\omega_{sp}t}-1}{(+2\hbar\omega_{sp})(E_{{\bf K},s}-E_{{\bf K-Q},f}+\hbar\omega_{sp})}
+
\frac{e^{i(E_{{\bf K},s}-E_{{\bf K-Q},f}-\hbar\omega_{sp})t/\hbar}-1}{(E_{{\bf K},s}-E_{{\bf K-Q},f})^{2}-(\hbar\omega_{sp})^{2}}\right)_{\mbox{abs}}\right],
\label{eq:C2omegas-}\\
&=&
\sum_{{\bf Q},f}\lambda_{\bf Q}^{2}|\langle f|V_{\bf Q}(z)|s\rangle|^{2}\nonumber\\
&\times&
\left[\frac{1}{\Delta_{{\bf K},s}(\mbox{em})}\left(\frac{1-e^{2i\omega_{sp}t}}{2\hbar\omega_{sp}}
+
\frac{1-e^{i\Delta_{{\bf K},s}(\mbox{abs})t/\hbar}}{(-\Delta_{{\bf K},s}(\mbox{abs}))}\right)                              
+
\frac{1}{\Delta_{{\bf K},s}(\mbox{abs})}\left(\frac{1-e^{-2i\omega_{sp}t}}{(-2\hbar\omega_{sp})}
+
\frac{1-e^{i\Delta_{{\bf K},s}(\mbox{em})t/\hbar}}{(-\Delta_{{\bf K},s}(\mbox{em}))}\right)\right],
\label{eq:C2Delta}
\earr
\end{widetext} 
where we have used the abbreviations
\barr
\Delta_{{\bf K},s}(\mbox{em})=E_{{\bf K},s}-E_{{\bf K-Q},f}-\hbar\omega_{sp},
\label{eq:Deltaem}\\
\Delta_{{\bf K},s}(\mbox{abs})=E_{{\bf K},s}-E_{{\bf K-Q},f}+\hbar\omega_{sp}.
\label{eq:Deltaabs}
\earr
The various terms in (\ref{eq:C2omegas+}), (\ref{eq:C2omegas-}) and (\ref{eq:C2Delta}) have been arranged such that (\ref{eq:C2omegas+}) and the first term in (\ref{eq:C2Delta}) signify the two-plasmon emission and  (\ref{eq:C2omegas-}) and the second term in (\ref{eq:C2Delta}) the two-plasmon absorption by the electron. 

The two terms in the square brackets of (\ref{eq:C2Delta}) are not simple complex conjugates of each other and to find the Floquet representation of $\exp[C_{2}^{(lg)}(2\omega_{sp};{\bf K},s,t)]$ analogous to (\ref{eq:expAB(t)}) we use the previous notation to write it in abbreviated form  
\begin{widetext}
\bq
\exp[C_{2}^{(lg)}(2\omega_{sp};{\bf K},s,t)]
=
e^{-2v_{{\bf K}.s}^{(lg)}(2\omega_{sp})}\exp\left[-\left({\cal P}_{{\bf K},s}^{(lg)}(\mbox{em}2)e^{2i\omega_{sp}t}-{\cal P}_{{\bf K},s}^{(lg)}(\mbox{abs}2)e^{-2i\omega_{sp}t}\right)+{\cal Q}_{{\bf K},s}^{(lg)}(t)(e^{i\omega_{sp}t}+e^{-i\omega_{sp}t})\right]
\label{eq:C2omega}
\eq
\end{widetext}
where the entries are easily deduced from (\ref{eq:C2omegas+}) and (\ref{eq:C2omegas-}):
\begin{widetext}
\barr
{\cal P}_{{\bf K},s}^{(lg)}(\mbox{em}2)&=&\sum_{{\bf Q},f}\lambda_{\bf Q}^{2}\frac{|\langle f|V{\bf Q}(z)|s\rangle|^{2}}{2\hbar\omega_{sp}(E_{{\bf K},s}-E_{{\bf K-Q},f}-\hbar\omega_{sp})}
=
\frac{\Omega_{{\bf K},s}^{(lg)}(\mbox{em})}{2\omega_{sp}}=\frac{\mbox{Re}\Sigma_{{\bf K},s}^{(lg)}(\mbox{em})}{2\hbar\omega_{sp}},
\label{eq:calPem}\\
{\cal P}_{{\bf K},s}^{(lg)}(\mbox{abs}2)&=&\sum_{{\bf Q},f}\lambda_{\bf Q}^{2}\frac{|\langle f|V{\bf Q}(z)|s\rangle|^{2}}{2\hbar\omega_{sp}(E_{{\bf K},s}-E_{{\bf K-Q},f}+\hbar\omega_{sp})}
=
\frac{\Omega_{{\bf K},s}^{(lg)}(\mbox{abs})}{2\omega_{sp}}=\frac{\mbox{Re}\Sigma_{{\bf K},s}^{(lg)}(\mbox{abs})}{2\hbar\omega_{sp}},
\label{eq:calPabs}\\
{\cal Q}_{{\bf K},s}^{(lg)}(t)&=&\sum_{{\bf Q},f}\lambda_{\bf Q}^{2}
\frac{|\langle f|V_{\bf Q}(z)|s\rangle|^{2}}{(E_{{\bf K},s}-E_{{\bf K-Q},f})^{2}-(\hbar\omega_{sp})^{2}}
\times
e^{i(E_{{\bf K},s}-E_{{\bf K-Q},f})t/\hbar},
\label{eq:calQ}\\
2v_{{\bf K},s}^{(lg)}(2\omega_{sp})&=&\left[-\left({\cal P}_{{\bf K},s}^{(lg)}(\mbox{em2})-{\cal P}_{{\bf K},s}^{(lg)}(\mbox{abs2})\right)+2{\cal Q}_{{\bf K},s}^{(\lg)}(0)\right].
\label{eq:v2omega}
\earr
\end{widetext}
The cumulant $C_{2}^{(lg)}(2\omega_{sp};{\bf K},s,t)$ in the exponent on the LHS of (\ref{eq:C2omega}) is the lowest order contribution quadratic in the interaction and contating the terms oscillating with $\pm 2\omega_{sp}$. 
To obtain the Floquet representation of full (\ref{eq:C2}) we again bring (\ref{eq:C2omega}) to the form of exponential function of trigonometric functions that are representable in terms of Bessel functions. Using (\ref{eq:C2lg1omega}) and (\ref{eq:C2omega}) we can write ($t_{0}=0$)
\begin{widetext}
\barr
e^{C_{2}^{(lg)}({\bf K},s,t)}&=&\exp\left[C_{2}^{(lg)}(1\omega_{sp};{\bf K},s,t)+ C_{2}^{(lg)}(2\omega_{sp};{\bf K},s,t)\right]=
e^{-i\Omega_{{\bf K},s}^{(lg)}t}e^{-2\left(v_{{\bf K},s}^{(lg)}(1\omega_{sp})+v_{{\bf K},s}^{(lg)}(2\omega_{sp})\right)}\nonumber\\
&\times&
\exp\left[\left({\cal M}_{{\bf K},s}^{(lg)}(\mbox{em1abs1};t)+{\cal Q}_{{\bf K},s}^{(lg)}(t)\right)e^{-i\omega_{sp}t}+ \left({\cal M}_{{\bf K},s}^{(lg)}(\mbox{abs1em1};t)+{\cal Q}_{{\bf K},s}^{(lg)}(t)\right)e^{i\omega_{sp}t}\right]\nonumber\\
&\times&
\exp\left[-\left({\cal P}_{{\bf K},s}^{(lg)}(\mbox{em}2)e^{2i\omega_{sp}t}-{\cal P}_{{\bf K},s}^{(lg)}(\mbox{abs}2)e^{-2i\omega_{sp}t}\right)\right]
\label{eq:expC2}
\earr
\end{widetext}

The Floquet representation of the $e^{\mp i\omega_{sp}t}$ components in the second exponential on the RHS of (\ref{eq:expC2}) can be immediately written down using the forms of (\ref{eq:expAB(t)}) and (\ref{eq:2vs}) with the replacement 
\bq
{\cal M}_{{\bf K},s}^{(lg)}(\mbox{em};t)\rightarrow {\cal M}_{{\bf K},s}^{(lg)}(\mbox{em};t)+{\cal Q}_{{\bf K},s}^{(lg)}(t),
\label{eq:MMQ}
\eq
and analogously so for ${\cal M}_{{\bf K},s}^{(lg)}(\mbox{abs};t)$. 

The last exponential on the RHS of (\ref{eq:expC2}) describing $2\omega_{sp}$-processes can be straightforwardly  brought to the form analogous to (\ref{eq:AB(t)}) describing $1\omega_{sp}$-processes. Using (\ref{eq:BesselJ}) we obtain the generating Bessel function representation
\begin{widetext}
\barr
& &\exp\left[-\left({\cal P}_{{\bf K},s}^{(lg)}(\mbox{em}2)e^{2i\omega_{sp}t}-{\cal P}_{{\bf K},s}^{(lg)}(\mbox{abs}2)e^{-2i\omega_{sp}t}\right)\right]\nonumber\\
&=&
\exp\left[-2i\sqrt{{\cal P}_{{\bf K},s}^{(lg)}(\mbox{em}2){\cal P}_{{\bf K},s}^{(lg)}(\mbox{abs}2)}\sin\left(2\omega_{sp}t-i\ln\sqrt{{\cal P}_{{\bf K},s}^{(lg)}(\mbox{em}2)/{\cal P}_{{\bf K},s}^{(lg)}(\mbox{abs}2)}\right)\right]\nonumber\\
&=&
\sum_{n=-\infty}^{\infty}e^{2i n\omega_{sp}t}\left(\sqrt{\frac{{\cal P}_{{\bf K},s}^{(lg)}(\mbox{em}2)}{{\cal P}_{{\bf K},s}^{(lg)}(\mbox{abs}2)}}\right)^{n}J_{n}\left(-2\sqrt{{\cal P}_{{\bf K},s}^{(lg)}(\mbox{em}2){\cal P}_{{\bf K},s}^{(lg)}(\mbox{abs}2)}\right).
\label{eq:C22omega}
\earr
\end{widetext}

Expression (\ref{eq:expC2}) appears as a product of two generating functions for Bessel functions that can be unified using (\ref{eq:JI}) by introducing the generalized Bessel function of two arguments (see Eq. \ref{eq:genJ}). After Fourier transforming to the $\omega$-space to obtain the spectrum of (\ref{eq:Gcumlength}) this becomes a convolution of repeated $1\omega_{sp}$ and $2\omega_{sp}$ processes. Hence, its form signifies an interplay of generic  self-energy- and vertex-like corrections appearing in Fig. \ref{C2lg}(a)-(c) and (d)-(e), respectively. The forms of these vertex corrections as expressed through (\ref{eq:calPem}) and (\ref{eq:calPabs}) are reminiscent of the Ward-Pitaevskii identities[\onlinecite{Nozieres}] but with the differentials replaced by energy differences.

A special limit of (\ref{eq:expC2}) in quasiparticle transitions between iso-energetic levels is outlined in Appendix \ref{sec:Isoenergetic}.


\subsection{Adiabatic switching on of the interaction ${\cal V}$}
\label{sec:adiabaticV}

\subsubsection{Adiabatic $1\omega_{sp}$-processes}
\label{sec:341}

Applying the SBC limit or adiabatic switching on of the interaction by letting $t_{0}=-\infty$ in (\ref{eq:C1omegas}) we obtain  
\begin{widetext}
\barr
C_{2}^{(lg)}(1\omega_{sp};{\bf K},s,t)&=&-\frac{i}{\hbar}\sum_{{\bf Q},f}|\langle f|V_{\bf Q}(z)|s\rangle|^{2}
\left[\frac{(1+\lambda_{\bf Q}^{2})}{E_{{\bf K},s}-E_{{\bf K-Q},f}-\hbar\omega_{sp}+i\delta}+\frac{\lambda_{\bf Q}^{2}}{E_{{\bf K},s}-E_{{\bf K-Q},f}+\hbar\omega_{sp}+i\delta}\right]t\nonumber\\
&=&
=-\frac{i}{\hbar}\left[\Sigma_{{\bf K},s}^{(lg)}(\mbox{vac})+\Sigma_{{\bf K},s}^{(lg)}(\mbox{em1})+\Sigma_{{\bf K},s}^{(lg)}(\mbox{abs1})\right]t
=-\frac{i}{\hbar}\Sigma_{{\bf K},s}^{(lg)}t=-i\Omega_{{\bf K},s}^{(lg)}t-\Gamma_{{\bf K},s}^{(lg)}(1\omega_{sp}) t.
\label{eq:C1ad}
\earr
\end{widetext}
Here the principal part of the expression in the square brackets on the  RHS side produces the RS energy shift $\Omega_{{\bf K},s}^{(lg)}$ of the $E_{{\bf K},s}$-level defined in (\ref{eq:Omegas}), whereas the imaginary part gives rise to the Fermi golden rule (FGR) type of  plasmonically induced decay of the $|{\bf K},s\rangle$ state by one-plasmon emission or absorption via the channels (\ref{eq:Gammavac})-(\ref{eq:Gamma1abs}).
%


\subsubsection{Adiabatic $2\omega_{sp}$-processes}
\label{sec:342}

The adiabatic limit of $2\omega_{sp}$-contribution to second order cumulant (\ref{eq:C2}) is obtained from the SBC limit of (\ref{eq:C2omegas})
\begin{widetext}
\bq
C_{2}^{(lg)}(2\omega_{sp};{\bf K},s,t)
=
-\left({\cal P}_{{\bf K},s}^{(lg)}(\mbox{em}2)e^{2i\omega_{sp}t}-{\cal P}_{{\bf K},s}^{(lg)}(\mbox{abs}2)e^{-2i\omega_{sp}t}\right)
\label{eq:C2lglambda}\\
=
-\left(\frac{\Sigma_{{\bf K},s}^{(lg)}(\mbox{em1})}{2\hbar\omega_{sp}}e^{2i\omega_{sp}t}-\frac{\Sigma_{{\bf K},s}^{(lg)}(\mbox{abs1})}{2\hbar\omega_{sp}}e^{-2i\omega_{sp}t}\right).
\label{eq:C2lgSigma}
\eq
\end{widetext}
Here the extension of ${\cal P}_{{\bf K},s}^{(lg)}(\mbox{em}2)$ and ${\cal P}_{{\bf K},s}^{(lg)}(\mbox{abs}2)$ to the complex plane through the appearance of $i\delta$ in the corresponding denominators in (\ref{eq:calPem}) and (\ref{eq:calPabs}) has resulted in  the appearance of full coherent state-generated self-energies $\propto \lambda_{\bf Q}^{2}$ in (\ref{eq:C2lglambda}). Hence, expression (\ref{eq:C2lgSigma})  can be represented by the same generating function yielding (\ref{eq:C22omega}). Evidently, the $2\omega_{sp}$-contributions lead to the same energy shifts and Floquet band structure both in the sudden and adiabatic limits. The difference appears in the absence of DWE in the latter case.  
Likewise in Sec. \ref{sec:332}, the Ward-Pitaevskii-like quantities $\Sigma_{{\bf K},s}^{(lg)}(\mbox{em1})/2\hbar\omega_{sp}$ and $\Sigma_{{\bf K},s}^{(lg)}(\mbox{abs1})/2\hbar\omega_{sp}$ have taken the role of vertex corrections generating the $2\omega_{sp}$ plasmon emissions and absorptions, respectively. Figs. \ref{C2lg}(d) and (e) are also indicative of this assignment.  
Since in the adiabatic limit of SBC we have assumed $t_{0}\rightarrow -\infty$, the unitarity condition on (\ref{eq:C1ad}) and (\ref{eq:C2lgSigma}) is achieved for $t\rightarrow -\infty$.  The rate of one-plasmon-induced electron emission $\Gamma_{{\bf K},s}^{(lg)}(\mbox{em})$ from the Floquet states derived from Ag(111) SS-band will be evaluated in Sec. \ref{sec:Gammalg}.





\section{Surface state electron transition rates from cumulant expansion in the length gauge}
\label{sec:DecayRates}
\subsection{On-the-energy-shell matrix elements}
\label{sec:enshell}

Applications of the cumulant expansion expressions for electron propagators developed in Secs. \ref{sec:EPdynlg} to the concrete problem of plasmoemission from Ag(111) surface bands requires the knowledge of the interaction matrix elements governing electron transitions. Once these matrix elements and the corresponding excitation energies are defined one can calculate the cumulant forms of the electron propagators.

We proceed by assuming that the interaction ${\cal V}_{S}(\tau)$  is turned on adiabatically and use (\ref{eq:Tlength}) to describe the electron excitation dynamics. For calculational convenience we further assume that the surface plasmon excitation mechanism produces indiscriminate occupation of the ${\bf Q}$-modes constituting the coherent state (\ref{eq:cohstate}). This renders the ${\bf Q}$-summations in ${\cal V}_{S}(\tau)$ and ${\cal V}_{S}^{'}(\tau)$ unrestricted by any additional distribution of excited plasmon modes. We also continue with neglecting the effects of weak surface plasmon dispersion which amounts to taking $\omega_{\bf Q}=\omega_{sp}$. 
With the same frequency of all plasmon modes propagating on planar surfaces one may also expect their indiscriminate role in forming the coherent states with isotropically distributed  $\lambda_{\bf Q}=\lambda_{Q}$. 

Now it is easily verified that the lowest order vertices generated by the matrix elements of the gauge-specific interaction taken between the states (\ref{eq:phi_s}) and (\ref{eq:phifin}) are constrained to the lateral momentum shell according to 
\bq
\langle\phi_{f}|{\cal V}|\phi_{i}\rangle\rightarrow \sum_{\bf Q}\left[\delta_{{\bf K}_{f},{\bf K}_{s}+{\bf Q}}(\dots)_{1}+\delta_{{\bf K}_{f},{\bf K}_{s}-{\bf Q}}(\dots )_{2}\right],
\label{eq:Vmomentum}
\eq
where $(\dots)_{n}$ are appropriate expressions obtained from the explicit forms of the here relevant potential ${\cal V}$. Such selection rules have already been assumed in Figs. \ref{C2lg} and \ref{C1&C2} and exploited in expressions (\ref{eq:Gamma0+-}).

In addition, special care must be taken in the concrete final state summations of expressions containing the products of transition matrix elements over the wavefunctions (\ref{eq:phi_s})-(\ref{eq:phifin}) and the corresponding energy-conserving $\delta$-functions. As noted after Eq. (\ref{eq:1plasPE}), the energy conservation involves the energy of the asymptotic wave state (\ref{eq:phifin}) whose energy zero is at the vacuum level. Hence $E_{f}=\frac{\hbar^{2}({\bf K}_{f}^{2}+k_{z,f}^{2})}{2m}=E_{{\bf K}_{f}}+E_{k_{z,f}}$ and the energy conservation in multiplasmon-induced electron emissions from the SS-band is expressed through
\bq
\delta(E_{{\bf K}_{s}}+E_{s}+n\hbar\omega_{sp}-E_{{\bf K}_{f}}-E_{k_{z,f}}),
\label{eq:deltamulti}
\eq
which should be applied in conjunction with the lateral momentum conservation ${\bf K}_{f}={\bf K}_{s}\pm {\bf Q}$. In surface scattering phenomena this combination of requirements is usually referred to as the scan curve. The final state summations along the scan curve are usually replaced by integrations complying with the normalization of (\ref{eq:phifin}) so that $\sum_{f}\rightarrow \left(\frac{L}{2\pi}\right)^{2}\int d^{2}{\bf K}_{f}\left(\frac{L_{z}}{2\pi}\right)\int dk_{f}$. Introducing 
\bq
k_{z,f}^{(n)}(\pm)=\sqrt{\frac{2m}{\hbar}(E_{{\bf K}_{s}}+E_{s}\pm n\hbar\omega_{sp}-E_{{\bf K}_{f}})},
\label{eq:kz+-}
\eq
expression (\ref{eq:deltamulti}) can be written as $\delta(\frac{\hbar^{2}k_{z,f}^{2}}{2m}-\frac{\hbar^{2}k_{z,f}^{(n)}(+)^{2}}{2m})$. We consider only real electron emission processes for which $k_{z,f}^{(n)}(+)^{2}>0$. Therefore the final state summations along the scan curve can be conveniently performed by making use of the following procedure (we use the short hand notation in which the indices $z$ are omitted)
\bq
\sum_{k'}(\dots)_{n}\delta(E_{k_{f}}-E_{k'})=\frac{L_{z}}{2\pi}\int dk'(\dots)_{n}\delta(E_{k_{f}}-E_{k'}) 
\longrightarrow 
\int dk'(\dots)_{n}\delta(k_{f}-k')\frac{L_{z}}{2\pi}\left/\left(\frac{\partial E_{k_{f}}}{\partial k_{f}}\right)\right.,
\label{eq:Kronecker}
\eq
where $E_{k_{f}}$ denotes the component of energy corresponding to the electron final state motion in the $z$-direction. Using (\ref{eq:Kronecker}) we may introduce the density of $k_{f}$-states around the electron final state energy $E_{k_{f}}$  through 
\bq
\rho(E_{k_{f}})=\frac{L_{z}}{2\pi}\left/\left(\frac{\partial E_{k_{f}}}{\partial k_{f}}\right)\right.,
\label{eq:rho}
\eq
so that
\bq
2\pi\hbar\rho(E_{k_{f}})\rightarrow 1/j_{z,f}
\label{eq:current}
\eq
where
\bq
j_{z,f}=(\hbar k_{z,f}/m)|\phi_{f}(z)|^{2}=v_{z,f}/L_{z}
\label{eq:jz}
\eq
is the electron current in the $n$-th emission channel that is fixed by the energy conservation appearing in (\ref{eq:Kronecker}). It carries the dimension of inverse time.

Alternatively,  one may  use the equivalent Kronecker symbol representation 
\bq
\sum_{k'}(\dots)\delta(E_{k_{f}}-E_{k'}) 
\longrightarrow 
\sum_{k'}(\dots)\delta_{k_{f},k'}\frac{L_{z}}{2\pi}\left/\left(\frac{\partial E_{k_{f}}}{\partial k_{f}}\right)\right.\nonumber\\
=
\sum_{k'}(\dots)\rho(E_{k_{f}})\delta_{k_{f},k'},
\label{eq:Kroneckersymbol}
\eq
to conveniently perform the summation over the final states $|k'\rangle$.
This fixes the prerequisites for calculations of cumulant representation of the electron propagators in both gauges.

\subsection{Transition rates from cumulant expansion in the length gauge}
\label{sec:Gammalg}

The most appropriate and convenient quantities characterizing the dynamics of coupled electron-plasmon system described by the propagator (\ref{eq:GKslength}) is the corresponding quasiparticle decay or transition rates (\ref{eq:C1ad})  obtained within the SBC or, equivalently, in the limit of adiabatic switching on of the interaction. We first address the excitation rate defined in (\ref{eq:C1ad}).

With the prerequisites developed above the plasmon absorption induced component of the full transition rate obtained from (\ref{eq:C1ad}) acquires the form ($+1\omega_{sp}=\mbox{abs1}$)
\begin{widetext}
\barr
2\Gamma_{{\bf K}_{s},s}^{(lg)}(\mbox{abs1})
&=&
\frac{2\pi}{\hbar}\sum_{{\bf Q},f}\lambda_{\bf Q}^{2}|\langle f|V_{\bf Q}(z)|s\rangle|^{2}
\delta(E_{{\bf K}_{s}+{\bf Q}}+E_{z,f}-E_{{\bf K}_{s}}-E_{s}-\hbar\omega_{sp})
\nonumber\\
&=&
\frac{2\pi}{\hbar}\sum_{{\bf Q},f}\lambda_{\bf Q}^{2}
V_{\bf Q}^{2}\left|f_{f,s}(Q)\right|^{2}
\delta\left(\frac{\hbar^{2}k_{z,f}(+)^{2}}{2m}-\frac{\hbar^{2}k_{z,f}^{2}}{2m}\right)=\sum_{\bf Q}\lambda^{2}V_{\bf Q}^{2}|f_{f,s}(Q)|^{2}\frac{L_{z}}{v_{z,f}(+)},
\label{eq:Gamma1+}
\earr
\end{widetext}
with $V_{\bf Q}$ defined in (\ref{eq:V_Q}), $E_{k_{z,f}}=\frac{\hbar^{2}k_{z,f}^{2}}{2m}$, ${\bf K}_{f}={\bf K}_{s}+{\bf Q}$ in $k_{z,f}^{(1)}(+)$  defined in (\ref{eq:kz+-}), and $f_{f,s}(Q)$ is the dimensionless generalized oscillator strength (cf. Eq. (A6) in ref. [\onlinecite{SGLD}])
\bq
f_{f,s}(Q)=\int dz u_{s}(z) e^{-Qz}\frac{e^{-ik_{z,f}z}}{\sqrt{L_{z}}}.
\label{eq:f_fs}
\eq

One way to perform the ${\bf Q}$-summation on the RHS of (\ref{eq:Gamma1+}) and  in similar subsequent expressions is to assume equal occupations of the plasmon modes in the coherent state, viz. to introduce linear and quadratic  Eliashberg-like equipartition ans\"{a}tze[\onlinecite{Eliashberg,Grimvall}]
\bq
\sum_{\bf Q}\lambda_{\bf Q}(\dots)_{\bf Q}=\rightarrow \lambda \sum_{\bf Q}(\dots)_{\bf Q},
\label{eq:lambdalin}
\eq 
\bq
\sum_{\bf Q}\lambda_{\bf Q}^{2}(\dots)_{\bf Q}=\rightarrow \lambda^{2} \sum_{\bf Q}(\dots)_{\bf Q},
\label{eq:lambdaqu}
\eq 
with $\lambda$ treated as an external parameter fixed by experimental conditions.
An artifact of this procedure is the assignment of the same phase to all plasmon modes constituting the coherent state (\ref{eq:cohstate}). 

\begin{figure}[tb]
\rotatebox{0}{\epsfxsize=8.5cm \epsffile{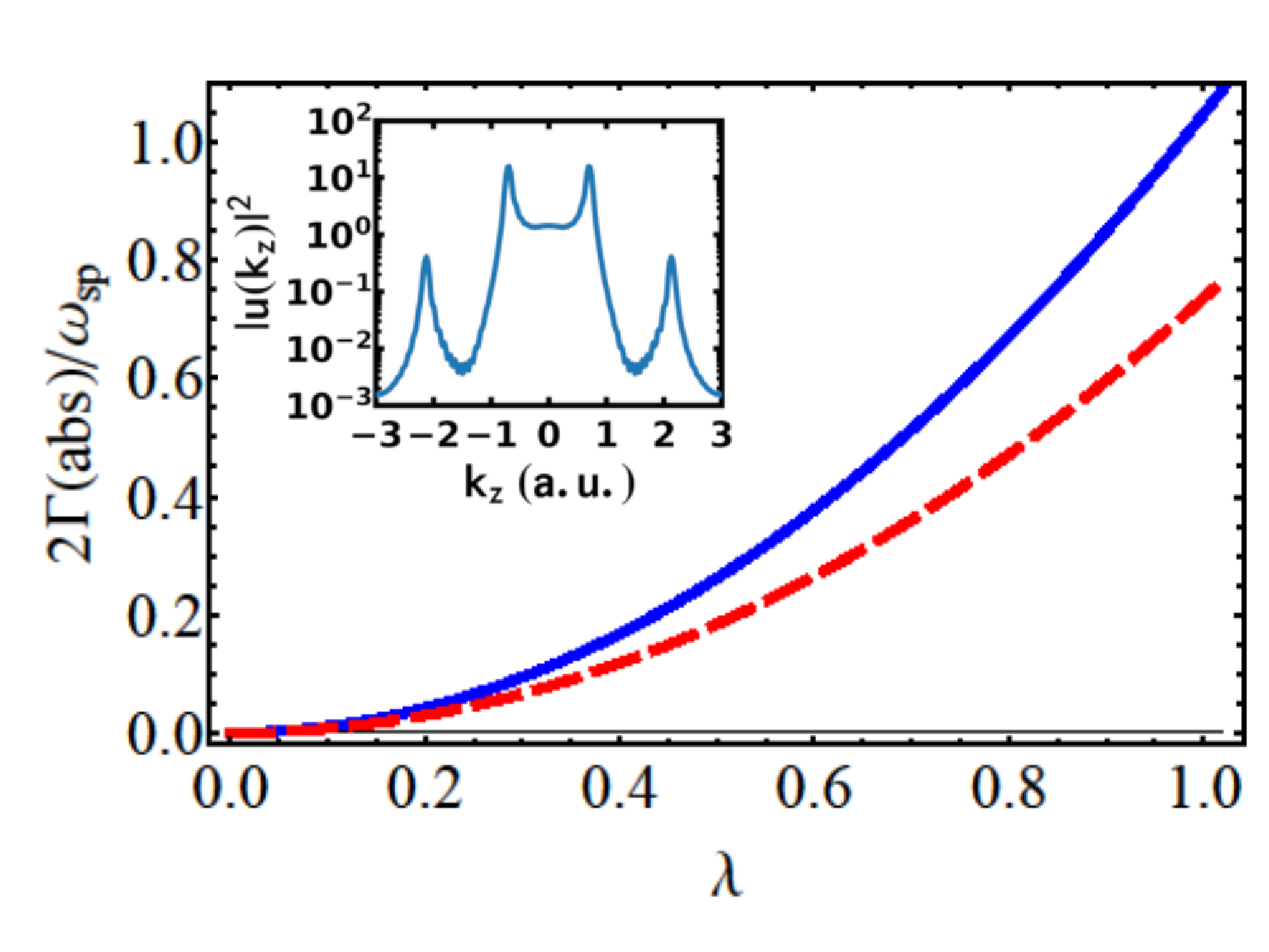}} 
\caption{Comparison of the one plasmon-induced excitation efficiency $2\Gamma_{{\bf K}=0,s}(\mbox{abs1})/\omega_{sp}$ calculated in the length and velocity gauges as functions of the coherent state parameter $\lambda$ and appropriate to one-plasmon iduced electron emission from SS-state band on Ag(111) surface with reduced work function.[\onlinecite{Horn}] Full blue curve: $2\Gamma_{{\bf K}=0,s}^{(lg)}(\mbox{abs1})/\omega_{sp}$ in the length gauge as given by Eq. (\ref{eq:Gammalg/omegas}). Dashed red curve: $2\Gamma_{{\bf K}=0,s}^{(vg)}(\mbox{abs1})/\omega_{sp}$ in the velocity gauge as given by Eq. (\ref{eq:Gammavg/omegasperp}). Same $z_{s}=1.2 a_{B}$ and $k_{z,f}(+)=0.274 a_{B}^{-1}$ used in both expressions are adopted from Refs. [\onlinecite{plasPE,plasFloquet}]. The two curves with different dependence on $z_{s}$, $k_{z,f}(+)$ and $\hbar\omega_{sp}$ coincide for $z_{s}=1 a_{B}$. 
They measure the efficiency of first order processes in which an SS-electron absorbs one surface plasmon from the plasmonic coherent state. These two concrete examples indicate that plasmoemission takes place during electron exposure to only few cycles of plasmon oscillation. 
Inset shows the input $|\tilde{u}_{s}(k_{z,f}(+)|^{2}/a_{B}$ for Eqs. (\ref{eq:Gammalg/omegas}) and (\ref{eq:Gammavg/omegasperp}) taken from Ref. [\onlinecite{plasFloquet}].}
\label{Gammaslgvg}
\end{figure}

Further analytical calculations of (\ref{eq:Gamma1+}) can be facilitated by observing that both the SP field and the wavefunction $u_{s}(z)$ are strongly localized around $z_{s}$ at the surface. Hence, without incurring much error we may substitute $e^{-Qz}\rightarrow e^{-Qz_{s}}$ in (\ref{eq:f_fs}) and obtain 
\bq
|f_{f,s}(Q)|^{2}\rightarrow e^{-2Qz_{s}} \frac{|\tilde{u}_{s}(k_{z,f})|^{2}}{L_{z}}
\label{eq:expzs}
\eq
where $z_{s}$ is the maximum of $u_{s}(z)$ relative to the image plane. Also, for nearly vertical transitions from the SS-band bottom (${\bf K}_{s}=0$) we may neglect $E_{\bf Q}$ relative to $E_{z,f}$ in (\ref{eq:kz+-}). 
Combining all that with (\ref{eq:Kronecker})-(\ref{eq:kz+-}) in (\ref{eq:Gamma1+}) yields the relevant dimensionless measure of electron emission induced from the bottom of SS-band by one plasmon absorption 
\bq
\frac{2\Gamma_{{\bf K}_{s}=0,s}^{(lg)}(\mbox{abs1})}{\omega_{sp}}=\frac{\lambda^{2}}{4(k_{z,f}(+)a_{B})}\left(\frac{z_{s}}{a_{B}}\right)^{-1}\frac{|\tilde{u}_{s}(k_{z,f}(+))|^{2}}{a_{B}},
\label{eq:Gammalg/omegas}
\eq
where $a_{B}$ is the Bohr radius and $|\tilde{u}_{s}(k_{z,f}(+))|^{2}$ plays the role of static form factor for the transition.
This result expressed in units of $\omega_{sp}$ is free from the quantization lengts $L$ and $L_{z}$ and its $z_{s}^{-1}$ dependence is equal to that of the image potential $\Phi({\bf r})$ in (\ref{eq:H0}). In the length gauge it does not exhibit explicit scaling with $\hbar\omega_{sp}$ which enters only through the energy conservation in $k_{z,f}(+)$. It also provides the emission component of the second order cumulant (\ref{eq:C2lgSigma}). 

To estimate the magnitude of (\ref{eq:Gammalg/omegas}) characteristic of one-plasmon induced electron emission from  Ag(111) surface we make use of the SS-band parametrization described in Sec. \ref{sec:SSenergetics}. To remain consistent  with experiments we evaluate (\ref{eq:Gammalg/omegas}) so as to pertain to Ag(111) surface with work function reduced by $\Delta\phi=1.3$ eV due to alkali submomolayer adsorption.[\onlinecite{Horn}]. The results for the excitation efficiency $2\Gamma_{{\bf K}=0,s}(\mbox{abs1})/\omega_{sp}$ are presented in Fig. \ref{Gammaslgvg} and demonstrate that plasmoemission may take place already within the first few cycles of plasmon oscillation even for modest average occupations $\lambda<1$ of the plasmonic coherent states. Thereby the quantity  
$2\Gamma_{f,s}^{(lg)}(\mbox{abs1})/\omega_{sp}$ lends itself as a natural measure of efficiency of SS-electron emission induced by a surface plasmon from the plasmonic coherent state.

Before closing the present analyses we recall the earlier findings[\onlinecite{Boyd2004,Reiss2013}] that the length gauge may not always prove suitable for nonperturbative treatment of transition amplitudes (\ref{eq:Tlength}) and the use of velocity gauge has been recommended instead. Hence, in the following we shall investigate this latter route in the development of quantal nonperturbative pump-probe picture of plasmoemission from surface bands.


\section{Velocity gauge representation of electron-surface plasmon interaction}
\label{sec:velocitygauge}
\subsection{Canonical transformation to the velocity gauge}
\label{sec:canonical}

The historically most frequently used form of the e-SP interaction (\ref{eq:V_Q}) corresponds to the length gauge or dipole representation. However, in nonperturbative treatments of the effects of boson fields on electron motion the choice of a different, velocity or radiation gauge, turns out more appropriate from the formal point of view as well as more advantageous because it enables pursuing analytic or closed form nonlinear solutions quite far in nonperturbative descriptions of electron dynamics.[\onlinecite{Boyd2004,Reiss2013}] The passage to this gauge for representation of e-SP interaction[\onlinecite{PlasGauge}] is achieved by performing on (\ref{eq:HsystSP}) a canonical (unitary) transformation generated by a nested commutator expansion
\bq
H'=\exp(i\hat{S})H\exp(-i\hat{S})=H+[i\hat{S},H]+\frac{1}{2}[i\hat{S},[i\hat{S},H]]+ \cdots
\label{eq:HS}
\eq
The hermitian operator $\hat{S}$ is chosen such that the transformation (\ref{eq:HS}) eliminates the length gauge interaction $V$ (\ref{eq:V}) from $H'$ and replaces it by the velocity gauge interaction $V'$. This requires that $\hat{S}$ be not explicitly time dependent and to depend only on the electron radius vector ${\bf r}$ and the plasmon field momenta $\propto(\hat{a}_{\bf Q}-\hat{a}_{\bf -Q}^{\dag})$, but not on the electron momentum ${\bf p}$, viz.
\bq
\hat{S}=i\sum_{\bf Q}\frac{V_{\bf Q}}{\hbar\omega_{\bf Q}}e^{i{\bf Q}\brho-Qz}(\hat{a}_{\bf Q}-\hat{a}_{\bf -Q}^{\dag}).
\label{eq:S}
\eq
Since $v({\bf r})$ and $\Phi({\bf r})$ commute with $\hat{S}$ we obtain for the transformed Hamiltonian[\onlinecite{PlasGauge}]
\bq
H'=\frac{({\bf p+A(r)})^2}{2m} +H_{0}^{pl}+v({\bf r})+\Phi({\bf r})
=
H_{0}+V',
\label{eq:H'}
\eq
with
\bq
H_{0}=\frac{{\bf p}^2}{2m}+v({\bf r})+\Phi({\bf r}) +H_{0}^{pl}
=
H_{0}^{el}+H_{0}^{pl},
\label{eq:H_0}
\eq
and the electron pseudopotential given by
\bq
v_{ps}({\bf r})=v({\bf r})+\Phi({\bf r}).
\label{eq:vps}
\eq
The set of electron eigenfunctions that diagonalize $H_{0}^{el}$ incorporates the SS-band states, in complete analogy with the length gauge situation. The transformed electron-plasmon interaction is
\bq
V'=V_{S}'=\frac{{\bf p\cdot A(r)+A(r)\cdot p}}{2m}+\frac{{\bf A}^{2}({\bf r})}{2m},
\label{eq:V'}
\eq
where subscript $_S$ denotes the Schr\"{o}dinger representation of the operator. 
Here ${\bf A(r)}$ with the dimension of momentum is a plasmonic field vector potential acting on the electron at ${\bf r}$, viz.
\barr
{\bf A(r)}&=&-\hbar\nabla \hat{S}=\sum_{\bf Q}\frac{V_{\bf Q}}{\omega_{\bf Q}}
({\bf Q},iQ)e^{i{\bf Q}\brho-Qz}(\hat{a}_{\bf Q}-\hat{a}_{\bf -Q}^{\dag}) 
\label{eq:A}\\
&=&
\sum_{\bf Q}\bcalA_{\bf Q}({\bf r})(\hat{a}_{\bf Q}-\hat{a}_{\bf -Q}^{\dag})\nonumber\\
&=&
\sum_{\bf Q}\left(\bcalA_{\bf Q}({\bf r})\hat{a}_{\bf Q}+\bcalA_{\bf Q}^{\dag}({\bf r})\hat{a}_{\bf Q}^{\dag}\right)\nonumber\\
&=& 
{\bf A}^{(-)}({\bf r})+ {\bf A}^{(+)}({\bf r})
\label{eq:A+-}
\earr
where 
\bq
\bcalA_{\bf Q}({\bf r})=\frac{V_{\bf Q}}{\omega_{\bf Q}}({\bf Q},iQ)e^{i{\bf Q}\brho-Qz}.
\label{eq:AandAdag}
\eq
Therefore the phases of lateral and perpendicular components of ${\bf A(r)}$ differ by a factor of $\pi/2$. The hermitian conjugated components ${\bf A}^{(-)}({\bf r})$ and ${\bf A}^{(+)}({\bf r})$ can be represented as 
\barr
{\bf A}^{(-)}({\bf r})&=&\sum_{\bf Q}\bcalA_{\bf Q}({\bf r})\hat{a}_{\bf Q}=\sum_{\bf Q}\frac{V_{\bf Q}}{\omega_{\bf Q}}({\bf Q},iQ)e^{i{\bf Q}\brho-Qz}\hat{a}_{\bf Q}
=
{\bf A}_{\parallel}^{(-)}({\bf r})+{\bf A}_{\perp}^{(-)}({\bf r}),
\label{eq:A-}\\
{\bf A}^{(+)}({\bf r})&=&\sum_{\bf Q}\bcalA_{\bf Q}^{\dag}({\bf r})\hat{a}_{\bf Q}^{\dag}=\sum_{\bf Q}\frac{V_{\bf Q}}{\omega_{\bf Q}}({\bf Q},-iQ)e^{-i{\bf Q}\brho-Qz}\hat{a}_{\bf Q}^{\dag}
=
{\bf A}_{\parallel}^{(+)}({\bf r})+{\bf A}_{\perp}^{(+)}({\bf r}),
\label{eq:A+}
\earr
where subscripts $\parallel$ and $\perp$ denote the directions of in-$(x,y)$-plane unit vector $\hat{\bf e}_{\bf Q}={\bf Q}/Q$ and the unit vector $\hat{\bf e}_{z}=\hat{\bf e}_{\perp}$ perpendicular to $(x,y)$-plane, respectively. Hence, the ${\bf Q}$-components of the plasmon vector potential ${\bf A}({\bf r})$ are polarized in the sagittal $(\hat{\bf e}_{\bf Q},\hat{\bf e}_{\perp})$-plane whereas the field itself is strongly localized in the surface region around $z=0$ (cf. Fig. 3 in Ref. [\onlinecite{Pitarke}]). Thereby the velocity gauge interaction (\ref{eq:V'}) displays the features of electron current coupling to the oscillating vector field of surface plasmon polaritons.[\onlinecite{Ritchie1973,Pitarke}] The decomposition of plasmon vector field operator (\ref{eq:A+-}) expressed through (\ref{eq:A-}) and (\ref{eq:A+}) proves particularly convenient in the studies of electron interactions with plasmonic coherent states.   

Exploiting the property (\ref{eq:A}) we may write
\bq
\exp(-i\hat{S})=\exp\left[\frac{i}{\hbar}\int^{\bf r} {\bf A(r)}d{\bf r}\right] ,
\label{eq:SA}
\eq
which is a quantum generalization of the gauge transformation established long ago by Pauli for classical translationally homogeneous vector fields.[\onlinecite{Pauli}] Here it brings the dynamical component of electron-plasmon interaction (\ref{eq:V'}) to the form of coupling in the velocity gauge.
The remaining e-SP interaction which is not explicitly time-dependent and has already been added to (\ref{eq:H0}) arises from
\bq
\Phi({\bf r})=\frac{1}{2}[i\hat{S},V]=-\sum_{\bf Q}\frac{V_{\bf Q}^{2}}{\hbar\omega_{\bf Q}}e^{-2Qz},
\label{eq:Phi}
\eq
and represents an instantaneous scalar potential depending only on the electron coordinates and not on plasmon operators. With the coupling (\ref{eq:V}) this is the standard electron image potential. Note that $\Phi({\bf r})$ arises from virtual excitation (creation and subsequent annihilation) of plasmons and therefore represents a shift of the plasmonic ground state energy induced by the electron. Therefore, the effect of  canonical transformation (\ref{eq:HS}) is to eliminate from the new gauge the  interaction (\ref{eq:V}) and replace the operator ${\bf p}$ and the potential $v({\bf r})$ by
\barr
{\bf p}&\rightarrow& {\bf p+A(r)}=\hat{\bf \pi},
\label{eq:p'}\\
 v({\bf r})&\rightarrow& v({\bf r})+\Phi({\bf r})=v_{ps}({\bf r}).
\label{eq:Vscal}
\earr
Thereby the dynamical e-SP plasmon interaction enters the transformed $H'$ only through the vector potential ${\bf A(r)}$. 

It is straightforward to verify that the vector potential ${\bf A(r)}$ given by (\ref{eq:A}) satisfies the Coulomb gauge outside the surface
\bq
{\bf \nabla\cdot A(r)}=0, \hspace{1cm} z>0.
\label{eq:nablaA}
\eq
This means that the operators ${\bf p}=-i\hbar\nabla$ and ${\bf A(r)}$ commute when acting on the various combinations of electron and plasmon operators as well as on the electron wavefunction. Lastly, it is important to note that the gauge transformation generated by (\ref{eq:S}) does not affect the interaction of electrons with the external pumping EM field, $W_{e-EM}(t)$, and therefore leaves the coherent state (\ref{eq:cohstate}) unchanged.


\subsection{Interaction representation of the potential $V'$ in the velocity gauge}
\label{sec:V'I}

The e-SP potential in the velocity gauge (\ref{eq:V'}) in the interaction representation is obtained from
\bq
V_{I}'(t)=e^{iH_{0}t/\hbar}V_{S}'e^{-iH_{0}t/\hbar}=
e^{iH_{0}^{pl}t/\hbar}e^{iH_{0}^{el}t/\hbar}V_{S}'e^{-iH_{0}^{e}t/\hbar}e^{-iH_{0}^{pl}t/\hbar},
\label{eq:VI}
\eq
since $[H_{0}^{el},H_{0}^{pl}]=0$. Were it not for the presence of $v_{ps}({\bf r})$ in $H_{0}^{el}$, the application of (\ref{eq:VI}) would lead to
\bq
V_{I}'(t)\longrightarrow e^{iH_{0}^{pl}t/\hbar}e^{i\frac{{\bf p}^2}{2m}t/\hbar}V_{S}'e^{-i\frac{{\bf p}^2}{2m}t/\hbar}e^{-iH_{0}^{pl}t/\hbar}
=
e^{iH_{0}^{pl}t/\hbar}V_{S}'e^{-iH_{0}^{pl}t/\hbar},
\label{eq:VIS}
\eq
and the explicit time dependence of $V_{I}'(t)$ would appear solely in the plasmon operators. We shall estimate the deviation of (\ref{eq:VIS}) from (\ref{eq:VI}) by using the identity (see p. 339 in Ref. [\onlinecite{MessiahI}])
\bq
B(t)=e^{iAt}Be^{-iAt}=B+i[A,\int_{0}^{t}B(\tau)d\tau]
=
B+i\int_{0}^{t}e^{iA\tau}[A,B]e^{-iA\tau}d\tau,
\label{eq:ABcomm}
\eq
where $A$ and $B$ are noncommuting operators. Substituting therein $A=H_{0}^{el}/\hbar$ and $B=V_{S}'/\hbar$ we obtain
\bq
V_{I}'(t)=V_{S}'+\frac{i}{\hbar}\int_{0}^{t}e^{iH_{0}^{el}\tau/\hbar}[H_{0}^{el},V_{S}']e^{-iH_{0}^{el}\tau/\hbar}d\tau,
\label{eq:HVcomm}
\eq
where for the sake of simplicity we have omitted the LHS and RHS operators $e^{iH_{0}^{pl}t/\hbar}$ and $e^{-iH_{0}^{pl}t/\hbar}$, respectively. Taking into account (\ref{eq:nablaA}), we find for the generating commutator in the integrand on the RHS of (\ref{eq:ABcomm})
\bq
\frac{1}{\hbar}[H_{0}^{el},V_{S}']= \frac{1}{m\hbar}[v_{ps}({\bf r}),{\bf p}\cdot{\bf A(r)}]=\frac{i}{m}{\bf A(r)}\cdot\left(\nabla v_{ps}({\bf r})\right).
\label{eq:corrVI}
\eq
This operator is system-specific and its transition matrix elements should be estimated in comparison with those of $V_{S}'$. Provided they are significantly smaller then the much simpler expression (\ref{eq:VIS}) should represent a good approximation to $V_{I}'(t)$, see Appendix \ref{sec:deviationV}.


\subsection{Evolution operators in the velocity gauge}
\label{sec:Uvel}

For later convenience we introduce the evolution operators $U_{S}(H',t,t_{0})$ and $U_{I}(H',t,t_{0})$ in the Schr\"{o}dinger and interaction representations, respectively, that are induced by a generally time dependent $H'(t)=H_{0}+V_{S}'(t)$.[\onlinecite{MessiahII,Davydov}] In the Schr\"{o}dinger representation and the velocity gauge this reads
\bq
i\hbar \frac{\partial}{\partial t}U_{S}(H',t,t_{0})=H'(t)U_{S}(H',t,t_{0}),
\label{eq:USeq}
\eq 
and the ensuing integral equation can be brought to the form
\bq
U_{S}(H',t,t_{0})=U_{S}^{0}(t,t_{0})-\frac{i}{\hbar}\int_{t_{0}}^{t}d\tau U_{S}^{0}(t,\tau)V_{S}'(\tau)U_{S}(H',\tau,t_{0}),
\label{eq:USint}
\eq
where $U_{S}^{0}(t,t_{0})=e^{-iH_{0}(t-t_{0})}$. 

For the evolution operator in the interaction representation one obtains
\bq
i\hbar \frac{\partial}{\partial t}U_{I}(H',t,t_{0})=V_{I}'(t)U_{I}(H',t,t_{0})
\label{eq:UIdiff}
\eq
where $V_{I}'(t)=e^{iH_{0}t}V_{S}'(t)e^{-iH_{0}t}$. This integrates to
\bq
U_{I}(H',t,t_{0})= 1-\frac{i}{\hbar}\int_{t_{0}}^{t} d\tau V_{I}'(\tau)U_{I}(H',\tau,t_{0})
=
1+\int_{t_{0}}^{t} d\tau \frac{\partial}{\partial\tau}U_{I}(H',\tau,t_{0}), 
\label{eq:UIint}
\eq

Expressions (\ref{eq:USint})-(\ref{eq:UIint}) are exact. Suitable approximations can be most conveniently implemented in the interaction representation following exponential forms (\ref{eq:UGross}) and (\ref{eq:G1}) whose truncated expansions are expected to provide a good description of the system dynamics in the regime of weakly correlated subsequent electron scattering events.       


\subsection{Surface plasmon-iduced electron transition amplitudes in the velocity gauge}
\label{sec:TVG}

The electron transition amplitudes in the velocity gauge are obtained by performing the canonical transformation generated by $e^{-i\hat{S}}$ on the operators and wavefunctions constituting the transition amplitudes (\ref{eq:Tlength}) in the length gauge. Inserting therein $1=e^{-i\hat{S}}e^{i\hat{S}}$ between operators and state vectors we find  
\begin{widetext}
\barr
T_{f,i}^{vel}(t,t_{0})&=&\langle\langle\mbox{coh}, \phi_{f},(t)||e^{-i\hat{S}}e^{i\hat{S}}U_{S}(H,t,t_{0})e^{-i\hat{S}}e^{i\hat{S}}||\phi_{i},\mbox{coh},(t_{0})\rangle\rangle \nonumber\\
&=&
\langle\langle\mbox{coh}, \phi_{f}||e^{iH_{0}t}e^{-i\hat{S}}e^{-iH_{0}t}e^{iH_{0}t}U_{S}(H',t,t_{0})e^{-iH_{0}t_{0}}
e^{iH_{0}t_{0}}e^{i\hat{S}}e^{-iH_{0}t_{0}}||\phi_{i},\mbox{coh}\rangle\rangle \nonumber\\
&=&
\langle\langle\mbox{coh}, \phi_{f}||e^{-i\hat{S}(t)}e^{iH_{0}t}U_{S}(H',t,t_{0})e^{-iH_{0}t_{0}}e^{i\hat{S}(t_{0})}||\phi_{i},\mbox{coh}\rangle\rangle\nonumber\\
&=&
\langle\langle\mbox{coh}, \phi_{f}||e^{-i\hat{S}(t)}U_{I}(H',t,t_{0})e^{i\hat{S}(t_{0})}||\phi_{i},\mbox{coh}\rangle\rangle.
\label{eq:Tvel}
\earr
\end{widetext}
Here the time dependence of the transformation generator $\hat{S}(t)$ obeys the interaction picture
\bq
\hat{S}(t)=e^{iH_{0}t}\hat{S} e^{-iH_{0}t},
\label{eq:Stau}
\eq
where the initial and final states, $e^{i\hat{S}(t_{0})}||\phi_{i},\mbox{coh}\rangle\rangle$ and $e^{i\hat{S}(t)}||\phi_{f},\mbox{coh}\rangle\rangle$, respectively, and the evolution operators $U_{S}(H',t,t_{0})=e^{i\hat{S}}U_{S}(H,t,t_{0})e^{-i\hat{S}}$ and $U_{I}(H',t,t_{0})=e^{i\hat{S}}U_{I}(H,t,t_{0})e^{-i\hat{S}}$, are expressed in the velocity gauge.


\subsection{Gauge dependence of initial states}
\label{sec:instates} 

 The different gauge formulations of the transition amplitudes (\ref{eq:Tlength}) and (\ref{eq:Tvel}) pose the question of the initial state appropriate to the temporal boundary conditions of the studied problem.[\onlinecite{Boyd2004}] In expression (\ref{eq:Tlength}) it was tacitly assumed that the interaction $V$ between electrons and surface plasmons from the prepumped coherent state cloud[\onlinecite{plasPE}] commences adiabatically from $t_{0}$ and terminates for  $t\rightarrow\infty$ so that both the initial and final states of the system could be represented by the Kronecker products of unperturbed electronic and plasmonic coherent states diagonalizing $H_{0}=H_{0}^{el}+H_{0}^{pl}$. This implies the scattering boundary conditions within which $V$ appearing in $U_{S}(H,t,t_{0})$ is effective only in the  time interval $(t,t_{0})$. Applying analogous reasoning to the action of  ${\bf A(r)}$ that generates $U_{S}(H',t,t_{0})$ we should also consider it effective only in the interval $(t,t_{0})$.  Consequently, in view of (\ref{eq:SA}) and the SBC we ascertain that the operator 
\bq
\lim_{SBC} e^{i\hat{S}(\tau)}=e^{iH_{0}\tau}\exp\left[ \frac{i}{\hbar}\int^{\bf r} {\bf A(r)}d {\bf r}\right]e^{-iH_{0}\tau}\rightarrow 1,
\label{eq:Ssbc}
\eq
for $\tau \geq t$ or $\tau \leq t_{0}$. Applying this in (\ref{eq:Tvel}) we obtain the SBC limit of the transition amplitude in the velocity gauge in the Schr\"{o}dinger and interaction pictures
\barr
\lim_{SBC}T_{f,i}^{vel}(t,t_{0})&=&\langle\langle\mbox{coh}, \phi_{f},(t)||U_{S}(H',t,t_{0})||\phi_{i},\mbox{coh},(t_{0})\rangle\rangle 
\label{eq:TvelSBCSchr}\\
&=&
\langle\langle\mbox{coh}, \phi_{f}||U_{I}(H',t,t_{0})||\phi_{i},\mbox{coh}\rangle\rangle,
\label{eq:TvelSBC}
\earr
which is taken over the same initial and final states as in (\ref{eq:Tlength}). Expressions (\ref{eq:Tlength}) and (\ref{eq:TvelSBC}) may seem gauge noninvariant. However, reintroducing (\ref{eq:Ssbc}) in (\ref{eq:TvelSBC}) brings the latter to the gauge invariant form (\ref{eq:Tvel}). Hence, it appears that within the scattering boundary conditions one should start from the same unperturbed state basis in the calculations of  transition amplitudes in the length and velocity gauges. This is also in accord with the conclusions reached in Sec. 2 of Ref. [\onlinecite{Boyd2004}].
%

%

\subsection{Semiclassical limit of the vector field ${\bf A}({\bf r})$}
\label{sec:calV'}

The connection between exact quantum expressions for the transition amplitudes (\ref{eq:TvelSch}) and (\ref{eq:TvelInt})  and the limit in which the electron is perturbed by the vector potential arising from a highly excited semiclassical plasmonic field is most conveniently established in the interaction picture.  Assuming that the pumped plasmonic coherent state represents a predetermined complete set fixed by the experimental conditions we may apply in all interaction operators in (\ref{eq:TvelSBC}) and sequels the coherent state averaging of ${\bf A}_{I}({\bf r},t)$ effectuated by using (\ref{eq:lambdaSP}) and the Baker-Hausdorff-Campbell formula to obtain 
\barr  
\bcalA_{I}({\bf r},t)&=&
\langle\mbox{coh}|{\bf A}_{I}({\bf r},t)|\mbox{coh}\rangle
=
\langle\mbox{coh}|e^{iH_{0}t}{\bf A(r)}e^{-iH_{0}t}|\mbox{coh}\rangle
=
e^{iH_{0}^{el}t}\langle 0|e^{i\hat{\cal P}_{\rm{\small pu}}}e^{iH_{0}^{pl}t}{\bf A(r)}e^{-iH_{0}^{pl}t}e^{-i\hat{\cal P}_{\rm{\small pu}}}|0\rangle e^{-iH_{0}^{el}t}\nonumber\\
&=&
e^{iH_{0}^{e}t}\bcalA({\bf r},t)e^{-iH_{0}^{e}t}.
\label{eq:Acoh}
\earr
Here $|0\rangle$ is the surface plasmon vacuum, and
\bq
\bcalA({\bf r},t)=\langle \mbox{coh}|e^{iH_{0}^{pl}t}{\bf A}({\bf r})e^{-iH_{0}^{pl}t}|\mbox{coh}\rangle
=
\sum_{\bf Q}\lambda_{\bf Q}[\bcalA_{\bf Q}({\bf r})e^{-i\omega_{\bf Q}t}+\bcalA_{\bf Q}^{\dag}({\bf r})e^{i\omega_{\bf Q}t}]
=
\bcalA_{\parallel}({\bf r},t)+\bcalA_{\perp}({\bf r},t),
\label{eq:Acohsum}
\eq
with real $\lambda_{\bf Q}$ and complex $\bcalA_{\bf Q}({\bf r})$ defined in Eqs. (\ref{eq:lambdaSP}) and (\ref{eq:AandAdag}), respectively, and 
\barr
\bcalA_{\parallel}({\bf r},t)&=&\sum_{\bf Q}\hat{\bf e}_{\bf Q}\lambda_{\bf Q}{\cal A}_{\bf Q}(z)\cos({\bf Q}\brho-\omega_{\bf Q}t), 
\label{eq:Apar} \\
\bcalA_{\perp}({\bf r},t)&=&-\sum_{\bf Q}\hat{\bf e}_{\perp}\lambda_{\bf Q}{\cal A}_{\bf Q}(z)\sin({\bf Q}\brho-\omega_{\bf Q}t),
\label{eq:Aperp}
\earr
where
\bq
{\cal A}_{\bf Q}(z)=2Q(V_{Q}/\omega_{\bf Q})e^{-Qz}.
\label{eq:defcalA}
\eq
We also note that ${\cal A}_{\bf Q}(z)$ and thereby $\bcalA_{\bf Q}({\bf r})$ scale as $\propto a_{B}/L$ due to the presence of this factor in $V_{\bf Q}$ defined in (\ref{eq:V_Q}).

 Due to the coherent state averaging each ${\bf Q}$-component of $\bcalA({\bf r},t)$ in (\ref{eq:Acoh}) and the ensuing transition amplitudes is free from plasmon field operators $\hat{a}_{\bf Q}(t)$ and $\hat{a}_{\bf Q}^{\dag}(t)$ whose role is now replaced by the factors $\lambda_{\bf Q}\exp(-i\omega_{\bf Q}t)$ and $\lambda_{\bf Q}\exp(i\omega_{\bf Q}t)$, respectively. This averaging does not affect the electron operators and their time dependence, as signified by the appearance of $e^{\pm iH_{0}^{e}t}$ in the last line of (\ref{eq:Acoh}). However, as it eliminates the effects of quantum fluctuations of the plasmonic field on the electron propagation, these should be restored separately (cf. Fig. \ref{C1&C2}b). The field $\bcalA({\bf r},t)$ defined in (\ref{eq:Acohsum}) is also divergence-free, viz.
\bq
\nabla\cdot\bcalA({\bf r},t)=0.
\label{eq:nablabcalA}
\eq
The coherent state generated dynamical scalar potential (\ref{eq:calV}), electric field (\ref{eq:bcalE}) and vector potential (\ref{eq:Acohsum}) satisfy Maxwellian-like relations
\bq
\frac{\partial}{\partial t}\bcalA({\bf r},t)=-\nabla {\cal V}({\bf r},t)=\bcalE({\bf r},t).
\label{calVcalA}
\eq
They illustrate the desired connection between the anisotropy of the scalar field ${\cal V}({\bf r},t)$ and the dynamical polarizations of SC vector fields $\bcalA({\bf r},t)$ and $\bcalE({\bf r},t)$. The absence of conventional "$-$" sign in front of $(\partial/\partial t)\bcalA({\bf r},t)$ is due to the choice of $\hat{S}$ in (\ref{eq:S}) that also produces "$+$" sign in front of the linear $\bcalA({\bf r},t)$ terms in (\ref{eq:V'}). Note also that this partial time derivative does not involve the quasi-static electron image potential $\Phi({\bf r})$ defined in (\ref{eq:Phi}) that arises from the same canonical transformation.

Expressions (\ref{eq:Acohsum})-(\ref{eq:Aperp}) establish the desired connection between quantized plasmon fields and their semiclassical counterparts, likewise in the theory of electromagnetic fields.[\onlinecite{GlauberPR}] It also gives the recipe for introducing the dipole approximation to the field in which $\brho{\bf Q}$ can be neglected relative to $\omega_{\bf Q}t$ in the arguments of the trigonometric functions in (\ref{eq:Apar}) and (\ref{eq:Aperp}).

Substitution of (\ref{eq:Acohsum}) into $V'$  defined in (\ref{eq:V'}) yields the Schr\"{o}dinger picture interaction of electron with semiclassical time dependent potential ${\cal V}_{S}'(t)$ in the form
\bq
{\cal V}_{S}'(t)=\frac{{\bf p}\cdot\bcalA({\bf r},t)}{m}+\frac{\bcalA^{2}({\bf r},t)}{2m}.
\label{eq:calV'S}
\eq
This is the velocity gauge analog of the interaction (\ref{eq:calV}). 
 The replacement of ${\bf A(r)}$ by $\bcalA({\bf r},t)$ in (\ref{eq:H'}) brings it to the form of a time-dependent one-electron Hamiltonian
\bq
{\cal H}'={\cal H}'(t)=H_{0}^{el}+{\cal V}_{S}'(t)
\label{eq:calH'}
\eq
which may be used in the calculations of electron transition amplitudes.
From this we derive the interaction picture potential ${\cal V}_{I}'(t)$ from the last line of Eq. (\ref{eq:Acoh}), viz.
\bq
{\cal V}_{I}'(t)=e^{iH_{0}^{e}t}\left(\frac{{\bf p}\cdot\bcalA({\bf r},t)}{m}+\frac{\bcalA^{2}({\bf r},t)}{2m}\right)e^{-iH_{0}^{e}t}.
\label{eq:calVI}
\eq
%



\section{Electron-surface plasmon interaction dynamics in the velocity gauge}
\label{sec:EPdynvg}

\subsection{Electron spectrum in the velocity gauge: Sudden switching on of the interaction}
\label{sec:Elspecvel}

\begin{figure}[tb]
\rotatebox{0}{\epsfxsize=6cm \epsffile{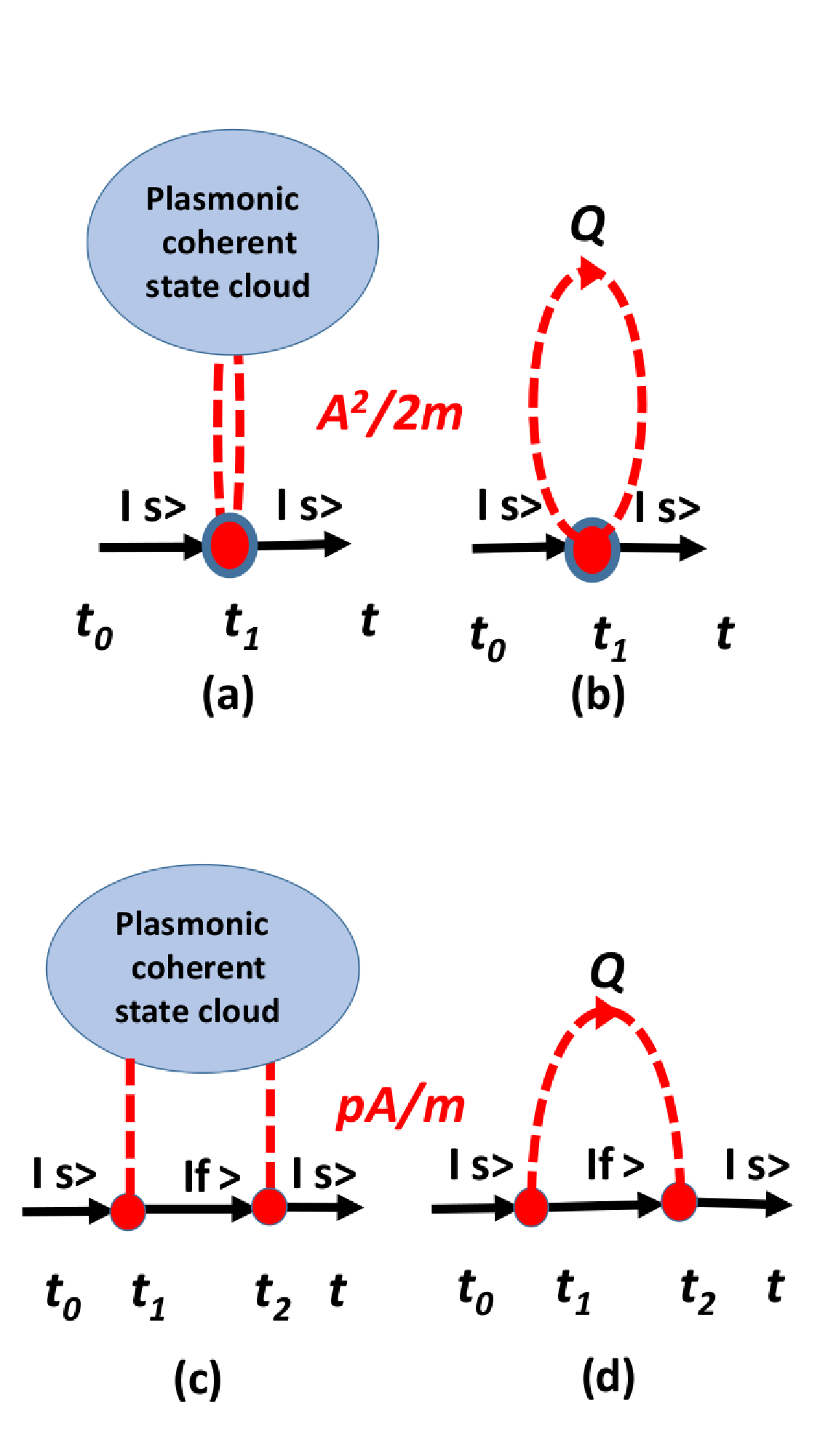}} 
\caption{(a) Sketch of the first order cumulant generated by the ${\bf A}({\bf r})^{2}/2m$ component of the interaction $V_{I}'(t_{1})$ in the velocity gauge (\ref{eq:V'})  over the plasmonic coherent state. (b) First order cumulant arising from quantum fluctuations of the plasmonic ground state that are mediated by the ${\bf Q}$-correlated component of the interaction ${\bf A}({\bf r})^{2}/2m$.  (c) Second order cumulant generated by the electron-plasmon interaction ${\bf p\cdot A}({\bf r})/m$. The latter excites the electron through plasmon emission (absorption) from the  initial state $|s\rangle$  into an intermediate state  $|f\rangle\neq |s\rangle$, and subsequently through plasmon reabsorption (emission) mediated by the coherent state back to the initial state. This generates four processes in the velocity gauge that are analogous to the ones depicted in Fig. \ref{C2lg}(a), (c), (d), (e) for the length gauge. (d) Second order cumulant describing ${\bf Q}$-correlated contribution induced by the interaction ${\bf p}\cdot {\bf A}({\bf r})/m$ without the mediation of coherent state cloud.  Here the  red dashed line symbolize plasmon propagators and filled dots ${\bf A}({\bf r})^{2}/2m$, (a) and (b), and ${\bf p}\cdot {\bf A}({\bf r})/m$, (c) and (d), interaction vertices. The contributions (a) and (c) are summed over $\lambda_{\bf Q}^{2}$ whereas (b) and (d) are free from these factors.}
\label{C1&C2}
\end{figure}

The most convenient point of departure for discussing the interplay of electronic relaxation energies (shifts) and transition amplitudes caused by electron-plasmon interaction expressed in the velocity gauge is again the spectrum of an electron introduced into the  $|{\bf K},s\rangle$-state.   
This is obtained as a Fourier transform of the {\it diagonal} single-electron propagator calculated in the velocity gauge. Denoting by caligraphic symbols this and the related quantities in the length gauge we write  
\bq
{\cal G}_{{\bf K},s}^{(vg)}(t-t_{0})=-i\langle\mbox{coh}|\langle 0|c_{{\bf K},s}(t)c_{{\bf K},s}^{\dag}(t_{0})|0\rangle |\mbox{coh}\rangle\Theta(t-t_{0})
=
-i\Theta(t-t_{0})e^{-iE_{{\bf K},s}(t-t_{0})/\hbar}
\langle\langle\mbox{coh},\phi_{{\bf K},s}||U_{I}(H',t,t_{0})||\phi_{{\bf K},s},\mbox{coh}\rangle\rangle,
\label{eq:Gvel}
\eq
where $|0\rangle$ is the electron vacuum, $c_{{\bf K},s}^{\dag}(t)$ and $c_{{\bf K},s}(t)$ are the electron creation and annihilation operators for the state $|\phi_{{\bf K},s}\rangle=|{\bf K},s\rangle$ in the Heisenberg picture, respectively, with the time dependence of the operators driven by $H'$ defined in the velocity gauge by (\ref{eq:H'}). Likewise in the length gauge  definition (\ref{eq:GKslength}), the imaginary part of the Fourier transform of (\ref{eq:Gvel}) gives the quasiparticle spectrum which contains information on the complete manifold of electron excitation processes induced in the system by the velocity gauge interaction $V'$.  
Note that the diagonal matrix element on the RHS of (\ref{eq:Gvel}) is in contrast to expressions (\ref{eq:Tlength}) and (\ref{eq:TvelSBC}) for the $T$-matrix which are obtained from the {\it non-diagonal} matrix elements of the evolution operator.

The average $\langle\langle\mbox{coh},\phi_{{\bf K},s}||U_{I}(H',t,t_{0})||\phi_{{\bf K},s},\mbox{coh}\rangle\rangle$ on the RHS of (\ref{eq:Gvel}) can be calculated using cumulant expansion [\onlinecite{Kubo}] in terms of specially arranged exponentiated  averages of powers of $V_{I}'(t')$ (\ref{eq:VIS}).[\onlinecite{GW+C}] This yields
\bq
{\cal G}_{{\bf K},s}^{(vg)}(t-t_{0})=-i\Theta(t-t_{0})e^{-iE_{{\bf K},s}(t-t_{0})/\hbar}\exp[{\cal C}_{{\bf K},s}^{(vg)}(t-t_{0})].
\label{eq:Gvelcum}
\eq
Here ${\cal C}_{{\bf K},s}^{(vg)}(t-t_{0})$ is the sum of all order cumulants generated on the state $|{\bf K},s\rangle|\mbox{coh}\rangle$ by the interaction $V_{I}'$   
\bq
{\cal C}_{{\bf K},s}^{(vg)}(t-t_{0})=\sum_{n=1}^{\infty} {\cal C}_{{\bf K},s}^{(vg,n)}(t-t_{0}).
\label{eq:Cvelcum}
\eq

Likewise in Sec. \ref{sec:EPdynlg} we shall first set $t_{0}=0$ in (\ref{eq:Gvelcum}) and ensuing expressions. Next we recall that in the weakly correlated scattering regime[\onlinecite{Dunn,Mahanbook,4cumul}] only the lowest order cumulants ${\cal C}_{{\bf K},s}^{(vg,1)}(t)$ and ${\cal C}_{{\bf K},s}^{(vg,2)}(t)$ that are linear and quadratic in $V_{I}(t')$, respectively, significantly contribute to the cumulant sum (\ref{eq:Cvelcum}). These are shown in Fig. \ref{C1&C2} (a)-(d). This brings expression (\ref{eq:Gvel}) to the form 
\bq
{\cal G}_{{\bf K},s}^{(vg)}(t)=
-i\Theta(t) e^{-i E_{{\bf K},s}t/\hbar}
\exp[{\cal C}_{{\bf K},s}^{(vg,1)}(t)+{\cal C}_{{\bf K},s}^{(vg,2)}(t)]. 
\label{eq:GKsVolkov}
\eq

To calculate the cumulants we first use the definitions (\ref{eq:A})-(\ref{eq:A+}) to express the interaction $V'$ defined in (\ref{eq:V'}) in the second quantization and interaction representation
\bq
V'_{I}(t)=V_{I}^{'(lin)}(t)+V_{I}^{'(qu)}(t)
\label{eq:Vlinqu}
\eq 
where
\begin{widetext}
\barr
V_{I}^{'(lin)}(t)&=&\sum_{j,i,{\bf Q}}
\left[\langle j|\frac{{\bf p}\cdot\bcalA_{\bf Q}({\bf r})}{m}| i\rangle c_{j}^{\dag}c_{i}\hat{a}_{\bf Q}e^{i(E_{j}-E_{i}-\hbar\omega_{sp})t}
+
\langle j|\frac{{\bf p}\cdot\bcalA_{\bf Q}^{\dag}({\bf r})}{m}|i\rangle c_{j}^{\dag}c_{i}\hat{a}_{\bf Q}^{\dag}e^{i(E_{j}-E_{i}+\hbar\omega_{sp})t}\right],
\label{eq:VIlin}\\
V_{I}^{'(qu)}(t)
&=&
\sum_{j,i}\sum_{\bf Q,Q'}\left[\langle j|\frac{\bcalA_{\bf Q}({\bf r})\bcalA_{\bf Q'}({\bf r})}{2m}|i\rangle  e^{i(E_{j}-E_{i}-2\hbar\omega_{sp})t_{1}}  
c_{j}^{\dag}c_{i}\hat{a}_{\bf Q}\hat{a}_{\bf Q'} 
+
\langle j|\frac{\bcalA_{\bf Q}^{\dag}({\bf r})\bcalA_{\bf Q'}^{\dag}({\bf r})}{2m}| i\rangle e^{i(E_{j}-E_{i}+2\hbar\omega_{sp})t_{1}} c_{j}^{\dag}c_{i}
\hat{a}_{\bf Q}^{\dag}\hat{a}_{\bf Q'}^{\dag}\right.\nonumber\\
&+&
\left.\langle j|\frac{\bcalA_{\bf Q}^{\dag}({\bf r})\bcalA_{\bf Q'}({\bf r})}{2m}|i\rangle e^{i(E_{j}-E_{i})t_{1}}c_{j}^{\dag}c_{i}
\hat{a}_{\bf Q}^{\dag}\hat{a}_{\bf Q'}
+
\langle j|\frac{\bcalA_{\bf Q}({\bf r})\bcalA_{\bf Q'}^{\dag}({\bf r})}{2m}|i\rangle e^{i(E_{j}-E_{i})t_{1}}c_{j}^{\dag}c_{i}
\hat{a}_{\bf Q}\hat{a}_{\bf Q'}^{\dag}\right]. 
\label{eq:VIqu}
\earr
\end{widetext}
In the first and second lines of (\ref{eq:VIqu}) we have retained for notational convenience the same order of dummy indices $i,j,{\bf Q}$ and ${\bf Q'}$ between the hermitian conjugate terms.
\subsubsection{First order cumulant in the velocity gauge and ponderomotive effects}
\label{sec:661}

The contribution of (\ref{eq:VIlin}) to the first order cumulant contains the average of the dipole operator over the initial state, i.e $\langle\langle \mbox{coh};{\bf K},s|{\bf p}\cdot\bcalA_{\bf Q}({\bf r})/m|{\bf K},s;\mbox{ coh}\rangle\rangle$. Hence, owing to the selection rules it becomes negligible in the long wavelength limit ${\bf Q}\rightarrow0$ that is assumed in the present model of electron-plasmon interaction satisfying the Coulomb gauge. By contrast, interaction (\ref{eq:VIqu}) gives a nonzero first order cumulant illustrated diagrammatically in Figs. \ref{C1&C2}(a) and \ref{C1&C2}(b). It takes the form
\barr
{\cal C}_{1}^{(vg)}({\bf K},s,t)
&=&
-\frac{i}{2m\hbar}\int_{0}^{t}dt_{1}\sum_{\bf Q,Q'}\lambda_{\bf Q}\lambda_{\bf Q'}
\left[\langle{\bf K},s|\bcalA_{\bf Q}^{\dag}({\bf r})\bcalA_{\bf Q'}({\bf r})|{\bf K},s\rangle\right.
+
\langle {\bf K},s|\bcalA_{\bf Q}({\bf r})\bcalA_{\bf Q'}^{\dag}({\bf r})|{\bf K},s\rangle \nonumber\\
&+&
\langle {\bf K},s|\bcalA_{\bf Q}({\bf r})\bcalA_{\bf Q'}({\bf r})|{\bf K},s\rangle  e^{-2i\hbar\omega_{sp}t_{1}}
+
\left.\langle {\bf K},s|\bcalA_{\bf Q}^{\dag}({\bf r})\bcalA_{\bf Q'}^{\dag}({\bf r})|{\bf K},s\rangle e^{2i\hbar\omega_{sp}t_{1}}\right]
\nonumber\\
&-&
\frac{i}{2m\hbar}\int_{0}^{t}dt_{1}\sum_{\bf Q}\langle {\bf K},s|\bcalA_{\bf Q}({\bf r})\bcalA_{\bf Q}^{\dag}({\bf r})|{\bf K},s\rangle 
\label{eq:calC1qu}
\earr
in which the lateral momentum conservation makes the four matrix elements in the square bracket in (\ref{eq:calC1qu}) proportional to  $\delta_{\bf Q,Q'}$, $\delta_{\bf Q,Q'}$, $\delta_{\bf Q,-Q'}$ and $\delta_{\bf Q,-Q'}$, respectively, as well as independent of ${\bf K}$. The last term in (\ref{eq:calC1qu}) is a quantum fluctuation correction arising from the noncommutativity of the operators $\hat{a}_{\bf Q}$ and $\hat{a}_{\bf Q}^{\dag}$. In a way, the diagrams in Figs. \ref{C1&C2}(a) and \ref{C1&C2}(b) and their transcriptions  in  (\ref{eq:calC1qu}) embody the processes that are analogs of the ones depicted in Fig. \ref{C1&C2} but shrunk to zero intermediate propagation times and momentum recoil. This makes possible the various afore mentioned combinations of multiplasmon (de)excitations. 

Assuming $\lambda_{\bf Q}=\lambda_{\bf -Q}$ and using (\ref{eq:AandAdag})-(\ref{eq:A+}) we obtain
\barr
{\cal C}_{1}^{(vg)}({\bf K},s,t)
&=&
-\frac{i}{\hbar}\int_{0}^{t}dt_{1}\sum_{\bf Q} 2\lambda_{\bf Q}^{2}\langle
{\bf K},s|\frac{\bcalA_{\bf Q}({\bf r})\bcalA_{\bf Q}^{\dag}({\bf r})}{2m}|{\bf K},s\rangle
\left[1-\cos(2\omega_{sp}t_{1})\right]
\label{eq:2omega}\\
&-&
\frac{i}{\hbar}\int_{t_{0}}^{t}dt_{1}\sum_{\bf Q}\langle {\bf K},s|\frac{\bcalA_{\bf Q}({\bf r})\bcalA_{\bf Q}^{\dag}({\bf r})}{2m}|{\bf K},s\rangle, 
\label{eq:cohfluc}
\earr
Here the factors $e^{i{\bf Q}\brho}$ and $e^{-i{\bf Q}\brho}$ from the matrix elements $\langle{\bf K}, s|\bcalA_{\bf Q}({\bf r})\bcalA_{\bf Q}^{\dag}({\bf r})|{\bf K},s\rangle$ cancel each other and therefore there is no further constraint on ${\bf Q}$-summation due to the lateral momentum conservation in the average $\langle {\bf K},s|\dots|{\bf K},s\rangle$.
Thus, the contributions to the lowest, first order cumulant ${\cal C}_{1}^{(vg)}({\bf K},s,t)$ can be decomposed as 
\bq
{\cal C}_{1}^{(vg)}({\bf K},s,t)
=
{\cal C}_{1}^{(vg)}(\mbox{pond};{\bf K},s,t)+{\cal C}_{1}^{(vg)}(2\omega_{sp};{\bf K},s,t)
+
{\cal C}_{1}^{(vg)}(\mbox{fluc};{\bf K},s,t)
\label{eq:C1full}
\eq
and all arise from quadratic electron coupling to plasmon field. 
The first contribution ${\cal C}_{1}^{(vg)}(\mbox{pond};{\bf K},s,t)$ originating  from the factor 1 in the square bracket in (\ref{eq:2omega}) gives rise to ponderomotive energy shift  $U_{s}$ of the $s$-state that is induced by fluctuations of the plasmonic coherent state. We can represent this $\lambda_{\bf Q}^{2}$-generated positive energy correction, the so-called ponderomotive shift,  as 
\bq
U_{s}=2\sum_{\bf Q} \lambda_{\bf Q}^{2}\langle {\bf K},s|\frac{\bcalA_{\bf Q}({\bf r})\bcalA_{\bf Q}^{\dag}({\bf r})}{2m}|{\bf K},s\rangle,
\label{eq:beta_s}
\eq
and write for the linear-in-$t$ contribution from (\ref{eq:2omega})
\bq
{\cal C}_{1}^{(vg)}(\mbox{pond};{\bf K},s,t)=-\frac{i}{\hbar}U_{s}t. 
\label{eq:Usp}
\eq
We observe that the contributions to (\ref{eq:2omega}) arise from instantaneous quadratic coupling whose time dependence $\propto\cos^{2}(\omega_{sp}t_{1})$ from the zero-duration intermediate state interval $(t_{1},t_{1})$ is integrated over the total propagation interval $(0,t)$, thereby producing the upward ponderomotive shift (\ref{eq:Usp}). In other words, the ponderomotive shift arises from a ${\bf K}$-diagonal and time-local interaction. There is no such a counterpart in the length gauge. 
In the case of spatially inhomogeneous vector potentials ${\bf A}({\bf r},t)$ this shift is according to (\ref{eq:beta_s}) state-specific.
Consecutive linear couplings discussed in the second part of this  subsection exhibit different polarization shifts associated with propagation in intermediate states and these can be either positive or negative [cf. Eqs. (\ref{eq:nus+}) and (\ref{eq:nus-})].     

Next, we represent the oscillatory contribution from (\ref{eq:2omega}) in the  form of a $2\omega_{sp}$-excitation component  
\bq
{\cal C}_{1}^{(vg)}(2\omega_{sp};{\bf K},s,t)=i\frac{U_{s}}{2\hbar\omega_{sp}}\sin(2\omega_{sp}t)=i\beta_{sp}\sin(2\omega_{sp}t),
\label{eq:C2omegasp}
\eq
where we have introduced 
\bq
\beta_{sp}=\frac{U_{s}}{2\hbar\omega_{sp}}.
\label{eq:betas}
\eq
Finally, the remaining term (\ref{eq:cohfluc}) integrates to 
\bq
{\cal C}_{1}^{(vg)}(\mbox{fluc};{\bf K},s,t)=-\frac{i}{\hbar}\int_{0}^{t}dt_{1}\sum_{\bf Q}\langle {\bf K},s|\frac{\bcalA_{\bf Q}({\bf r})\bcalA_{\bf Q}^{\dag}({\bf r})}{2m}|{\bf K},s\rangle
=
-\frac{i}{\hbar}\Phi_{s}t
\label{eq:C1Phi}
\eq
which describes an upward $\lambda_{\bf Q}$-independent shift of the $s$-level that is induced by quantum fluctuations of the plasmonic ground state.  Expressions (\ref{eq:Usp}) and (\ref{eq:C1Phi}) can be viewed as instantaneous (time local) lowest order self-energy corrections whereas (\ref{eq:C2omegasp}) bears the features of vertex corrections.

The lowest order result for the spectrum ${\cal N}^{(vg)}({\bf K},s,\hbar\omega)$ of ${\cal G}_{{\bf K},s}^{(vg)}(t)$ defined in (\ref{eq:Gvel}) is obtained if only (\ref{eq:C1full}) is retained in the calculation. In this case we may use the generating function formula (\ref{eq:BesselJ}) to write $\exp[i\beta_{sp}\sin(2\omega_{sp}t)]=\sum_{n=-\infty}^{\infty}J_{n}(\beta_{sp})\exp(2in\omega_{sp}t)$ to obtain the lowest order contribution to the spectrum
\begin{widetext}
\barr
{\cal N}_{1}^{(vg)}({\bf K},s,\hbar\omega)&=&\frac{1}{2\pi\hbar}\int_{-\infty}^{\infty}dt e^{i\left(\hbar\omega -E_{{\bf K},s}-U_{s}-\Phi_{{\bf K},s}\right)t/\hbar} \exp\left[i\beta_{sp}\sin(2\omega_{sp}t)\right]\nonumber\\
&=&
\frac{1}{2\pi\hbar}\int_{-\infty}^{\infty}dt e^{i\left(\hbar\omega -E_{{\bf K},s}-U_{s}-\Phi_{s}\right)t/\hbar}\sum_{n=-\infty}^{\infty}J_{n}\left(\beta_{sp}\right)e^{-i2n\omega_{sp}t}\nonumber\\
&=&
\sum_{n=-\infty}^{\infty}J_{n}(\beta_{sp})\delta\left(\hbar\omega -E_{{\bf K},s}-U_{s}-\Phi_{s}-2n\hbar\omega_{sp}\right).  
\label{eq:calN1}
\earr
\end{widetext} 

The quantities $U_{s}$, $\beta_{sp}$ and $\Phi_{s}$ determining expression (\ref{eq:calN1}) will be explicitly evaluated in the dipole approximation in the context of plasmoemission from Ag surface bands in Sec. \ref{sec:vgdynamical}. Since in all ensuing velocity gauge expressions involving  $U_{s}$ and $\Phi_{s}$ these two shifts appear additive, we shall replace their sum by 
\bq
\tilde{U}_{s}=U_{s}+\Phi_{s}.
\label{eq:U+Phi}
\eq
The separate notations for $U_{s}$ and $\Phi_{s}$ will be restored in the final scaling expressions (\ref{eq:Ulambda}) and (\ref{eq:Phislambda1}).

\subsubsection{Second order cumulants in the velocity gauge}
\label{sec:662}

Contributions to the second order cumulant that are also quadratic in $\lambda_{\bf Q}$ arise from linear electron-plasmon coupling (\ref{eq:VIlin}), i.e. they are quadratic in the interaction vertices generated by the interaction ${\bf p\cdot A}/m$ defined in the interaction representation in (\ref{eq:VIlin}). The five contributions arising from the coherent state averages of the plasmon creation and annihilation operators (\ref{eq:cohaverage1})-(\ref{eq:cohaverage4}) are completely analogous to the ones sketched in Fig. \ref{C2lg}, the only difference being in the meaning of the dots denoting the interaction vertices. Here they are associated with the matrix elements $\langle {\bf K}_{f},f|{\bf p}\cdot\bcalA_{\bf Q}({\bf r})/m|{\bf K},s\rangle$. Hence we can repeat step by step the derivation of the second order cumulants from Sec. \ref{sec:EspectrumLG}, but now in terms of the linear velocity gauge component of the interaction (\ref{eq:VIlin}). The momentum selection rules in the $s\rightarrow f$ transitions restrict this phase space to the one discussed in Sec. \ref{sec:EspectrumLG}. Thus we obtain 
\begin{widetext}
\barr
C_{2}^{(vg)}(1\omega_{sp};{\bf K},s,t)=
-
2v_{{\bf K},s}^{(vg)}(1\omega_{sp})
+
{\cal M}_{{\bf K},s}^{(vg)}(\mbox{em};t)e^{-i\omega_{sp}t}
+
{\cal M}_{{\bf K},s}^{(vg)}(\mbox{abs};t)e^{i\omega_{sp}t}
-
i\Omega_{{\bf K},s}^{(vg)}t.
\label{eq:C1C2vg}
\earr 
\end{widetext}
Here the normalization constant or the DWE for the elastic line of the spectrum reads
\bq
-2v_{{\bf K},s}^{(vg)}(1\omega_{sp})=-\left[{\cal M}_{{\bf K},s}^{(vg)}(\mbox{em};0)+{\cal M}_{{\bf K},s}^{(vg)}(\mbox{abs};0)\right] 
\label{eq:DWEvg1}
\eq
and the probabilities of one plasmon creation followed by reabsorption are given by 
\bq
{\cal M}_{{\bf K},s}^{(vg)}(\mbox{em};t)=\sum_{f,{\bf Q}}(\lambda_{\bf Q}^{2}+1)\frac{|\langle f|{\bf p}\cdot \bcalA_{\bf Q}({\bf r})^{\dag}/m|{\bf K},s\rangle|^{2}}{(E_{{\bf K},s}-E_{f}-\hbar\omega_{sp})^{2}}
e^{i(E_{{\bf K},s}-E_{f})t/\hbar},
\label{eq:Us+}
\eq
and for absorption followed by creation by
\bq
{\cal M}_{{\bf K},s}^{(vg)}(\mbox{abs};t)=\sum_{f,{\bf Q}}\lambda_{\bf Q}^{2}\frac{|\langle f|{\bf p}\cdot \bcalA_{\bf Q}({\bf r})/m|{\bf K},s\rangle|^{2}}{(E_{{\bf K},s}-E_{f}+\hbar\omega_{sp})^{2}}
e^{i(E_{{\bf K},s}-E_{f})t/\hbar}.
\label{eq:Us-}
\eq
In (\ref{eq:Us+}) and (\ref{eq:Us-}) and hereafter we conveniently use the abbreviated notation for the excited state $|f\rangle=|{\bf K}_{f},f\rangle$, and analogously so for the corresponding energies and summation indices.

The energy relaxation shifts from these processes are 
\barr
\hbar\Omega_{{\bf K},s}^{(vg)}&=&\hbar\Omega_{{\bf K},s}^{(vg)}(\mbox{em})+\hbar\Omega_{{\bf K},s}^{(vg)}(\mbox{abs})=\mbox{Re}\Sigma_{{\bf K},s}^{(vg)},
\label{eq:nus}\\
\Omega_{{\bf K},s}^{(vg)}(\mbox{em})&=&\sum_{f,{\bf Q}}(\lambda_{\bf Q}^{2}+1)\frac{|\langle f|{\bf p}\cdot\bcalA_{\bf Q}({\bf r})^{\dag}/m|{\bf K},s\rangle |^{2}}{\hbar(E_{{\bf K},s}-E_{f}-\hbar\omega_{sp})},
\label{eq:nus+}\\
\Omega_{{\bf K},s}^{(vg)}(\mbox{abs})&=&\sum_{f,{\bf Q}}\lambda_{\bf Q}^{2}\frac{|\langle f|{\bf p}\cdot\bcalA_{\bf Q}({\bf r})/m|{\bf K},s\rangle |^{2}}{\hbar(E_{{\bf K},s}-E_{f}+\hbar\omega_{sp})}.
\label{eq:nus-}
\earr

It is seen that the above quantities ${\cal M}_{{\bf K},s}^{(vg)}(+1\omega_{sp})$, ${\cal M}_{{\bf K},s}^{(vg)}(-1\omega_{sp})$, $\Omega_{{\bf K},s}^{(vg)}(\mbox{em})$ and $\Omega_{{\bf K},s}^{(vg)}(\mbox{abs})$ play the same role as do (\ref{eq:Ms+}), (\ref{eq:Ms-}), (\ref{eq:Omegas+}) and (\ref{eq:Omegas-}), respectively, in the formulation of the electron spectrum in the length gauge in Sec. \ref{sec:EspectrumLG}. Therefore they generate functionally equivalent contributions to the electron spectrum also in the velocity gauge. In both gauges their form reflects nonadiabatic electron-plasmon interaction dynamics due to the sudden switching on of the interaction at $t_{0}=0$.

The spectrum of (\ref{eq:Gvel}) arising from (\ref{eq:C1C2vg}) alone is
\begin{widetext}
\barr
{\cal N}_{2}^{(vg)}({\bf K},s,\hbar\omega)&=&\int_{-\infty}^{\infty}\frac{dt}{2\pi\hbar} e^{i\left(\hbar\omega -E_{{\bf K},s}\right)t/\hbar}\exp\left[C_{2}^{(vg)}(1\omega_{sp};{\bf K},s,t)\right]\nonumber\\ 
&=& 
\int_{-\infty}^{\infty}\frac{dt}{2\pi\hbar} e^{i\left(\hbar\omega -E_{{\bf K},s}-\Omega_{{\bf K},s}^{(vg)}\right)t/\hbar}e^{-2v_{{\bf K},s}^{(vg)}(1\omega_{sp})}
\exp\left[{\cal M}_{{\bf K},s}^{(vg)}(\mbox{em};t)e^{-i\omega_{sp}t}+{\cal M}_{{\bf K},s}^{(vg)}(\mbox{abs};t)e^{i\omega_{sp}t}\right]\nonumber\\
&=&
\int_{-\infty}^{\infty}\frac{dt}{2\pi\hbar} e^{i\left(\hbar\omega -E_{{\bf K},s}-\Omega_{{\bf K},s}^{(vg)}\right)t/\hbar}e^{-2v_{{\bf K},s}^{(vg)}(1\omega_{sp})}\nonumber\\
&\times&
\sum_{n=-\infty}^{\infty}e^{i n\omega_{sp}t}\left(\sqrt{\frac{{\cal M}_{{\bf K},s}^{(vg)}(\mbox{abs};t)}{{\cal M}_{{\bf K},s}^{(vg)}(\mbox{em};t)}}\right)^{n}I_{n}\left(2\sqrt{{\cal M}_{{\bf K},s}^{(vg)}(\mbox{em};t){\cal M}_{{\bf K},s}^{(vg)}(\mbox{abs};t)}\right)
\label{eq:calN2}
\earr
\end{widetext} 
where the last line has been obtained in analogy with (\ref{eq:expAB(t)}).  

Next we must also account for the $\pm 2\omega_{sp}$ contributions induced by the interaction ${\bf p}\cdot{\bf A}/m$ and mediated by the plasmonic coherent state cloud. They are derived in a completely analogous fashion as expressions (\ref{eq:C2omegas+}) and (\ref{eq:C2omegas-}) and are described by 
\begin{widetext}
\bq
\exp[C_{2}^{(vg)}(2\omega_{sp};{\bf K},s,t)]
=
e^{-2v_{{\bf K},s}^{(vg)}(2\omega_{sp})}\exp\left[-\left({\cal P}_{{\bf K},s}^{(vg)}(\mbox{em2})e^{2i\omega_{sp}t}-{\cal P}_{{\bf K},s}^{(vg)}(\mbox{abs2})e^{-2i\omega_{sp}t}\right)+2{\cal Q}_{{\bf K},s}^{(vg)}(t)\cos(\omega_{sp}t)\right],
\label{eq:Cvg2omegas}
\eq
\end{widetext}
where for $\mbox{em2}=+2\omega_{sp}$ and $\mbox{abs2}=-2\omega_{sp}$ we have
\bq
{\cal P}_{{\bf K},s}^{(vg)}(\pm 2\omega_{sp})=\sum_{{\bf Q},f}\lambda_{\bf Q}^{2}\frac{|\langle f|{\bf p}\cdot \bcalA_{\bf Q}({\bf r})^{\dag}/m|{\bf K},s\rangle|^{2}}{2\hbar\omega_{sp}(E_{{\bf K},s}-E_{f}\mp\hbar\omega_{sp})},
\label{eq:calPvg-+}
\eq
\bq
{\cal Q}_{{\bf K},s}^{(vg)}(t)=\sum_{{\bf Q},f}\lambda_{\bf Q}^{2}
\frac{|\langle f|{\bf p}\cdot \bcalA_{\bf Q}({\bf r})^{\dag}/m|{\bf K},s\rangle|^{2}}{(E_{{\bf K},s}-E_{f})^{2}-(\hbar\omega_{sp})^{2}}
e^{i(E_{{\bf K},s}-E_{f})t/\hbar},
\label{eq:calQvg}
\eq
\bq
2v_{{\bf K},s}^{(vg)}(2\omega_{sp})=-\left({\cal P}_{{\bf K},s}^{(vg)}(\mbox{em2})-{\cal P}_{{\bf K},s}^{(vg)}(\mbox{abs2})\right)+
2{\cal Q}_{{\bf K},s}^{(vg)}(0).
\label{eq:v2omegavg}
\eq

We can now group the various $\pm 1\omega_{sp}$ and $\pm 2\omega_{sp}$, as well as the relaxation energy contributions from (\ref{eq:C1full}), (\ref{eq:C1C2vg}) and (\ref{eq:Cvg2omegas}) to obtain the RHS of (\ref{eq:GKsVolkov}) in the form ($t_{0}=0$)
\begin{widetext}
\barr
{\cal G}_{{\bf K},s}^{(vg)}(t)
&=&
-i\Theta(t)e^{-iE_{{\bf K},s}t/\hbar}\exp[{\cal C}_{1}^{(vg)}({\bf K},s,t)+{\cal C}_{2}^{(vg)}(1\omega_{sp};{\bf K},s,t)+{\cal C}_{2}^{(vg)}(2\omega_{sp};{\bf K},s,t)]\nonumber\\
&=& 
-i\Theta(t)e^{-i\left(E_{{\bf K},s}+\tilde{U}_{s}+\Omega_{{\bf K},s}^{(vg)}\right)t/\hbar} e^{-\left(2v_{{\bf K},s}^{(vg)}(1\omega_{sp})+2v_{{\bf K},s}^{(vg)}(2\omega_{sp})\right)}\nonumber\\
&\times&
\exp\left[\left({\cal M}_{{\bf K},s}^{(vg)}(\mbox{em};t)+{\cal Q}_{{\bf K},s}^{(vg)}(t)\right)e^{-i\omega_{sp}t}
+
\left({\cal M}_{{\bf K},s}^{(vg)}(\mbox{abs};t)+{\cal Q}_{{\bf K},s}^{(vg)}(t)\right)e^{i\omega_{sp}t}
\right]
\label{eq:Cvg11}\\
&\times&
\exp\left[\left(\frac{\beta_{sp}}{2}-{\cal P}_{{\bf K},s}^{(vg)}(\mbox{em2})\right)e^{2i\omega_{sp}t}-\left(\frac{\beta_{sp}}{2}-{\cal P}_{{\bf K},s}^{(vg)}(\mbox{abs2})\right)e^{-2i\omega_{sp}t}\right].
\label{eq:Cvg12}
\earr
\end{widetext}

Representing in expressions (\ref{eq:Cvg11}) and (\ref{eq:Cvg12}) the exponential functions of exponential functions $e^{\mp i\omega_{sp}t}$ and $e^{\mp 2i\omega_{sp}t}$ through the generating functions of Bessel functions  we find that the spectrum of ${\cal G}_{{\bf K},s}^{(vg)}(t)$ is given by the convolution of the spectra with the structure similar to (\ref{eq:calN1}) and (\ref{eq:calN2}) and involving the Bessel functions $J_{n}(x)$ and $I_{n}(x)$,  thereby revealing the effects of juxtaposed multiples of lowest order vertex and self-energy corrections, respectively. This leads to the Floquet band structure  illustrated schematically in Fig. \ref{SPFloquet}.

So far we have restricted our discussion to contributions of the first and second order cumulants only (cf. Fig. \ref{C1&C2}(a)-(d)) induced by the commuting interactions ${\bf p\cdot A(r)}/m$ and ${\bf A(r)}^{2}/2m$ that  yield contributions $\propto \lambda_{\bf Q}^{2}$. A typical example of higher order cumulants is a third order one comprising two consecutive one-plasmon interactions ${\bf p\cdot A(r)}/m$ terminating in the two-plasmon one ${\bf A(r)}^{2}/2m$. If proceeding  via the coherent state they are integrated over the factors $\lambda_{\bf Q}^4$, in contrast to processes from Fig. \ref{C1&C2}(a) and (c) which are integrated over $\lambda_{\bf Q}^2$. 

\subsection{Adiabatic switching on of the interaction}
\label{sec:vgSBC}

Using the results of the previous subsection we can now inspect the form of electron spectrum in the case of adiabatic switching on of the electron-plasmon interaction in the velocity gauge that is appropriate to the scattering boundary conditions. The form of (\ref{eq:C1full}) leading to (\ref{eq:calN1}) remains unchanged but differences appear in the second order cumulant. Here we can follow step by step the derivations of the corresponding quantities in the length gauge to obtain the velocity gauge counterparts
\begin{widetext}
\barr
{\cal C}_{{\bf K},s}^{(vg,2)}(1\omega_{sp};t)
&=&
-\frac{i}{\hbar}\sum_{{\bf Q},f}|\langle f|\frac{{\bf p}\cdot {\bf A}^{\dag}({\bf Q,r})}{m}|{\bf K},s\rangle|^{2}
\left[\frac{(\lambda_{\bf Q}^{2}+1)}{E_{{\bf K},s}-E_{f}-\hbar\omega_{sp}+i\delta}+\frac{\lambda_{\bf Q}^{2}}{E_{{\bf K},s}-E_{f}+\hbar\omega_{sp}+i\delta}\right]t\nonumber\\
&=&
-i\Omega_{{\bf K},s}^{(vg)}t-\Gamma_{{\bf K},s}^{(vg)}(1\omega_{sp})t=-\frac{i}{\hbar}\Sigma_{{\bf K},s}^{(vg)}(1\omega_{sp})t,
\label{eq:C1advg}
\earr
where
\barr
\Gamma_{{\bf K},s}^{(vg)}(1\omega_{sp})
&=&
\frac{\pi}{\hbar}\sum_{{\bf Q},f}|\langle f|\frac{{\bf p}\cdot {\bf A}^{\dag}({\bf Q,r})}{m}|{\bf K},s\rangle|^{2}
\left[(\lambda_{\bf Q}^{2}+1)\delta({E_{{\bf K},s}-E_{f}-\hbar\omega_{sp}})+\lambda_{\bf Q}^{2}\delta(E_{{\bf K},s}-E_{f}+\hbar\omega_{sp})\right]\nonumber\\
&=&
\Gamma_{{\bf K},s}^{(vg)}(\mbox{vac})+\Gamma_{{\bf K},s}^{(vg)}(\mbox{em})+\Gamma_{{\bf K},s}^{(vg)}(\mbox{abs}),
\label{eq:Gamma1vg}
\earr
\end{widetext}
with analogous divisions as in (\ref{eq:Gammavac})-(\ref{eq:Gamma1abs}). 

In a similar fashion, the $\pm 2\omega_{sp}$ processes give rise to
\begin{widetext}
\barr
{\cal C}_{2}^{(vg)}(2\omega_{sp};{\bf K},s,t)
&=&
-\left(\sum_{{\bf Q},f}\lambda_{\bf Q}^{2}\frac{|\langle f|{\bf p}\cdot {\bf A}^{\dag}({\bf Q,r})/m|{\bf K},s\rangle|^{2}}{2\hbar\omega_{sp}(E_{{\bf K},s}-E_{f}-\hbar\omega_{sp}+i\delta)}e^{2i\omega_{sp}t}
- 
\sum_{{\bf Q},f}\lambda_{\bf Q}^{2}\frac{|\langle f|{\bf p}\cdot {\bf A}^{\dag}({\bf Q,r})/m|{\bf K},s\rangle|^{2}}{2\hbar\omega_{sp}(E_{{\bf K},s}-E_{f}+\hbar\omega_{sp}+i\delta)}e^{-2i\omega_{sp}t}\right)\nonumber\\
&=&
-\left({\cal P}_{{\bf K},s}^{(vg)}(\mbox{em2})e^{2i\omega_{sp}t}-{\cal P}_{{\bf K},s}^{(vg)}(\mbox{abs2})e^{-2i\omega_{sp}t}\right)
=
-\left(\frac{\Sigma_{{\bf K},s}^{(vg)}(\mbox{em1})}{2\hbar\omega_{sp}}e^{2i\omega_{sp}t}-\frac{\Sigma_{{\bf K},s}^{(vg)}(\mbox{abs1})}{2\hbar\omega_{sp}}e^{-2i\omega_{sp}t}\right),
\label{eq:Cvg22omega}
\earr
\end{widetext}
which is again reminiscent of the Ward-Pitaevskii relations for vertex corrections with differentials replaced by energy differences. Note also the absence of the factors $\frac{\beta_{sp}}{2}$ (i.e. Debye-Waller exponents) present in the case of transient switching on of the interaction. 
Likewise in expression (\ref{eq:C2lgSigma}), the same argument applies here regarding the contributions from the real and imaginary terms from the energy denominators in (\ref{eq:Cvg22omega}). Hence, the same representation in terms of generating functions of Bessel functions can be used as well. 
Lastly, we note that $2\hbar\omega_{sp}$-contributions to second order cumulants in either gauge, viz. (\ref{eq:C2omegas}) and (\ref{eq:Cvg2omegas}), scale with $\lambda_{\bf Q}^{2}$ and therefore do not appear in the case of cumulant averaging over the ground state.

\subsection{Transition rates from cumulant expansion in the velocity gauge}
\label{sec:Gammavg}

To calculate the transition rate $\Gamma_{{\bf K}_{s},s}^{(vg)}(\mbox{abs})$ from (\ref{eq:Gamma1vg}) that corresponds to plasmoemission from the SS-band on Ag(111) surface (i.e. processes corresponding to the plasmon line arrows in Fig. \ref{C2lg}(c) weighted by $\lambda^{2}$) we now introduce the partition of ${\bf A}({\bf r})$ into paralel and perpendicular components following definitions (\ref{eq:A-}) and (\ref{eq:A+}). This produces the one-plasmon absorption components of the matrix elements of the interaction
\bq
\langle \mbox{coh}|\langle {\bf K}_{f},k_{z,f}|\left(\frac{{\bf A}_{\parallel}\cdot{\bf p}_{\parallel}}{m^{*}}+\frac{{\bf A}_{\perp}\cdot{\bf p}_{\perp}}{m} \right)|{\bf K}_{s},u_{s}\rangle|\mbox{coh}\rangle
=
\sum_{\bf Q}\lambda_{\bf Q} \frac{V_{\bf Q}}{\hbar\omega_{sp}}\left(\frac{\hbar^2{\bf K}_{s}{\bf Q}}{m^{*}}+i\frac{\hbar^{2}Q k_{z,f}}{m}\right)f_{f,s}(Q)
\delta_{{\bf K}_{f},{\bf K}_{s}+{\bf Q}},
\label{eq:<Ap>}
\eq
where we have denoted the effective mass $m^{*}$ in the operators acting on the wavefunctions describing lateral electron motion in SS-bands.
The corresponding one-plasmon emission components are given by the hermitian conjugate of expression (\ref{eq:<Ap>}) and the replacement $\lambda_{\bf Q}\rightarrow \lambda_{\bf Q}+1$. Therefore, the plasmon absorption and re-emission transition rate with the dimension of inverse time takes the form

\begin{widetext}
\barr
2\Gamma_{{\bf K}_{s},s}^{(vg)}(\mbox{abs1})&=&\frac{2\pi}{\hbar}\sum_{{\bf K}_{f},k_{z,f}}\sum_{\bf Q}\lambda_{\bf Q}^{2}\left(\frac{V_{\bf Q}}{\hbar\omega_{sp}}\right)^{2}\left[\left(\frac{\hbar^{2}{\bf K}_{s}{\bf Q}}{m^{*}}\right)^{2}+\left(\frac{\hbar^{2}k_{z,f}Q}{m}\right)^{2}\right]\left|f_{f,s}^{\dag}(Q)\right|^{2}
\delta_{{\bf K}_{f},{\bf K}_{s}+{\bf Q}}\delta(E_{{\bf K}_{f},k_{z,f}}-E_{{\bf K}_{s},s}-\hbar\omega_{sp}).
\label{eq:Gammavg1}\\
&=&
2\Gamma_{{\bf K}_{s},s}^{(vg)\parallel}(\mbox{abs1})+
2\Gamma_{{\bf K}_{s},s}^{(vg)\perp}(\mbox{abs1})
\label{eq:Gammapar+perp}
\earr
\end{widetext}
Here we may combine (\ref{eq:kz+-}), (\ref{eq:lambdaqu}) and (\ref{eq:expzs})   which allows a straightforward evaluation of (\ref{eq:Gammavg1}). For electron emission from the bottom of the SS-band at which ${\bf K}_{s}=0$ this yields the on-the-energy-shell quantities
\begin{widetext}
\barr
2\Gamma_{{\bf K}_{s}=0,s}^{(vg)\parallel}(\mbox{abs1})&=& 0,
\label{eq:Gammavg/omegaspar}\\
2\Gamma_{{\bf K}_{s}=0,s}^{(vg)\perp}(\mbox{abs1})&=&
\frac{\lambda^{2}}{\hbar^{2}}\sum_{{\bf Q}}\left(\frac{V_{\bf Q}}{\hbar\omega_{sp}}\right)^{2}e^{-2Q|z_{s}|}\left(\frac{\hbar^{2}k_{z,f}(+)Q}{m}\right)^{2}\frac{|\tilde{u}_{s}(k_{z,f}(+))|^{2}}{v_{z,f}(+)}
\label{eq:Gammavg2}
\earr
\end{widetext}
where $v_{z,f}(+)=\hbar k_{z,f}(+)/m$ and $V_{\bf Q}$ is defined in (\ref{eq:V_Q}). Neglecting the weak ${\bf Q}$-dependence of $k_{z,f}(+)$ we obtain for electron emission from the SS-band bottom  
\begin{widetext}
\bq
\frac{2\Gamma_{{\bf K}_{s}=0,s}^{(vg)\perp}(\mbox{abs1})}{\omega_{sp}}
=
\frac{\lambda^{2}(k_{z,f}(+)a_{B})}{8}\left(\frac{\hbar\omega_{sp}}{1H}\right)^{-2}\left(\frac{z_{s}}{a_{B}} \right)^{-3}\frac{|\tilde{u}_{s}(k_{z,f}(+))|^{2}}{a_{B}},
\label{eq:Gammavg/omegasperp}
\eq
\end{widetext}
where $1H$  denotes the atomic unit of energy ($1H$=1 Hartree=27.2 eV).
The result (\ref{eq:Gammavg/omegasperp}) explicitly scales with the plasmon frequency as $\omega_{sp}^{-2}$ which may indicate different aspects of nonadiabaticity of decay regimes described by (\ref{eq:Gammalg/omegas}) and (\ref{eq:Gammapar+perp}). These differences will be further elaborated in the next sections. Likewise in the length gauge, expressions (\ref{eq:Gammavg/omegaspar}) and (\ref{eq:Gammavg/omegasperp}) provide entries to the emission components of the second order cumulant (\ref{eq:Cvg22omega}) for ${\bf K}_{s}=0$.

 The velocity gauge expression (\ref{eq:Gammavg/omegasperp}) is compared in Fig. \ref{Gammaslgvg} with the analogous one (\ref{eq:Gammalg/omegas}) calculated in the length gauge. It is seen that the magnitudes of two expressions are very close for the choice of parameters characteristic of Ag(111) surface with reduced work function. In particular, they coincide upon a small reduction of $z_{s}$ from $1.2 a_{B}$ to $1 a_{B}$, thereby demonstrating the importance of a proper choice of the parameters characteristic of the studied system. Likewise the length gauge counterpart  (\ref{eq:Gammalg/omegas}), expression (\ref{eq:Gammavg/omegasperp}) may serve as a natural measure of the efficiency of SS-electron emission induced by absorption of surface plasmon from the plasmonic coherent state. Analogous expressions for one-plasmon emissions associated with electron transitions into the SS-state are obtained by microreversibility.

\begin{figure}[tb]
\rotatebox{0}{\epsfxsize=8cm \epsffile{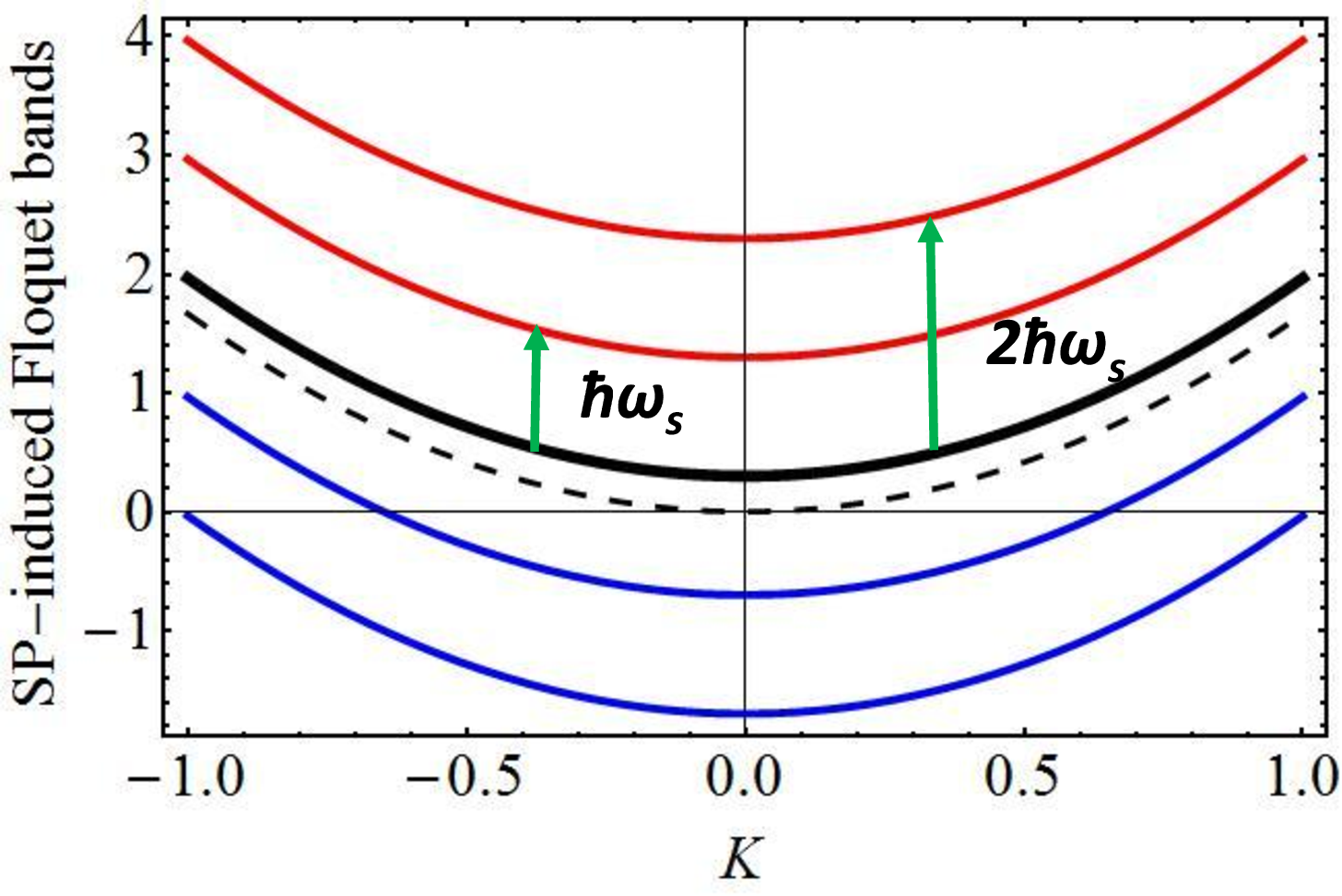}} 
\caption{Schematic of the SP-induced 2D Floquet band structure (arbitrary units) described in the effective mass approximation, as  manifesting in Eqs. (\ref{eq:Cvg11}), (\ref{eq:Cvg12}) and (\ref{eq:Wvg}). Thin dashed black curve denotes the parent 2D band, e.g. the SS-band on Ag(111) (cf. Fig. \ref{SSband}), and full thick black curve its ponderomotive shifted replica ($n=0$). Relative energy displacement is characteristic of the value of nonadiabaticity parameter  $\beta_{sp}<1$. Red and blue curves denote plasmonic Floquet side bands for positive and negative $n$, respectively.  Vertical arrows exemplify one and two SP-assisted electronic transitions induced by the action of potential ${\cal V}'(t)$ on the Volkov-dressed electronic state $|\psi_{{\bf K},s}(t)\rangle$. (After Ref. [\onlinecite{plasFloquet}]) }
\label{SPFloquet}
\end{figure}

This completes our formal analyses of electron-plasmon scattering dynamics based on nonperturbative cumulant approach in which the role of small expansion parameter is played by the correlations in momentum and energy transfer between successive scattering events. These correlations arise in the evolution operator from the noncommutativity of electron-plasmon interactions at different times. Their neglect enables the derivation of closed form nonperturbative solutions to second order in $\lambda^{2}$ demonstrated in the preceding sections.


\subsection{Applicability of second order cumulant expansion to multiplasmon processes}
\label{sec:discusscumulants}

The main advantage of calculating the quasiparticle propagators by cumulant expansion is a relatively feasible nonperturbative treatment of multiple electron excitation and de-excitation processes.  
The use of second order cumulant expansion to describe multiplasmon-induced electron emission from SS-bands on Ag(111) surfaces introduced in Secs. \ref{sec:Gammalg} and \ref{sec:Gammavg} requires the discussion of applicability of such an approximation to this concrete system. Truncation of cumulant series beyond the second order terms is justified in the regime of small correlations between the action of successive interactions $V$ or $V'$, as indicated in (\ref{eq:trans_E}) and (\ref{eq:trans_V}). In this case multiple excitation processes may be represented as a repeated and appropriately weighted combination of elementary processes obtained by the exponentiation of $C_{2}^{(lg)}(1\omega_{sp};{\bf K},s,t)$, or $C_{1}^{(vg)}(2\omega_{sp};{\bf K},s,t)$ and $C_{2}^{(vg)}(1\omega_{sp};{\bf K},s,t)$. All cumulants are free from the plasmon quantization lengths $L$ which are cancelled in summations over the intermediate ${\bf Q}$.

However, on Ag(111) surface the requirement of weak lateral momentum correlations in (\ref{eq:trans_E}) may hold provided the effective masses of electrons propagating in succesive electron states are nearly equal.[\onlinecite{4cumul}] The potentially critical issue is the lack of translational invariance (\ref{eq:trans_E}) and (\ref{eq:trans_V}) in the space of quantum numbers associated with quasiparticle motion in the direction normal to the surface.
 Therefore, the validity of interpretation of $2\hbar\omega_{sp}$ peaks in Fig. \ref{AllAgFloquet} in terms of the action of exponentiated $C_{2}^{(lg,vg)}(1\omega_{sp};{\bf K},s,t)$ without critical estimates of errors involved may be unclear. The same applies to $C_{2}^{(lg,vg)}(2\omega_{sp};{\bf K},s,t)$ that are expressed in terms of $\Sigma_{{\bf K},s}^{(lg,vg)}(\mbox{em1/abs1})/2\hbar\omega_{sp}$. Another important issue is that by construction the diagonal one-particle propagators do not directly yield asymptotic currents measured in electron emission processes. This may pose interpretational difficulties. Thus, in order to clarify these controversies we shall resort in the next sections to the description of multiplasmon absorption and emission processes within the complementary nonperturbative $T$-matrix approach based on the Volkov operator ansatz for the evolution operator.

\begin{figure}[tb]
\rotatebox{0}{\epsfxsize=9cm \epsffile{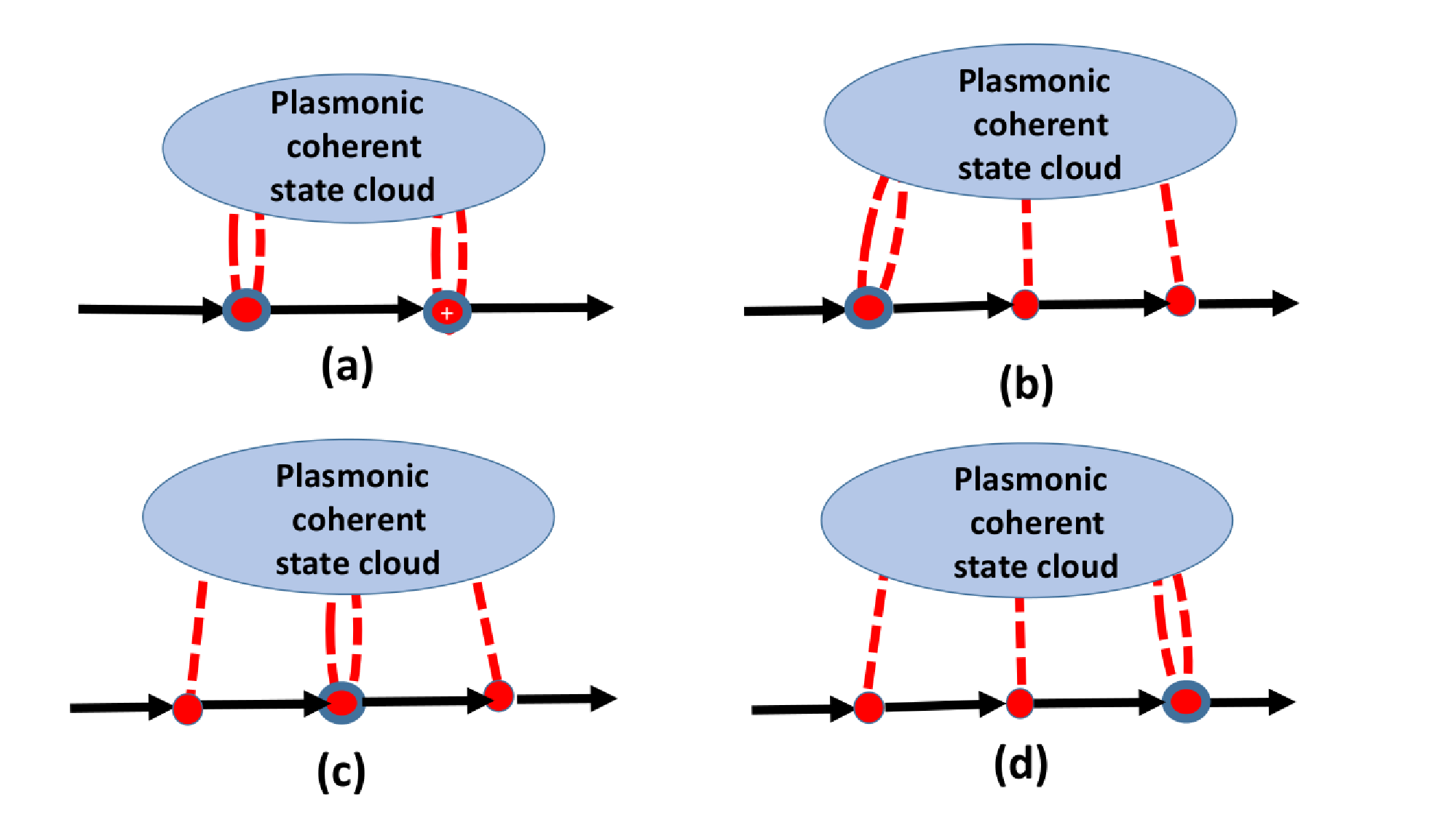}} 
\caption{(a) Sketch of the second order cumulant generated by the ${\bf A}({\bf r})^{2}/2m$ component (red filled blue circles) of the interaction $V_{I}'(t_{1})$ in the velocity gauge (\ref{eq:V'})  over the plasmonic coherent state. Here the contributing ${\bf Q}$-terms carry the weight $\propto \lambda_{\bf Q}^{4}$. (b)-(d) Third order cumulants arising from the interplay of interactions ${\bf A}({\bf r})^{2}/2m$ and ${\bf A}({\bf r})\cdot{\bf p}/m$ (full red dots) whose contributing ${\bf Q}$-terms are also $\propto\lambda_{\bf Q}^{4}$. More terms that are $\propto \lambda_{\bf Q}^{4}$ appear also in the fourth order cumulants generated by the interaction ${\bf A}({\bf r})\cdot{\bf p}/m$}
\label{C2C3}
\end{figure}


\section{Nonperturbative $T$-matrix formulation of surface plasmoemission in the length gauge}
\label{sec:Tlength}

\begin{figure}[tb]
\rotatebox{0}{\epsfxsize=11cm \epsffile{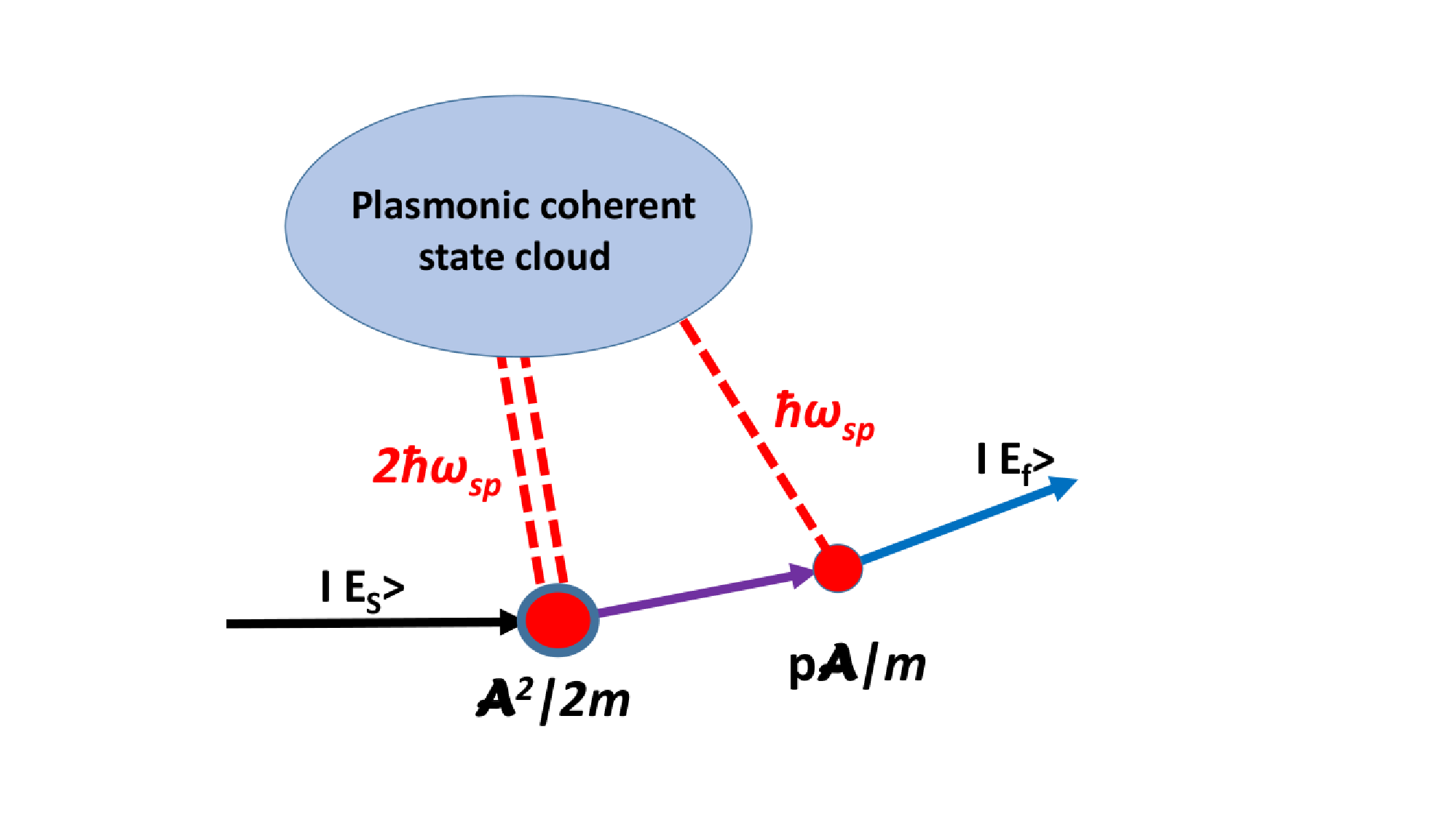}} 
\caption{(a) Schematic of a second order SP-induced electron excitation from a surface state $|E_{S}\rangle$ as induced by the length gauge interaction ${\cal V}_{S}(t)$ arising from a plasmonic coherent state environment. Full arrows denote the electron propagators and dashed lines the SP propagators.
 (b) Schematic of a third order SP-induced electron excitation from a surface state $|E_{S}\rangle$ as induced by the linear and quadratic components of velocity gauge interaction ${\cal V'}_{S}(t)$ of electron with plasmonic coherent state environment. There are two more third order contributions, one with the order of one- and two-plasmon interactions inverted, and the other comprising three one-plasmon interactions. }
\label{CohState}
\end{figure}

In the next sections we proceed with direct calculations of transition amplitudes in nonperturbative approach which takes into account all orders of the elementary processes that arise in the length gauge from the action of ${\cal V}_{S,I}(t)$ and in the velocity gauge from ${\cal V}_{S,I}'(t)$. An example of such low order processes induced in the length and velocity gauge is illustrated in Fig. \ref{CohState}. In both cases one may follow two alternative routes. Standard methods for nonperturbative description of electron motion in strong spatially homogeneous external fields are: {\it (i)} constructions of the wavefunction based on the Volkov ansatz[\onlinecite{Volkov}] introduced to describe free electron interactions with strong EM fields[\onlinecite{Truscott1,Truscott2}] and  elaborated in the calculations of transition rates characteristic of photoionization of atomic[\onlinecite{Keldysh1965,Faisal,Reiss1980,Madsen,Faisal2016,Keldysh2017}] and condensed matter  systems[\onlinecite{Keldysh1965,Reiss1977,Yalunin,Kidd}], or {\it (ii)} Fourier analysis of the underlying time dependent Schr\"{o}dinger equation.[\onlinecite{FaisalKaminski97,FaisalKaminski05,Park,PengZhang}] 
In accord with the previous sections we resort to method {\it (i)} elaborated in Refs. [\onlinecite{Keldysh1965,Faisal,Reiss1980,Madsen,Keldysh2017,Reiss1977,Kidd}]. In doing so we shall ignore the dissipative electronic environment because it is not expected to be of significance on the energy scale of plasmon dynamics[\onlinecite{NuskePRR2020}] discussed below. This aspect is demonstrated in Appendix D od Ref. [\onlinecite{plasFloquet}].

In Secs. \ref{sec:EPdynlg} and \ref{sec:EPdynvg} we have studied the electron excitation spectra in the coupled electron-plasmon system using cumulant expansion. We applied this method to calculate special averages of the evolution operator over plasmonic coherent states. The averages of these operators were transformed to the form of exponential functions of cumulant averages so that all summations over the excitation matrix elements appeared in the exponent of fastly converging exponential functions and their generalizations. This provided valuable although in some aspects restricted information on multiplasmon electron emission processes.    

In the next sections we shall develop a more direct nonperturbative approach for gaining information on the amplitudes of multiexcitation processes by calculating the offdiagonal matrix elements of the corresponding scattering operator or the $T$-matrix.  We shall show that in this approach the Volkov operator ansatz for nonperturbative modelling of the $T$-matrix and wavefunctions[\onlinecite{Volkov}] plays the role analogous to that of cumulant expansion in the calculations of propagators in Secs. \ref{sec:EPdynlg} and \ref{sec:EPdynvg}. Our goal in implementation of this method is to investigate the $\lambda_{\bf Q}$-dependence of the $T$-matrix which directly gives the plasmoemission rates and currents.

Depending on the boundary conditions specific to a particular problem, the Volkov ansatz-derived electron states may participate in emission or scattering processes as either initial, intermediate or final states.[\onlinecite{Keldysh1965,Faisal,Reiss1980,Madsen,Yalunin,Park,Gedik}] The initial field-dressed band states are conventionally termed Floquet or Bloch-Floquet states whereas the final outgoing field-dressed electron states are designated Volkov states.[\onlinecite{Yalunin,Park,Gedik}] In the present problem of electron emission from surface localized bands we assume their strong Floquet renormalization by the equally surface localized and gauge-specific electron-plasmon interaction (\ref{eq:V}) or (\ref{eq:V'}). On the other hand, we assume the outgoing delocalized electron emission states negligibly affected by the surface localized SP field (\ref{eq:A}), likewise in the case of multiple absorption of photons by atoms.[\onlinecite{Faisal}] 

\subsection{Volkov operator ansatz in the length gauge}
\label{sec:lgT}

In the length gauge and the SBC limit we obtain from (\ref{eq:Tlength}) exact expressions
\begin{widetext}
\barr
T_{f,i}^{(lg)}=\lim_{t\rightarrow\infty,t_{0}\rightarrow-\infty}T_{f,i}^{(lg)}(t,t_{0})
&=&
-\frac{i}{\hbar}\int_{t_{0}\rightarrow-\infty}^{t\rightarrow\infty} d\tau \langle\langle\mbox{coh}, \phi_{f},(\tau)||V_{S}U_{S}(H,\tau,t_{0})||\phi_{i},\mbox{coh},(t_{0})\rangle\rangle
\label{eq:TlgSch}\\
&=&
-\frac{i}{\hbar}\int_{t_{0}\rightarrow -\infty}^{t\rightarrow\infty} d\tau \langle\langle\mbox{coh}, \phi_{f}||V_{I}(\tau)U_{I}(H,\tau,t_{0})||\phi_{i},\mbox{coh}\rangle\rangle
\label{eq:TlgInt}
\earr
\end{widetext}
 Here $||\phi_{j},\mbox{coh},(\tau)\rangle\rangle=e^{-iH_{0}\tau/\hbar}||\phi_{j},\mbox{coh}\rangle\rangle$ and $V_{I}(\tau)=e^{iH_{0}\tau}V_{S}e^{-iH_{0}\tau}$, where  consistent with (\ref{eq:Ssbc}) $V_{S}=V$ vanishes for $t\rightarrow -\infty$ and reaches full strength at the coincidence time of the two pictures $\tau=0$.  
The lateral momentum selection rules enter (\ref{eq:TlgSch}) and (\ref{eq:TlgInt}) through the matrix elements (\ref{eq:Vmomentum}). This leads to
%
\barr
T_{f,i}^{(lg)}(t,t_{0})
&=&
-\frac{i}{\hbar}\int_{t_{0}}^{t} d\tau \langle\phi_{f}(\tau)|{\cal V}_{S}(\tau){\cal U}_{S}({\cal H},\tau,t_{0})|\phi_{i}(t_{0})\rangle
\label{eq:TlgSchcoh}\\
&=&
-\frac{i}{\hbar}\int_{t_{0}}^{t} d\tau \langle \phi_{f}|{\cal V}_{I}(\tau){\cal U}_{I}({\cal H},\tau,t_{0})|\phi_{i}\rangle,
\label{eq:TlgIntcoh}
\earr
%
with $ {\cal H}=\langle\mbox{coh}|H|\mbox{coh}\rangle$ and ${\cal U}$ generated by ${\cal H}$. Using ${\cal H}$ we take into account only electron interactions with plasmons excited by the pump interaction into the coherent state cloud and omit contributions from plasmon ground state fluctuations.

The offdiagonal matrix elements $\langle\phi_{f}|\dots|\phi_{i}\rangle$ that constitute $T_{f,i}^{(lg)}(t,t_{0})$ defined in (\ref{eq:TlgSchcoh}) and (\ref{eq:TlgIntcoh}) involve the evolution operators ${\cal U}_{S}$ and ${\cal U}_{I}$, respectively. These operators are amenable to various useful representations. A particularly convenient representation is in the strictly exponential form described in Appendix  \ref{sec:Gross}. The required exponent appears as a sum of low order uncorrelated and higher order mutually correlated scattering contributions, see expression (\ref{eq:UGross}). Hence, in the dominantly uncorrelated scattering regime the representations (\ref{eq:calH}), (\ref{eq:calH'}), (\ref{eq:calVI}) and the uncorrelated scattering (UCS) approximation (\ref{eq:Uuncorr}) suggest the introduction of the following Volkov-like operator ansatz for the evolution operator that is expressed in terms of commuting $H_{0}^{el}$ and ${\cal V}_{I}(\tau)={\cal V}_{S}(\tau)$ 
\barr
{\cal U}_{S}({\cal H},\tau,t_{0})&=&
e^{-iH_{0}^{el}\tau}{\cal U}_{I}({\cal H},\tau,t_{0})e^{iH_{0}^{el}t_{0}}\nonumber\\
&=&
\exp\left[-\frac{i}{\hbar}H_{0}^{el}(\tau-t_{0})\right]\exp\left[-\frac{i}{\hbar}\int_{t_{0}}^{\tau}dt_{1}{\cal V}_{S}(t_{1})\right]
=
\exp\left[-\frac{i}{\hbar}H_{0}^{el}(\tau-t_{0})-\frac{i}{\hbar}\int_{t_{0}}^{\tau} d t_{1} {\cal V}_{S}(t_{1})\right],
\label{eq:calU_S}
\earr
where due to the time dependence of semiclassical plasmonic field
\bq
{\cal U}_{I}({\cal H},\tau,t_{0})=\exp\left[-\frac{i}{\hbar}\int_{t_{0}}^{\tau}dt_{1}{\cal V}_{S}(t_{1})\right].
\label{eq:calU_I}
\eq
The Volkov operator ansatz representations (\ref{eq:calU_S}) and (\ref{eq:calU_I}) become exact in the long wavelength limit of quantized homogeneous plasmon field. This reflects the property that the fast convergence of the expansion in the exponent of (\ref{eq:UGross}) depends on the smallness of correlations between successive scattering events rather than on the smallness of the coupling strength, likewise in cumulant expansion. Following Ref. [\onlinecite{Faisal}] we now apply (\ref{eq:calU_S}) to (\ref{eq:TlgSchcoh}) to bring it to the form
\begin{widetext}
\barr
T_{f,i}^{(lg)}(t,t_{0})
&=&
-\frac{i}{\hbar}\int_{t_{0}}^{t} d\tau  e^{i(E_{f}-E_{i})\tau/\hbar}\langle \phi_{f}| {\cal V}_{S}(\tau)\exp\left[-\frac{i}{\hbar}\int_{t_{0}}^{\tau}dt_{1}{\cal V}_{S}(t_{1})\right]|\phi_{i}\rangle
\label{eq:TlgVolkov}\\
&=&
-\frac{i}{\hbar}(E_{f}-E_{i})\int_{t_{0}}^{t} d\tau  e^{i(E_{f}-E_{i})\tau/\hbar}\langle \phi_{f}|\exp\left[-\frac{i}{\hbar}\int_{t_{0}}^{\tau}dt_{1}{\cal V}_{S}(t_{1})\right]|\phi_{i}\rangle
\label{eq:TlgVolkovSBC}\\
&=&
-\frac{i}{\hbar}(E_{f}-E_{i})\int_{t_{0}}^{t} d\tau  e^{iE_{f}\tau/\hbar}\langle \phi_{f}|\psi_{i}^{(lg)}(\tau)\rangle_{Volkov}
\label{eq:TlgVolkovPsi}
\earr
\end{widetext}
where expression (\ref{eq:TlgVolkovSBC}) follows from partial integration with implied $t_{0}\rightarrow -\infty$. Note that although the matrix element $\langle \phi_{f}|\exp\left[-i\int_{t_{0}}^{\tau}dt_{1}{\cal V}_{S}(t_{1})/\hbar\right]|\phi_{i}\rangle$ in (\ref{eq:TlgVolkovSBC}) may bear resemblance to cumulant form, it is distinctively different. This represents an offdiagonal matrix element of exponential operator of the interaction whereas cumulant expansion gives just the reverse, an exponential series of diagonal matrix elements of the interactions.   

In expression (\ref{eq:TlgVolkovPsi}) we have introduced the Volkov ansatz wavefunction  
\bq
|\psi_{i}^{(lg)}(\tau)\rangle_{Volkov}=\exp\left[-\frac{i}{\hbar}E_{i}^{el}(\tau-t_{0})\right]
\exp\left[-\frac{i}{\hbar}\int_{t_{0}}^{\tau}dt_{1}{\cal V}_{S}(t_{1})\right]|\phi_{i}(t_{0})\rangle.
\label{eq:VolkovPsilg}
\eq
Thus, the Volkov ansatz wavefunction in the length gauge  can be readily determined for particular surface bands once $|\phi_{i}\rangle$ is known.

In order to simplify the calculations we assume nondispersive plasmons $\omega_{\bf Q}=\omega_{sp}$ implying $V_{\bf Q}=V_{Q}$, isotropic $\lambda_{\bf Q}=\lambda_{\bf -Q}$ and observe that $V^{\dag}({\bf Q},{\bf r})=V({\bf -Q},{\bf r})$ so that we can use (\ref{eq:Vsc}) to write ${\cal V}_{S}(t)$ in the forms
%
\barr
{\cal V}_{S}(t)&=&\sum_{\bf Q}\lambda_{\bf Q}V({\bf Q},{\bf r})e^{-i\omega_{sp}t} + 
2\sum_{\bf Q}\lambda_{\bf Q}V({\bf -Q},{\bf r})e^{i\omega_{sp}t}
=
2\sum_{\bf Q}\lambda_{\bf Q}V_{Q}e^{-Qz}\cos({\bf Q}\brho-\omega_{sp}t)
\label{eq:calVdip}\\
&=&
2\sum_{\bf Q}\lambda_{\bf Q}V({\bf Q},{\bf r})\cos(\omega_{sp}t)
=
{\cal W}({\bf r})\cos(\omega_{sp}t),
\label{eq:calVomegas}
\earr
%
where 
\bq
{\cal W}({\bf r})={\cal W}(\brho,z)=2\sum_{\bf Q}\lambda_{\bf Q}V({\bf Q},{\bf r}),
\label{eq:calW}
\eq
with $V({\bf Q},{\bf r})=
V_{Q}e^{-Qz}e^{i{\bf Q}\brho}$ as defined in (\ref{eq:VQr}).  The form (\ref{eq:calVdip}) is particularly convenient as it pinpoints the use of dipole approximation in which ${\bf Q}\brho$ can be neglected relative to $\omega_{sp}t$, particularly in view of subsequent implementation of the equipartition ans\"{a}tze (\ref{eq:lambdalin}) and (\ref{eq:lambdaqu}).

 Upon substituting (\ref{eq:calW}) in (\ref{eq:VolkovPsilg}), integrating over $t_{1}$ and $\tau$, setting $t_{0}\rightarrow -\infty$, and using (\ref{eq:BesselJ}) to expand (\ref{eq:TlgVolkovPsi}) we obtain in the length gauge
\begin{widetext}
\barr
T_{f,i}^{(lg)}=\lim_{t\rightarrow\infty,t_{0}\rightarrow -\infty}T_{f,i}^{(lg)}(t,t_{0})&=&-\frac{i}{\hbar}(E_{f}-E_{i})\int_{-\infty}^{\infty}d\tau e^{i(E_{f}-E_{i})\tau/\hbar}\langle \phi_{f}|\exp\left[-i\frac{{\cal W}({\bf r})}{\hbar\omega_{sp}}\sin(\omega_{sp}\tau)\right]|\phi_{i}\rangle
\label{eq:TlgSBCtau}\\
&=&
-2\pi i(E_{f}-E_{i})\sum_{n=-\infty}^{\infty}\langle\phi_{f}|J_{n}\left({\cal W}({\bf r})/\hbar\omega_{sp}\right)|\phi_{i}\rangle \delta(E_{f}-E_{i}-n\hbar\omega_{sp}),
\label{eq:TlgSBC}
\earr
\end{widetext}
where the operator ${\cal W}({\bf r})/\hbar\omega_{sp}$ in the argument of the Bessel function depends on the coherent state parameters $\lambda_{\bf Q}$. 
The scattering amplitudes given by its on-the-energy-shell matrix elements also display the appearance of Ward-Pitaevskii identities for vertex corrections. 
The processes with $n\geq 1$ describe plasmoemission of electron. The first order Born approximation result is obtained for $n=1$ (one-plasmon absorption) and $n=-1$ (one-plasmon emission).  



\subsection{Volkov ansatz in electron transition amplitudes calculated in the length gauge}
\label{sec:Volkovdipole}

The length gauge interaction ${\cal V}_{S}={\cal V}_{S}({\bf r},t_{1})={\cal V}_{S}(\brho,z,t_{1})$, and therefore ${\cal W}(\brho,z)/\hbar\omega_{sp}$ in the argument of the Bessel function in (\ref{eq:TlgSBC}), is a function of electron coordinates only. Hence, the action of (\ref{eq:calU_I}) on $u_{s}({\bf r})$ defined in (\ref{eq:phi_s}) amounts to its multiplication by a time dependent function of electron coordinates. In this respect the calculation of the matrix elements in (\ref{eq:TlgVolkovSBC}) with the wavefunctions (\ref{eq:phi_s}) and (\ref{eq:phifin}) should pose no conceptial problem.
We proceed with the calculations of plasmoemission from Ag(111) surface bands  by substituting (\ref{eq:VolkovPsilg}) into (\ref{eq:TlgVolkovPsi}) to obtain the relevant rates in the length gauge. Using the wavefunctions (\ref{eq:phi_s}) and (\ref{eq:phifin}) we obtain for the state-to-state transition (\ref{eq:TlgSBC}) 
\begin{widetext}
\bq
T_{{\bf K}_{f},k_{z,f}\leftarrow{\bf K}_{s},s}^{(lg)}=
-2\pi i(E_{{\bf K}_f,k_{z,f}}-E_{{\bf K}_s,s})\sum_{n=-\infty}^{\infty}\langle{\bf K}_f,k_{z,f}|J_{n}\left({\cal W}(\brho,z)/\hbar\omega_{sp}\right)|{\bf K}_s,s\rangle \delta(E_{{\bf K}_f,k_{z,f}}-E_{{\bf K}_s,s}-n\hbar\omega_{sp}),
\label{eq:TlgAg}
\eq
\end{widetext}
where  $E_{{\bf K}_f,k_{z,f}}=\frac{\hbar^{2}{\bf K}_{f}^{2}}{2m}+\frac{\hbar^{2}k_{z,f}^{2}}{2m}$ for excited state bands and $E_{{\bf K}_s,s}=\frac{\hbar^{2}{\bf K}_{s}^{2}}{2m^{*}}+E_{s}$ for surface bands. Therefore, the energy component that determines $k_{z,f}^{(n)}(+)$ in (\ref{eq:kz+-}) is
\bq
E_{k_{z,f}}^{(n)}=\frac{\hbar^{2}{\bf K}_{s}^2}{2m^{*}}+E_{s}+n\hbar\omega_{sp}-\frac{\hbar^{2}{\bf K}_{f}^2}{2m}.
\label{eq:Ekn}
\eq

The measurements of electron emission current described in the Introduction section were performed for nearly vertical transitions from the SS-band states $|{\bf K}_{s},s\rangle$ to final states of constant energy. Hence, the quantity of interest for description of the corresponding transition rates is obtained by taking the absolute square of (\ref{eq:TlgAg}) and summing it over a narow window $[{\bf K}_{f},k_{z,f}]$ determining the electron final state momentum and energy. In view of the nonoverlapping $\delta$-functions for different values of $n$ and relations (\ref{eq:Kronecker})-(\ref{eq:kz+-}), this quantity takes the form
\begin{widetext}
\bq
\sum_{[{\bf K}_{f},k_{z,f}]}|T_{{\bf K}_{f},k_{z,f}\leftarrow{\bf K}_{s},s}^{(lg)}|^{2}
=
\sum_{n=-\infty}^{\infty}\sum_{[{\bf K}_{f},k_{z,f}]} W_{{\bf K}_{f},k_{z,f}\leftarrow{\bf K}_{s},s}^{(lg)}(n)2\pi\hbar\rho(E_{k_{z,f}}^{(n)})\delta_{k_{z,f},k_{z,f}^{(n)}}.
\label{eq:Tlg^2}
\eq
Here the multiplasmon induced electron transition rate reads
\bq
 W_{{\bf K},k_{f}\leftarrow {\bf K}_{s}, s}^{(lg)}(n)
=
\frac{2\pi}{\hbar}(n\hbar\omega_{sp})^{2} \left|\langle {\bf K}_{f},k_{z,f}|J_{n}\left(\frac{{\cal W}(\brho,z)}{\hbar\omega_{sp}}\right)|{\bf K}_{s},s\rangle\right|^{2}
\delta\left(\frac{\hbar^{2}{\bf K}_{f}^2}{2m}+\frac{\hbar^{2}k_{z,f}^2}{2m}-\frac{\hbar^{2}{\bf K}_{s}^2}{2m^{*}}-E_{s}-n\hbar\omega_{sp}\right).
\label{eq:Wlgn}
\eq
\end{widetext}
and has the dimension of inverse time. Conversely, the $n$-th channel $k_{z,f}$-component of the final electron density of states is according to (\ref{eq:rho})-(\ref{eq:jz}) given by $2\pi\hbar\rho(E_{k_{f}}^{(n)})=1/j_{z,f}^{(n)}$, with the normal current $j_{z,f}^{(n)}=\hbar k_{z,f}^{(n)}(+)/mL_{z}=v_{z,f}^{(n)}/L_{z}$ corresponding to the energy $E_{k_{z,f}}^{(n)}$ defined in (\ref{eq:Ekn}), which has the dimension of time. This makes the quantum mechanical transition probability (\ref{eq:Tlg^2}) a dimensionless quantity and free from the quantization length $L_{z}$, as it should be.  
 
This is as far as one can go with the closed form solution of (\ref{eq:Tlg^2}) based on the exponential representation of the evolution operator (\ref{eq:UGross}) in which one neglects the momentum and energy correlations among successive plasmoemission and/or plasmoabsorption events described by the operator $G_{corr}^{I}$ in the exponent on the RHS of this expression. Thereby the basic quantum character of the transitions contained in each vertex has been retained. Specifically, expression (\ref{eq:Wlgn}) correctly reproduces the first order Born approximation result upon expanding $J_{n=\mp 1}({\cal W}/\hbar\omega_{sp})$ component of the sum in (\ref{eq:TlgAg}) into a power series, whereby the factor $1/2$ from the linear expansinon term is canceled by the prefactor $2$ on the RHS of (\ref{eq:calW}) yielding expressions (\ref{eq:Gamma1em}) and (\ref{eq:Gamma1abs}) found earlier using cumulant expansion (see Fig. \ref{Gammaslgvg}). Further progress on the representation of (\ref{eq:Tlg^2}) will now be made upon exploiting this fact.

\subsection{Model calculations of multiplasmon-induced transition rates in the length gauge}
\label{sec:Wlgn}

To obtain the plasmoemission rates in the length gauge we start from expressions (\ref{eq:TlgAg}) and (\ref{eq:Tlg^2})). We consider dominantly vertical transitions from the initial state at the SS-band bottom $|{\bf K}_{s}=0,s\rangle$ with unperturbed energy $E_{s}$ to the final outgoing waves $|\phi_{{\bf K}_{f}},f\rangle$ with energy $E_{{\bf K}_{f}}+E_{f}$. Thereby we obtain the dimensionless probability of transition into a narrow interval $[{\bf K}_{f},k_{z,f}]$
\bq
\sum_{[{\bf K}_{f},k_{z,f}]}|T_{({\bf K}_{f},f)\leftarrow( {\bf K}_{s}=0,s)}^{(lg)}|^{2}
=
\sum_{[{\bf K}_{f},k_{z,f}]}\sum_{n=-\infty}^{\infty} |T_{({\bf K}_{f},f)\leftarrow( {\bf K}_{s}=0,s)}^{(lg)}(n)|^{2}
=
\sum_{n=-\infty}^{\infty}\sum_{[{\bf K}_{f}]}W_{({\bf K}_{f},f)\leftarrow( {\bf K}_{s}=0,s)}^{(lg)}(n)2\pi\hbar\rho(E_{k_{z,f}}^{(n)}),
\label{eq:Tlgsquared}
\eq
where  $2\pi\hbar\rho(E_{k_{z,f}^{(n)}})=L_{z}/v_{z,f}^{(n)}$ arises from the transformation (\ref{eq:Kroneckersymbol}) of one of the $\delta$-functions constituting $|T_{({\bf K}_{f},f)\leftarrow( {\bf K}_{s}=0,s)}^{(lg)}(n)|^{2}$. Note in the second line only the summation over $[{\bf K}_{f}]$ since the $k_{z,f}$-summation has been absorbed in $\rho(E_{k_{z,f}}^{(n)})$. Hence the corresponding transition rate reads
\begin{widetext}
\bq
 \sum_{[{\bf K}_{f}]}W_{({\bf K}_{f},f)\leftarrow({\bf K}_{s}=0,s)}^{(lg)}(n)=\frac{2\pi}{\hbar}(n\hbar\omega_{sp})^{2} \sum_{[{\bf K}_{f}]}\left|\langle {\bf K}_{f},k_{z,f}|J_{n}\left(\frac{{\cal W}(\brho,z)}{\hbar\omega_{sp}}\right)|{\bf K}_{s}=0,s\rangle\right|^{2}
\delta\left(E_{{\bf K}_f}+E_{k_{z,f}}-E_{s}-n\hbar\omega_{sp}\right).
\label{eq:Wlg}
\eq 
\end{widetext}
This nonperturbative closed form expression is exact insofar the representation of the evolution operator (\ref{eq:Uuncorr}) may be considered exact.

To proceed feasibly we shall in view of the previous approximations introduce 
\bq
\frac{{\cal W}(\brho,z)}{\hbar\omega_{sp}}\longrightarrow \frac{{\cal W}(\brho,z_{s})}{\hbar\omega_{sp}}.
\label{eq:xilambda}
\eq
This substitution immediately reduces the matrix element of the Bessel function to 
\bq
\left|\langle {\bf K}_{f},k_{z,f}|J_{n}\left(\frac{{\cal W}(\brho,z_{s})}{\hbar\omega_{sp}}\right)|{\bf K}_{s},s\rangle\right|^{2}
\longrightarrow
\left|\langle {\bf K}_{f}|J_{n}\left(\frac{{\cal W}(\brho,z_{s})}{\hbar\omega_{sp}}\right)|{\bf K}_{s}\rangle\right|^{2} \frac{|\tilde{u}_{s}(k_{z,f})|^{2}}{L_{z}}. 
\label{eq:Jxi}
\eq
Thereby $L_{z}^{-1}$ from the absolute square in (\ref{eq:Jxi}) cancels $L_{z}$ from $2\pi\hbar\rho(E_{k_{z,f}^{(n)}})$ in (\ref{eq:Tlgsquared}), rendering the latter free from the perpendicular quantization length.
This square gives the intensity of the sum of all interfering transition amplitudes constituting the $|{\bf K}_{s},s\rangle \rightarrow[|{\bf K}_{f},k_{z,f}^{(n)}\rangle]$ excitation probabilities described by (\ref{eq:Wlg}). Upon expanding $J_{n}(x)$ into a power series the resulting matrix elements contain the ascending perturbative factors of the form $\propto \langle {\bf K}_{f}|\exp(i\sum_{l}{\bf Q}_{l}\brho)|{\bf K}_{s}\rangle=\delta_{{\bf K}_{f}-{\bf K}_{s},\sum_{l}{\bf Q}_{l}}$ which keep track of the total lateral momentum conservation in the intermediate interaction vertices which are the sources of intermediate ${\bf Q}_{l}$ emissions and absorptions that must also be summed over.
Therefore, $W_{({\bf K}_{f},f)\leftarrow( {\bf K}_{s}=0,s)}^{(lg)}(n)$ given by (\ref{eq:Wlg}) plays the role of transition rate per unit time obtained in the length gauge for nonperturbative description of electron transitions induced by the interaction (\ref{eq:V}) between the parent SS-state and the $n$-th Floquet band channel.

\begin{figure}[tb]
\rotatebox{0}{\epsfxsize=8.5cm \epsffile{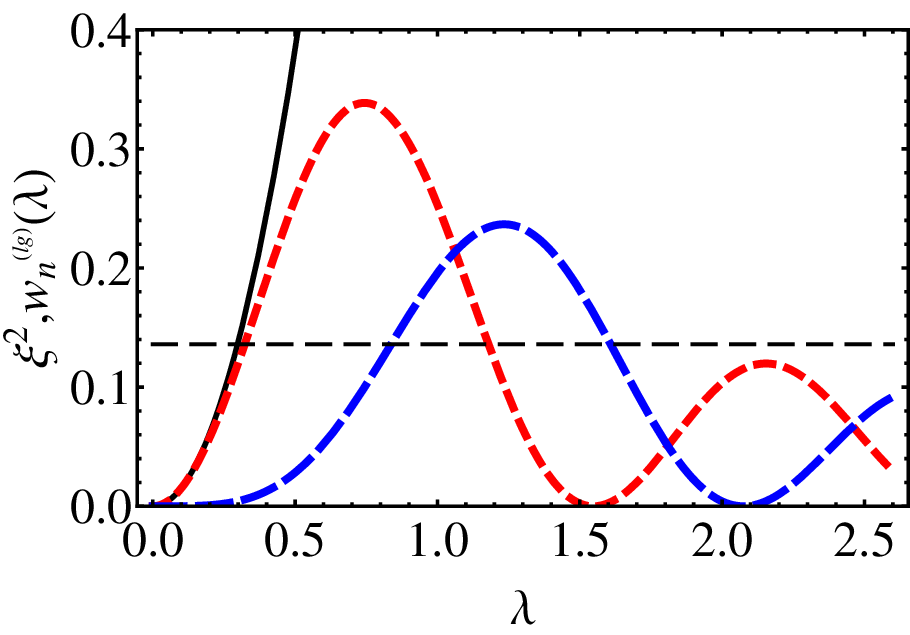}} 
\caption{$\lambda$-dependence of expression $w_{n}^{(lg)}(\lambda)$, Eq. (\ref{eq:ansatzJn}), constituting the transition probability (\ref{eq:Wlg}) appropriate to plasmoemission from SS-band bottom as calculated in the length gauge. Dashed red curve: $n=1$; long dashed blue curve: $n=2$; black curve: Born approximation result $|\xi_{s}^{(off)}(\lambda,z_{s}))|^{2}$. Horizontal thin dashed line denotes the surface plasmon energy on Ag(111) surface measured in a.u.}
\label{wnlglambda}
\end{figure}

The estimate of the lowest order contribution to (\ref{eq:Wlg}) in powers of $\lambda$, i.e. the first order Born approximation (BA) result $\propto\lambda^{2}$, is illustrated in Fig. \ref{Gammaslgvg}. Such an estimate of the full (\ref{eq:Wlg}) would be prohibitive but we can nevertheless build on the BA result by following the analogy with  second order cumulant expansion. We first observe that the limited $[{\bf K}_{f}]$-summation has little effect on the energy conserving $\delta$-function on the RHS of (\ref{eq:Wlg}) because of the flat density of two-dimensional ${\bf K}_{f}$-states. Therefore, without incurring much error, we can effectuate the $[{\bf K}_{f}]$-summation only onto the square of the matrix element (\ref{eq:Jxi}). Then, following the analogy with the Ward-Pitaevskii identities we introduce in (\ref{eq:Wlg}) and (\ref{eq:Jxi}) off-the-energy-shell and on-the-momentum-shell ansatz consistent with the vertices in Fig. \ref{C2lg}. This "equivalent BA substitution" naturally arises from cumulant-derived expressions (\ref{eq:Gamma1em}) and (\ref{eq:Gamma1abs}) and reads
%
\barr
\sum_{[{\bf K}_{f}]}\left|\langle {\bf K}_{f}|J_{n}\left(\frac{{\cal W}(\brho,z_{s})}{\hbar\omega_{sp}}\right)|{\bf K}_{s}=0\rangle\right|^{2}
&\longrightarrow& 
\left|J_{n}\left(2\sqrt{\sum_{[{\bf K}_{f}],{\bf Q}}\lambda^{2}\frac{\left|\langle{\bf K}_{f}|V_{\bf Q}(\brho,z_{s})|{\bf K}_{s}=0\rangle\right|^{2}}{(\hbar\omega_{sp})^{2}}}\right)\right|^{2}\nonumber\\
&=&
|J_{n}(2\xi_{s}^{(off)}(\lambda,z_{s}))|^{2}=w_{n}^{(lg)}(\lambda),
\label{eq:ansatzJn}
\earr
%
where $\xi_{s}^{(off)}(\lambda,z_{s})$ exhibits the scaling structure
\bq
\xi_{s}^{(off)}(\lambda,z_{s})=\lambda\sqrt{\frac{e^{2}}{4z_{s}\hbar\omega_{sp}}}=\frac{\lambda}{2}\left(\frac{z_{s}}{a_{B}}\right)^{-\frac{1}{2}}\left(\frac{\hbar\omega_{sp}}{1\mbox{H}}\right)^{-\frac{1}{2}}.
\label{eq:xioff}
\eq
Here the superscript $^{(off)}$ denotes the off-the-energy-shell character of the underlying quantity since this property is controlled separately by the $\delta$-function on the RHS od (\ref{eq:Wlg}). This makes it different relative to the on-shell quantity (\ref{eq:Gammalg/omegas}). 
The rationale underlying the use of this ansatz is the elimination of the quantization lengths $L$ from the dimensionless argument $2\xi_{s}^{(off)}(\lambda,z_{s})$ of the Bessel function in (\ref{eq:ansatzJn}).
The first order BA result is then retrieved from the first term in the expansion of $J_{n=1}$ in (\ref{eq:ansatzJn}) and its substitution into (\ref{eq:Wlg}). This is consistent with (i.e. equal to) $2\Gamma_{{\bf K}_{s},s}^{(lg)}(\mbox{abs1})/\omega_{sp}$ obtained with expression (\ref{eq:Gamma1abs}). With this ansatz the term $n=0$ should be excluded from the sum over $n$ in (\ref{eq:Wlg}) in order to avoid overcounting through violation of orthogonality. The plots of (\ref{eq:ansatzJn}) for $n=1$ and $n=2$ as function of $\lambda$ and with parameters typical of Ag(111) surface are displayed in Fig. \ref{wnlglambda}. Notable are the maximum values of $w_{1}^{(lg)}(\xi(\lambda))$ and $w_{2}^{(lg)}(\xi(\lambda))$ that are reached already for relatively small occupations (as measured by $\lambda$) of the coherent plasmonic states. This indicates a relatively high efficiency of plasmoemission processes. The minima at higher values of $\lambda$ occur due to the strong interference between the actions of plasmon absorptions and emissions.

\section{$T$-matrix and Volkov operator ansatz for plasmoemission from surface bands in the velocity gauge}
\label{sec:TwithcalV}

Next we turn to the studies of plasmoemission in the velocity gauge formulation of electron-plasmon interaction.
The use of the velocity gauge in the calculations of transition amplitudes usually provides faster converging results relative to the electron-eletromagnetic field coupling strength,[\onlinecite{Boyd2004,Reiss2013}] particularly because of the automatic account of quadratic electron-field interaction that gives rise to ponderomotive potential. Hence, we rewrite (\ref{eq:TvelSBCSchr}) and (\ref{eq:TvelSBC}) starting from integral representation of the evolution operator (\ref{eq:UIint}) in the velocity gauge to obtain the SBC limit of transition amplitudes in either Schr\"{o}dinger or  interaction pictures  
\begin{widetext}
\barr
T_{f,i}^{(vg)}=\lim_{t\rightarrow\infty,t_{0}\rightarrow-\infty}T_{f,i}^{(vg)}(t,t_{0})
&=&
-\frac{i}{\hbar}\int_{t_{0}\rightarrow-\infty}^{t\rightarrow\infty} d\tau \langle\langle\mbox{coh}, \phi_{f},(\tau)||V_{S}^{'}U_{S}(H',\tau,t_{0})||\phi_{i},\mbox{coh},(t_{0})\rangle\rangle
\label{eq:TvelSch}\\
&=&
-\frac{i}{\hbar}\int_{t_{0}\rightarrow -\infty}^{t\rightarrow\infty} d\tau \langle\langle\mbox{coh}, \phi_{f}||V_{I}^{'}(\tau)U_{I}(H',\tau,t_{0})||\phi_{i},\mbox{coh}\rangle\rangle
\label{eq:TvelInt}
\earr
\end{widetext}
Here $V_{I}'(\tau)=e^{iH_{0}\tau}V_{S}'e^{-iH_{0}\tau}$, where  consistent with (\ref{eq:Ssbc}) the interaction $V_{S}'$ defined in (\ref{eq:V'}) vanishes for $\tau\rightarrow -\infty$ and reaches full strength at the coincidence time of the two pictures $\tau=0$. Fully quantum expressions (\ref{eq:TvelSch}) and (\ref{eq:TvelInt}) are exact as no approximations have been made in their derivation from (\ref{eq:TvelSBCSchr}) and (\ref{eq:TvelSBC}).

To proceed with the calculations of the transition amplitudes by using (\ref{eq:TvelSch}) or (\ref{eq:TvelInt}) we again assume that primary excitation of the system by the laser field always gives rise to the same plasmonic coherent state cloud $|\mbox{coh}\rangle$. In this case we may average all interactions in generic expressions (\ref{eq:TvelSch}) and (\ref{eq:TvelInt}) over  $|\mbox{coh}\rangle$ to obtain
%
\barr
T_{f,i}^{(vg)}(t,t_{0})
&=&
-\frac{i}{\hbar}\int_{t_{0}}^{t} d\tau \langle\phi_{f}(\tau)|{\cal V}_{S}^{'}(\tau)U_{S}({\cal H}',\tau,t_{0})|\phi_{i}(t_{0})\rangle
\label{eq:TvelSchcoh}\\
&=&
-\frac{i}{\hbar}\int_{t_{0}}^{t} d\tau \langle \phi_{f}|{\cal V}_{I}^{'}(\tau)U_{I}({\cal H}',\tau,t_{0})|\phi_{i}\rangle,
\label{eq:TvelIntcoh}
\earr
%
with ${\cal V}_{S}^{'}$ and  ${\cal V}_{I}^{'}$ defined in (\ref{eq:calV'S}) and (\ref{eq:calVI}), respectively. Using this we take into account only the electron interactions with plasmons excited by the pump interaction into the coherent state cloud and omit the contributions from plasmon ground state fluctuations depicted in Fig. \ref{C1&C2}(b) and (d) that are independent of $\lambda_{\bf Q}$. This contribution can be easily restored wherever necessary by adding the modified plasmon emission term in which $\lambda_{\bf Q}=1$.

\subsection{Volkov operator ansatz in the velocity gauge}
\label{sec:Volkovvg}

 Solutions of general expressions (\ref{eq:TvelSchcoh}) and (\ref{eq:TvelIntcoh}) are very complex. Simplifications in the calculations may arise in connection with the forms of interactions and evolution operators. Thus, in the currently studied case of e-SP interaction in the velocity gauge we shall retain the full quantum character of the electron dynamics in the vector potential ${\bf A}({\bf r})$ constituting $V'$ whereas the plasmon field itself will be modeled by its quasiclassical high excitation limit (\ref{eq:Acohsum}). This fixes the interaction in the form (\ref{eq:calVI}). However, in order to enable contact with fully quantum descriptions of electron-plasmon interactions in the velocity gauge in Sec. \ref{sec:Elspecvel} we must associate with the thus obtained ${\cal V'}_{I,S}$ the same momentum selection rules appropriate to the considered processes, like those described by the diagrams in Fig. \ref{C2lg}. 

In the second step we make use of the evolution operator in the interaction picture in the lowest, uncorrelated form $U_{I}(H',\tau,t_{0})=\exp[-iG_{1}^{I}(\tau,t_{0})]$[\onlinecite{Gross}] with $G_{1}^{I}$ defined in (\ref{eq:G1}). This 
yields
\bq
U_{I}({\cal H}',\tau,t_{0})\rightarrow {\cal U}_{I}({\cal H}',\tau,t_{0})=
\exp\left[-\frac{i}{\hbar}\int_{t_{0}}^{\tau} d t_{1} {\cal V}_{I}'(t_{1})\right].
\label{eq:calUI}
\eq
This operator ansatz should generally hold in the UCS regime in which the subsequent scattering events in higher order processes may be treated independently.[\onlinecite{PhysRep}] This can be extended to both pictures to give
\begin{widetext}
\barr
T_{f,i}^{(vg)}(t,t_{0})|_{UCS}
&=&
-\frac{i}{\hbar}\int_{t_{0}}^{t} d\tau \langle \phi_{f}(\tau)|{\cal V}_{S}'(\tau){\cal U}_{S}'({\cal H}',\tau,t_{0})|\phi_{i}(t_{0})\rangle.
\label{eq:TvelSchUCS}\\
&=&
-\frac{i}{\hbar}\int_{t_{0}}^{t} d\tau \langle \phi_{f}|{\cal V}_{I}'(\tau){\cal U}_{I}'({\cal H}',\tau,t_{0})|\phi_{i}\rangle,
\label{eq:TvelIntUCS}
\earr
\end{widetext}
where in accord with (\ref{eq:calUI}) we have replaced $H'$ by ${\cal H'}$ and consequently $U_{S,I}'$ by ${\cal U}_{S,I}'$.

In the next step of approximation we shall neglect the contribution of (\ref{eq:corrVI}) to $V_{I}^{'}(t)$. A heuristic argument for neglecting this contribution is outlined in Appendix \ref{sec:deviationV}. This leads to the relation  
\bq
[H_{0}^{el},{\cal V'}_{S}(t)]\rightarrow 0, 
\label{eq:commHV'}
\eq
which implies an important property of the electron interaction with plasmonic coherent state field $\bcalA ({\bf r},t)$ defined in (\ref{eq:Acohsum}), viz.
\bq
{\cal V'}_{I}(t)\approx {\cal V'}_{S}(t).
\label{eq:calVIS}
\eq
Note in passing that this relation is exactly satisfied in the length gauge, cf. (\ref{eq:calV_S}). The property (\ref{eq:calVIS}) allows introducing in (\ref{eq:TvelSchUCS}) the Volkov evolution operator ansatz in equivalent forms
%
\barr
{\cal U}_{S}'({\cal H}',\tau,t_{0})=
e^{-iH_{0}^{el}\tau}{\cal U}_{I}'({\cal H}',\tau,t_{0})e^{iH_{0}^{el}t_{0}}
&=&
\exp\left[-\frac{i}{\hbar}H_{0}^{el}(\tau-t_{0})\right]\exp\left[-\frac{i}{\hbar}\int_{t_{0}}^{\tau}dt_{1}{\cal V}_{S}'(t_{1})\right]
\nonumber\\
&=&
\exp\left[-\frac{i}{\hbar}H_{0}^{el}(\tau-t_{0})-\frac{i}{\hbar}\int_{t_{0}}^{\tau} d t_{1} {\cal V}_{S}'(t_{1})\right]
\label{eq:calUS}
\earr
%

In view of (\ref{eq:commHV'}) and (\ref{eq:calVIS}) this operator ansatz is intuitive and plausible. However, its present formulation rests upon nontrivial and specific conditions of the Coulomb gauge (\ref{eq:nablaA}),  smallness of the commutator (\ref{eq:corrVI}) satisfied by the e-SP interaction, and the exponential representation of the evolution operators described in Sec. \ref{sec:Uvel} and Appendix \ref{sec:Gross}. This explains its rather tedious detour derivation from exact closed form solutions (\ref{eq:TvelSchUCS}) and (\ref{eq:TvelIntUCS}).

We shall use (\ref{eq:calUS}) for construction of the wavefunction $|\psi(t)\rangle$ of an electron whose dynamics is determined by ${\cal H}'(t)$, viz.
\bq
|\psi(t)\rangle={\cal U}_{S}({\cal H}',t,t_{0})|\phi_{i}(t_{0})\rangle.
\label{eq:PsiS}
\eq
Note that this wavefunction ansatz becomes exact if (\ref{eq:commHV'}) vanishes but this generally does not hold for other types of vector potentials.

Consistent with (\ref{eq:commHV'}) the major effect of the action of ${\cal U}_{S}({\cal H'},\tau,t_{0})$ from (\ref{eq:calUS}) on the initial state wavefunction is the change of its phase. This favours a particular subset of the scattering channels complying with the SBC regime in which the phase change is induced by $-\frac{i}{\hbar}\int_{t_{0}}^{\tau} d t_{1} {\cal V}_{S}'(t_{1})$ from the exponent of (\ref{eq:calUS}). Such a restricted channel wavefunction is represented by  
\bq
|\psi_{i}^{(vg)}(\tau)\rangle_{Volkov}=\exp\left[-\frac{i}{\hbar}E_{i}(\tau-t_{0})\right]
\exp\left[-\frac{i}{\hbar}\int_{t_{0}}^{\tau}dt_{1}{\cal V}_{S}'(t_{1})\right]|\phi_{i}(t_{0})\rangle,
\label{eq:VwavePsi}
\eq
where $E_{i}|\phi_{i}\rangle=H_{0}^{el}|\phi_{i}\rangle$.
This form is in accord with the standard Volkov wavefunction ansatz.[\onlinecite{Volkov}] It embodies all uncorrelated intermediate state processes induced by ${\cal V}_{S}'(t_{1})$.
 The validity and applicability of this procedure is discussed in Ref. [\onlinecite{Faisal2016}].        
Substituting (\ref{eq:VwavePsi}) into (\ref{eq:TvelSchUCS}) we obtain
\begin{widetext}
\bq
T_{f,i}^{(vg)}(t,t_{0})
=
-\frac{i}{\hbar}\int_{t_{0}}^{t} d\tau  e^{i(E_{f}-E_{i})\tau/\hbar}\langle \phi_{f}| {\cal V}_{S}^{'}(\tau)\exp\left[-\frac{i}{\hbar}\int_{t_{0}}^{\tau}dt_{1}{\cal V}_{S}'(t_{1})\right]|\phi_{i}\rangle.
\label{eq:TVolkov}
\eq
\end{widetext}
Derivation of this expression has involved several restrictive assumptions but none of them have been truly classical in the sense that despite the introduced simplifications there exist quantum systems amenable to such model descriptions. Note also that it fully corresponds to expression (\ref{eq:TlgVolkov}) derived in the length gauge.

Exploiting the assumption that ${\cal V}_{S}^{'}(\tau)$ switches on and off adiabatically during the interval $(t,t_{0})$ and writing 
\bq
-\frac{i}{\hbar} {\cal V}_{S}^{'}(\tau)\exp\left[-\frac{i}{\hbar}\int_{t_{0}}^{\tau}dt_{1}{\cal V}_{S}'(t_{1})\right]
=
\frac{\partial}{\partial t} \exp\left[-\frac{i}{\hbar}\int_{t_{0}}^{\tau}dt_{1}{\cal V}_{S}'(t_{1})\right],
\label{eq:partialcalV}
\eq
expression (\ref{eq:TVolkov}) can be integrated by parts to yield

\begin{widetext}
\barr
T_{f,i}^{(vg)}=\lim_{t\rightarrow\infty,t\rightarrow -\infty} T_{f,i}^{(vg)}(t,t_{0})
&=&
-\frac{i}{\hbar}(E_{f}-E_{i})\int_{-\infty}^{\infty} d\tau  e^{i(E_{f}-E_{i})\tau/\hbar}\langle \phi_{f}|\exp\left[-\frac{i}{\hbar}\int_{-\infty}^{\tau}dt_{1}{\cal V}_{S}'(t_{1})\right]|\phi_{i}\rangle\label{eq:TVolkovSBC}\\
&=&
-\frac{i}{\hbar}(E_{f}-E_{i})\int_{-\infty}^{\infty} d\tau  e^{iE_{f}\tau/\hbar}\langle \phi_{f}|\psi_{i}^{(vg)}(\tau)\rangle_{Volkov}
\label{eq:TVolkovPsi}
\earr
\end{widetext}
 This nonperturbative result is formally equivalent 
 to Eq. (36) of Ref. [\onlinecite{plasFloquet}] based on the earlier expression (14) of Faisal,[\onlinecite{Faisal}] and can be treated using analogous methods.[\onlinecite{Reiss1980,Madsen}] The proportionality $T_{f,i}^{vg}\propto -i(E_{j}-E_{i})/\hbar$ reflects the adiabatic SBC limit.[\onlinecite{Davydov}]
 
%


Implementation of the Volkov ansatz becomes more complicated  in the present case of velocity gauge because of the noncommuting operators ${\bf r}$ and ${\bf p}$ appearing in $V'$.
Semiclassical $\bcalA({\bf r},t)$ that replaces ${\bf A(r)}$ in $H'$ remains to satisfy (\ref{eq:nablaA}), implying $[{\bf p},\bcalA({\bf r},t)]=0$, and therefore the order of ${\bf p}$ and $\bcalA({\bf r},t)$ is irrelevant in the action of their products on the electron wavefunction. Moreover, since Ag(111)  surface state wavefunctions $\phi_{{\bf K},s}(\brho,z)$ are strongly localized around $z_{s}$, and thus restrict the perpendicular range of $\bcalA(\brho,z,t)$, we shall follow the arguments leading to (\ref{eq:xilambda}) and for computational convenience make analogous replacement 
\bq
\bcalA({\bf r},t)\rightarrow\bcalA(\brho,z_{s},t).
\label{eq:Az_s}
\eq
This is equivalent to the use of the dipole approximation for the field.
 Together with (\ref{eq:VolkovPsilg}), this substitution enables the use of Volkov-like ansatz in description of electron dynamics in Q2D surface bands.

Plasmoemission yield is most conveniently calculated in the velocity  gauge from expression (\ref{eq:TVolkovPsi}). On the symbolic level, this off-diagonal expression describes vertex corrections with intermittent insertions of self-energy corrections. As their lowest order contributions also constitute cumulant expansion of (\ref{eq:Gcumlength}) and (\ref{eq:Gvelcum}) used to calculate the electron spectra, we shall exploit this circumstance and construct $|\psi_{{\bf K},s}(\tau)\rangle_{Volkov}$ in (\ref{eq:TVolkovPsi}) so as to comprise analogous elementary processes.

As we have already pointed out, within the discussed conditions the linear and quadratic field interactions constituting ${\cal V'}_{S}$ in (\ref{eq:TVolkovSBC}) may be treated as commuting operators. To take into account different electron masses $m^{*}$ and $m$ in the lateral ($\parallel$) and perpendicular ($\perp$) to the surface directions, respectively, we first use the decompositions
\barr
& &\hat{\bf p}=-i\hbar\nabla=-i\hbar(\nabla_{\parallel}+\nabla_{\perp}),
\label{eq:p_z}\\
& &\bcalA(\brho,z_{s},t)=\bcalA_{\parallel}(\brho,z_{s},t)+\bcalA_{\perp}(\brho,z_{s},t),
\label{eq:Aparperp}
\earr
which enable us to represent the action of interaction (\ref{eq:calV'S}) on the SS-band states using a mass tensor 
\begin{widetext}
\bq
\frac{\left(\hat{\bf p}+\bcalA(\brho,z_{s},t)\right)^{2}}{2m}\rightarrow \left(\frac{\hat{\bf p}_{\parallel}^{2}}{2m^{*}}+\frac{\hat{\bf p}_{\perp}^{2}}{2m}\right)
+
\left(\frac{\bcalA_{\parallel}(\brho,z_{s},t)\hat{\bf p}_{\parallel}}{m^{*}}+\frac{\bcalA_{\perp}(\brho,z_{s},t)\hat{\bf p}_{\perp}}{m}\right)
+
\left(\frac{\bcalA_{\parallel}^{2}(\brho,z_{s},t)}{2m^{*}}+\frac{\bcalA_{\perp}^{2}(\brho,z_{s},t)}{2m}\right).
\label{eq:pAseparation}
\eq
\end{widetext}
Employing this and (\ref{eq:p_z}) we represent the transition amplitude $\langle\phi_{f}|\psi_{{\bf K},s}(\tau)\rangle$ from (\ref{eq:TVolkovPsi}) in the form
\begin{widetext}
\barr
\langle\phi_{f}|\psi_{{\bf K},s}^{(vg)}(\tau)\rangle 
= 
\exp\left[-i\left(\frac{(\hbar{\bf K})^2}{2m^{*}\hbar}+\frac{E_{s}}{\hbar}\right)\tau\right]
&\times&
\int d^{2}\brho \int dz \phi_{f}^{*}(\brho,z)
\left\{\exp\left[-\frac{i}{\hbar}\int_{t_{0}}^{\tau}dt'
\left(\frac{\bcalA_{\parallel}^{2}(\brho,z_{s},t')}{2m^{*}}+
\frac{\bcalA_{\perp}^{2}(\brho,z_{s},t')}{2m}\right)\right]\right.\nonumber\\
&\times&
\left.\exp\left[-\int_{t_{0}}^{\tau}dt'
\left(\bcalA_{\parallel}(\brho,z_{s},t')\frac{\nabla_{\parallel}}{m^{*}}+\bcalA_{\perp}(\brho,z_{s},t')\frac{\nabla_{\perp}}{m}\right)\right]\right\}\phi_{{\bf K},s}(\brho,z),
\label{eq:psiVolkovA2}
\earr
\end{widetext}
where we use the notation $\langle\brho,z|\phi_{f}\rangle=\phi_{f}(\brho,z)=\phi_{f}(\brho)\phi_{f}(z)$.
Expression (\ref{eq:psiVolkovA2}) explicitly demonstrates the roles of effective electron masses in  interactions with the surface plasmon field. 
Expression (\ref{eq:psiVolkovA2}) is analogous to the earlier derived result describing multiphoton ionization of atoms.[\onlinecite{Faisal}] 

In order to identify and collect the relevant nonperturbative contributions to offdiagonal or vertex-like form $\langle\phi_{f}|\psi_{{\bf K},s}(\tau)\rangle$ in (\ref{eq:psiVolkovA2}) we shall trace the analogy with the diagrams of Figs. \ref{C2lg} and \ref{C1&C2} that can be cut in two conjugated vertex corrections. 
 First, the action of the exponential operator containing $\bcalA^{2}$ in the matrix element on the RHS of (\ref{eq:psiVolkovA2}) on the initial state wavefunction $|\phi_{{\bf K},s}\rangle$ will be implemented   
following the arguments used in the calculation of the first cumulant shown in Fig. \ref{C1&C2}(a), i.e. we assume it diagonal on the initial state and thereby eliminate in the exponent all contributions higher than those $\propto \lambda_{\bf Q}^{2}$. Following the result (\ref{eq:2omega}) we obtain 
\begin{widetext}
\barr
& &\exp\left[-\frac{i}{\hbar}\int_{-\infty}^{t}dt'
\left(\frac{\bcalA_{\parallel}^{2}(\brho,z_{s},t')}{2m^{*}}+
\frac{\bcalA_{\perp}^{2}(\brho,z_{s},t')}{2m}\right)\right]\phi_{{\bf K},s}({\bf r})
\nonumber\\
&=&
\exp\left[-\frac{i}{\hbar}\int_{-\infty}^{\tau}dt' \sum_{\bf Q} \lambda_{\bf Q}^{2}\left(\langle {\bf K},s|\frac{\bcalA_{\parallel,{\bf Q}}({\bf r})\bcalA_{\parallel,{\bf Q}}^{\dag}({\bf r})}{2m^{*}}|{\bf K},s\rangle
+
\langle {\bf K},s|\frac{\bcalA_{\perp,{\bf Q}}({\bf r})\bcalA_{\perp,{\bf Q}}^{\dag}({\bf r})}{2m}|{\bf K},s\rangle\right)
2(1-\cos(2\omega_{sp}t'))\right]\phi_{{\bf K},s}({\bf r})\nonumber\\
&=&
\exp\left[-i\left(\frac{(U_{s}^{\parallel}+U_{s}^{\perp})}{\hbar}\tau-\frac{(U_{s}^{\parallel}+U_{s}^{\perp})}{2\hbar\omega_{sp}}\sin(2\omega_{sp}\tau)\right)\right]\phi_{{\bf K},s}({\bf r})
=
\exp\left[-i\left(\frac{U_{s}}{\hbar}\tau-\beta_{sp}\sin(2\omega_{sp}\tau)\right)\right]\phi_{{\bf K},s}({\bf r}),
\label{eq:A2phi}
\earr
\end{widetext}
where
\barr
U_{s}&=&U_{s}^{\parallel}+U_{s}^{\perp},
\label{eq:Up}\\
\beta_{sp}&=&\frac{U_{s}}{2\hbar\omega_{sp}},
\label{eq:betap}
\earr
and the lateral, $U_{s}^{\parallel}=2{\cal E}_{\parallel}$,  and perpendicular component $U_{s}^{\perp}=2{\cal E}_{\perp}$, of the total ponderomotive shift (\ref{eq:Up}) are according to (\ref{eq:beta_s}) obtained from
\barr
U_{s}^{\parallel}&=& 2{\cal E}_{\parallel}=\sum_{\bf Q} \lambda_{\bf Q}^{2}\langle {\bf K},s|\frac{\bcalA_{\parallel,{\bf Q}}({\bf r})\bcalA_{\parallel,{\bf Q}}^{\dag}({\bf r})}{m^{*}}|{\bf K},s\rangle,
\label{eq:calEpar}\\
U_{s}^{\perp}&=& 2{\cal E}_{\perp}=\sum_{\bf Q} \lambda_{\bf Q}^{2}\langle {\bf K},s|\frac{\bcalA_{\perp,{\bf Q}}({\bf r})\bcalA_{\perp,{\bf Q}}^{\dag}({\bf r})}{m}|{\bf K},s\rangle.
\label{eq:calEperp}
\earr
 Here the factors $1/L$ arising from the scaling of $V_{\bf Q}$ (\ref{eq:V_Q}) and present in $\bcalA$ and $\bcalA^{\dag}$ are cancelled by $L^{2}$ from the two-dimensional density of ${\bf Q}$-states $(L/2\pi)^{2}$ when going from summation to integration over ${\bf Q}$. 
Finally, the vacuum fluctuation contribution $\Phi_{s}$ to the total shift  introduced in (\ref{eq:U+Phi}) is given by
\bq
\Phi_{s}=\sum_{\bf Q}\langle {\bf K},s|\frac{\bcalA_{\parallel,{\bf Q}}({\bf r})\bcalA_{\parallel,{\bf Q}}^{\dag}({\bf r})}{2m^{*}}
+
\frac{\bcalA_{\perp,{\bf Q}}({\bf r})\bcalA_{\perp,{\bf Q}}^{\dag}({\bf r})}{2m}|{\bf K},s\rangle.
\label{eq:Phis_parperp}
\eq

The result in the exponent in (\ref{eq:A2phi}) is completely analogous to expression (\ref{eq:2omega}) from cumulant expansion which therefore offers a physical interpretation of the lowest order two-plasmon contribution to the phase of (\ref{eq:psiVolkovA2}). 
Hence, the effect of the interaction quadratic in $\bcalA$ is to add a time dependent phase to the wavefunction $|\phi_{{\bf K},s}\rangle$. The total ponderomotive shift $U_{s}$ is within the present theory intrisically positive as it arises from the quadratic coupling term  in the Hamiltonian.  The oscillating term is the double plasmon frequency Floquet term[\onlinecite{ReissPRA42}] weighted by the dimensionless two-plasmon absorption and emission amplitude $\beta_{sp}$. Observe also that $\beta_{sp}$ measures the adiabaticity of plasmon field action on electrons and is an analog of the Keldysh's parameter $1/\gamma$.[\onlinecite{Keldysh1965,HommelhoffPRA104}]

The single plasmon Floquet term in the wave function $|\psi_{{\bf K},s}^{(vg)}\rangle$ results from the action of the exponential operator containing $\bcalA\cdot{\bf p}/m$ on the RHS of (\ref{eq:psiVolkovA2}).[\onlinecite{ReissPRA42}] This action is best visualized in the coordinate representation. Invoking (\ref{eq:p_z}) and the factorization of the phase (\ref{eq:A2phi}) we obtain  
\begin{widetext}
\barr
& &
\exp\left[-\left(\int_{t_{0}}^{\tau}dt'
\frac{\bcalA_{\parallel}(\brho,z_{s},t')}{m^{*}}\nabla_{\parallel}+
\int_{t_{0}}^{\tau}dt'
\frac{\bcalA_{\perp}(\brho,z_{s},t')}{m}\nabla_{\perp}\right)\right]\phi_{{\bf K},s}({\bf r})\nonumber\\
&=&
\exp\left[-\left({\bf R}(\brho,z_{s},\tau)\nabla_{\parallel}+Z_{s}(\brho,z_{s},\tau)\nabla_{\perp}\right)\right]\phi_{{\bf K},s}({\bf r}).
\label{eq:Apphi}
\earr
\end{widetext}
Hence, the exponential operator in front of $\phi_{{\bf K},s}({\bf r})$ in the second line on the RHS of (\ref{eq:Apphi}) represents a translation operator in the coordinate space conveniently separated into lateral and perpendicular to the surface components.[\onlinecite{Faisal,MadsenPRA}] The time dependent lateral and perpendicular displacement vectors are obtained by using (\ref{eq:Apar}) and (\ref{eq:Aperp}) as 
\bq
{\bf R}(\brho,z_{s},\tau)=\int_{t_{0}\rightarrow -\infty}^{\tau}dt'
\frac{\bcalA_{\parallel}(\brho,z_{s},t')}{m^{*}}
=
-\sum_{\bf Q}{\bf \hat{e}_{Q}}\lambda_{\bf Q}\frac{2QV_{Q}}{m^{*}\omega_{sp}^{2}}e^{-Q|z_{s}|}\sin({\bf Q}\brho-\omega_{sp}t),
\label{eq:Rpart}
\eq
\bq
Z_{s}(\brho,z_{s},\tau){\bf \hat{e}}_{\perp}=\int_{t_{0}\rightarrow -\infty}^{\tau}dt'
\frac{\bcalA_{\perp}(\brho,z_{s},t')}{m}
=
-{\bf \hat{e}_{\perp}}\sum_{\bf Q}\lambda_{\bf Q}\frac{2QV_{Q}}{m\omega_{sp}^{2}}e^{-Q|z_{s}|}\cos({\bf Q}\brho-\omega_{sp}t).
\label{eq:Zst}
\eq
Both quantities are real and mutually phase shifted by $\frac{\pi}{2}$ that signifies circular polarization of the interaction. Since $\bcalA_{\perp}(\brho,z_{s},t')$ is localized strongly around $z_{s}$, it is reasonable to apply the dipole approximation to (\ref{eq:Zst}) by neglecting ${\bf Q}\brho$ relative to $\omega_{sp}t$. Using (\ref{eq:lambdalin}) this yields
\bq
Z_{s}(\brho,z_{s},\tau)|_{dip}
=
-\sum_{\bf Q}\lambda_{\bf Q}\frac{2QV_{Q}}{m\omega_{sp}^{2}}e^{-Q|z_{s}|}\cos(\omega_{sp}\tau)
\longrightarrow
Z_{s}(z_{s},\tau)=-Z_{s}(\lambda)\cos(\omega_{sp}\tau),
\label{eq:Zsdip}
\eq
where
\bq
Z_{s}(\lambda)=\lambda\sum_{\bf Q}\frac{2QV_{Q}}{m\omega_{sp}^{2}}e^{-Q|z_{s}|}.
\label{eq:Zslambda}
\eq
This approximation may not be justified for (\ref{eq:Rpart}) whose full form is required in (\ref{eq:psiVolkovA2}) for establishing the lateral momentum conservation in the interaction vertices, as illustrated for second order processes in Fig. \ref{C1&C2}c.
However, here the lateral momentum conservation appears in the exponent of an {\em offdiagonal} expression (\ref{eq:psiVolkovA2}) which should be distinguished from cumulant expansion of {\em diagonal} propagators discussed in Sec. \ref{sec:Elspecvel}.

Combining (\ref{eq:A2phi}) and (\ref{eq:Apphi}) with (\ref{eq:Zsdip}) and appropriately restoring the plasmon vacuum fluctuation shift $\Phi_{s}$ we can write (\ref{eq:psiVolkovA2}) as
\begin{widetext}
\barr
\langle\phi_{f}|\psi_{{\bf K}_{s},s}^{(vg)}(\tau)\rangle_{Volkov}&=&\exp\left[-i\left(\frac{(\hbar{\bf K}_{s})^2}{2m^{*}\hbar}+\frac{E_{s}}{\hbar}\right)\tau\right]
\exp\left[-i\left(\frac{\tilde{U}_{s}}{\hbar}\tau-\frac{U_{s}}{2\hbar\omega_{sp}}\sin(2\omega_{sp}\tau)\right)\right]\nonumber\\
&\times&
\left[\int d^{2}\brho \phi_{f}^{*}(\brho)
 \exp\left(-{\bf R}(\brho,z_{s},\tau)\nabla_{\parallel}\right)\phi_{{\bf K}_{s}}(\brho)\right]
\left[\int dz \phi_{f}^{*}(z)\exp\left(-Z_{s}(z_{s},\tau)\frac{\partial}{\partial z}\right) u_{s}(z)\right],
\label{eq:phipsi}
\earr
\end{widetext}
where $\tilde{U}_{s}=U_{s}+\Phi_{s}$ as defined in (\ref{eq:U+Phi}), and $\nabla_{\parallel}=\nabla_{\brho}$.
The main characteristic of this scattering amplitude are the factorized electron coordinate jitterings induced via (\ref{eq:Rpart}) and (\ref{eq:Zsdip}) that in the linear coupling terms are driven by the single plasmon frequency. This time dependence is then additionally modulated by the double plasmon frequency of quadratic coupling from the phase of the second exponential in the first line on the RHS of (\ref{eq:phipsi}). The factorization in the second line on the RHS of (\ref{eq:phipsi}) follows from the commutativity of two exponentials with fixed $z_{s}$. On noticing that the exponential operators in this line represent the translation operators we can write the two spatial integrals in the form of "jittering overlaps"
\begin{widetext}
\bq
\int d^{2}\brho \phi_{f}^{*}(\brho) \phi_{{\bf K}_{s}}\left(\brho-{\bf R}(\brho,z_{s},\tau)\right)
\int dz \phi_{f}^{*}(z) u_{s}\left(z-Z_{s}(z_{s},\tau)\right)
\rightarrow
\left[\int \frac{d^{2}\brho}{\sqrt{L^{2}}} e^{i({\bf K}_{s}-{\bf K}_{f})\brho -i{\bf K}_{s}{\bf R}(\brho,z_{s},\tau)}\right]
\frac{\tilde{u}_{s}(k_{z,f})}{\sqrt{L_{z}}}e^{-ik_{z,f}Z_{s}(z_{s},\tau)}
\label{eq:translatedphi}
\eq
\end{widetext}
where the RHS follows from substituting model wavefunctions (\ref{eq:phi_s})-(\ref{eq:phifin}) into (\ref{eq:phipsi}) and the "surface local" ${\bf R}(\brho,z_{s},\tau)$ was defined in (\ref{eq:Rpart}) prior to the application of dipole approximation.
Since for the electron initial state wave function $\phi_{s}(\brho)\propto e^{i{\bf K}_{s}\brho}$ at the SS-band bottom ${\bf K}_{s}=0$, the same must hold also for the final state $\phi_{f}(\brho)\propto e^{i{\bf K}_{f}\brho}$ which, in turn, produces the $\brho$-integral equal to unity.

\begin{figure}[tb]
\rotatebox{0}{\epsfxsize=8cm \epsffile{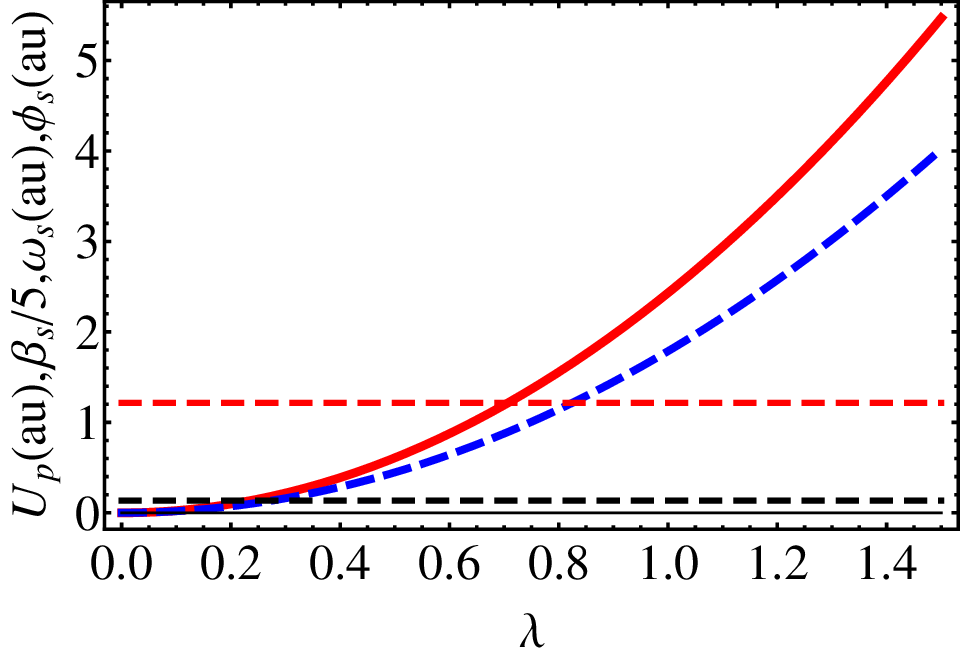}} 
\caption{$\lambda$-scaling of the energy parameters in expression (\ref{eq:phipsi}) determining transition matrix (\ref{eq:TVolkovPsi}) in the velocity gauge. Ponderomotive shift $U_{s}=U_{s}^{\parallel}+U_{s}^{\perp}$ (full red line), Eqs. (\ref{eq:calE||}) and (\ref{eq:calE=}); ground state fluctuation shift $\Phi_{s}$ (upper dashed red horizontal line), Eq. (\ref{eq:Phitot}); adiabaticity parameter $\beta_{s}(\lambda)/5$ (dashed blue line), Eq. (\ref{eq:betaslambda}). Lower dashed black horizontal line denotes the surface plasmon energy $\hbar\omega_{sp}$. All entries shown as functions of plasmonic coherent state parameter $\lambda$ and other parameters fixed at the values characteristic of Ag(111) surface (see Sec. \ref{sec:SSenergetics}). Horizontal dashed black line shows the value of SP energy $\hbar\omega_{sp}$. Vertical scale in atomic units. The values of $\beta_{s}(\lambda)<1$ for $\lambda<0.5$ signify the high plasmon frequency regime.}
\label{ponderomotives}
\end{figure}

Using the parameters characteristic of the SS-band on Ag(111) surface 
the two-plasmon-induced ponderomotive shifts, vacuum fluctuation shift and the adiabaticity parameter, all arising from quadratic coupling, after application of (\ref{eq:lambdaqu}) read  
\barr
U_{s}^{\parallel}(\lambda)&=&\lambda^{2}\sum_{\bf Q}\frac{Q^{2}}{m^{*}}\left(\frac{V_{Q}}{\omega_{sp}}\right)^{2}e^{-2Q|z_{s}|}\nonumber\\
&=&
\frac{\lambda^{2}}{8}\left(\frac{m}{m^{*}}\right)\left(\frac{z_{s}}{a_{B}}\right)^{-3}\left(\frac{\hbar\omega_{sp}}{1 H}\right)^{-1}\times 1H,
\label{eq:calE||}\\
U_{s}^{\perp}(\lambda)&=&\lambda^{2}\sum_{\bf Q}\frac{Q^{2}}{m}\left(\frac{V_{Q}}{\omega_{sp}}\right)^{2}e^{-2Q|z_{s}|}\nonumber\\
&=&
\frac{\lambda^{2}}{8}\left(\frac{z_{s}}{a_{B}}\right)^{-3}\left(\frac{\hbar\omega_{sp}}{1 H}\right)^{-1}\times 1H,
\label{eq:calE=}\\
\Phi_{s}
&=&
\frac{1}{16}\left(\frac{m+m^{*}}{m^{*}}\right)\left(\frac{z_{s}}{a_{B}}\right)^{-3}\left(\frac{\hbar\omega_{sp}}{1 H}\right)^{-1}\times 1H,
\label{eq:Phitot}\\
\beta_{s}(\lambda)&=&\frac{U_{s}(\lambda)}{2\hbar\omega_{sp}}=\frac{\lambda^{2}}{16}\left(\frac{m+m^{*}}{m^{*}}\right)\left(\frac{z_{s}}{a_{B}}\right)^{-3}\left(\frac{\hbar\omega_{sp}}{1 H}\right)^{-2}.
\label{eq:betaslambda}
\earr
 These quantities fully parametrize the contributions from quadratic coupling in (\ref{eq:A2phi}) and (\ref{eq:phipsi}). For the set of parameters introduced in Sec. \ref{sec:SSenergetics} their scaling with $\lambda$ is illustrated in Fig. \ref{ponderomotives}.

Calculations of the contributions from the linear coupling terms appearing in the second line of (\ref{eq:phipsi}) are more complicated. For the sake of simplicity we shall again consider electron emission from the SS-band bottom so that ${\bf K}={\bf K}_{s}=0$. This fixes ${\bf K}_{f}=0$ in $\phi_{f}^{*}(\brho)=\phi_{{\bf K}_{f}}^{*}(\brho)$ in the last line on the RHS of (\ref{eq:translatedphi}). Together with (\ref{eq:Zsdip}) and (\ref{eq:translatedphi}) this yields
\begin{widetext}
\bq
\langle\phi_{f}|\psi_{{\bf K}_{s}=0,s}^{(vg)}(\tau)\rangle_{Volkov}=
\exp\left[-i\left(E_{s}+\tilde{U}_{s}\right)\tau/\hbar+i\beta_{s}(\lambda)\sin(2\omega_{sp}\tau)\right]
\exp\left[ik_{z,f}Z_{s}(\lambda)\cos(\omega_{sp}\tau)\right]\frac{\tilde{u}_{s}(k_{z,f})}{\sqrt{L_{z}}},
\label{eq:phipsiK=0}
\eq
\end{widetext}

Upon writing 
\bq
\exp\left[ik_{z,f}Z_{s}(\lambda)\cos(\omega_{sp}\tau)\right]
=
\exp\left[-ik_{z,f}Z_{s}(\lambda)\sin\left(\omega_{sp}\tau-\frac{\pi}{2}\right)\right]
=
\sum_{n=-\infty}^{\infty}i^{n}e^{in\omega_{sp}\tau}J_{n}(k_{z,f}Z_{s}(\lambda)),
\label{eq:expZs+}
\eq
we can bring the whole expression (\ref{eq:phipsiK=0}) to the form representable in terms of generating functions of the Bessel functions.
Combining  the formulas from Appendix  \ref{sec:alternativeT} with the  formulas  (\ref{eq:BesselJ}) and (\ref{eq:BesselI}) and the definition of generalized Bessel functions[\onlinecite{Reiss1980}]
\bq
 J_{n}(x,y)=\sum_{k=-\infty}^{\infty}J_{n-2k}(x)J_{k}(y),
\label{eq:genJ}
\eq
we can represent the oscillating phase component of (\ref{eq:phipsiK=0}) for ${\bf K}_{s}=0$ in the form (cf. Appendix \ref{sec:alternativeT}) 
\begin{widetext}
\barr
\exp\left[i\beta_{s}(\lambda)\sin(2\omega_{sp}\tau)+ik_{z,f}Z_{s}(\lambda)\cos(\omega_{sp}\tau)\right]
&\rightarrow&
\exp\left[-i\beta_{s}(\lambda)\sin\left(2(\omega_{sp}\tau-\pi/2)\right)-ik_{z,f}Z_{s}(\lambda)\sin(\omega_{sp}\tau-\pi/2)\right]\nonumber\\
&=&
\sum_{m=-\infty}^{\infty}(i)^{m}e^{im\omega_{sp}\tau}J_{m}\left(-k_{z,f}Z_{s}(\lambda),-\beta_{s}(\lambda)\right).
\label{eq:oscphase+}
\earr
\end{widetext}
 The rationale underlying the representation of expression on the LHS of (\ref{eq:oscphase+}) in the form displayed on the RHS  is the multiple $m\omega_{sp}$  in the exponent of $e^{im\omega_{sp}\tau}$ that for $m>0$ is associated with plasmon absorption by the electron.   
Substituiting (\ref{eq:oscphase+}) back in (\ref{eq:phipsi}) we can bring the latter to the Floquet form
\begin{widetext}
\bq
\langle\phi_{f}|\psi_{{\bf K}_{s}=0,s}^{(vg)}(\tau)\rangle=
\exp\left[-i\left(\frac{E_{s}+\tilde{U}_{s}(\lambda)}{\hbar}\right)\tau\right]
\sum_{n=-\infty}^{\infty}(i)^{n} e^{in \omega_{sp}\tau}
J_{n}\left(-k_{z,f}Z_{s}(\lambda),-\beta_{s}(\lambda)\right)\frac{\tilde{u}_{s}(k_{z,f})}{\sqrt{L_{z}}}.
\label{eq:fullVolkov}
\eq
\end{widetext}
Here the repeated minus signs in the argument of $J_{n}(x,y)$ keep track of the transformation of trigonometric functions in the first line of (\ref{eq:oscphase+}). 
Hence, the total time-dependent multiplasmon emission/absorption amplitude appears as an interplay between the elementary one-plasmon and two-plasmon excitation amplitudes controlled by $Z_{s}(\lambda)$ and $\beta_{s}(\lambda)$, respectively. 
This finally gives the velocity gauge analog of (\ref{eq:TlgSBC}) in the dipole approximation and for initial ${\bf K}_{s}=0$ in the closed form
\begin{widetext}
\bq
 T_{{\bf K}_{s}=0,f\leftarrow s}^{(vg)}= -2\pi i\left(E_{f}-E_{s}\right)
\sum_{n=-\infty}^{\infty} (i)^{n}J_{n}\left(-k_{z,f}Z_{s}(\lambda),-\beta_{s}(\lambda)\right)\frac{\tilde{u}_{s}(k_{z,f})}{\sqrt{L_{z}}}
\delta\left(E_{f}-E_{s}-\tilde{U}_{s}(\lambda)-n\hbar\omega_{sp}\right),
\label{eq:Tfinvg}
\eq
\end{widetext}
The $\pm$ signs of the arguments of $J_{n}(x,z)$ can be handled using the identities (B7) from Ref. [\onlinecite{Reiss1980}], viz. 
\barr
J_{n}(-x,y)=(-1)^{n}J_{n}(x,y),
\label{eq:Jn-xy}\\
J_{n}(x,-y)=(-1)^{n}J_{-n}(x,y),
\label{eq:Jnx-y}
\earr
and the same remark made after Eq. (\ref{eq:TlgSBC}) that concerns the Born approximation limit of the $T$-matrix (\ref{eq:Tfinvg}) applies here as well.

 The sums on the RHS of (\ref{eq:TlgSBCtau}) and (\ref{eq:fullVolkov}) have the appearance of Fourier series of functions periodic in the time interval $2\pi/\omega_{sp}$.  Such components of the wave function may lead to periodic structures in the electron excitation spectra. Their intensities are determined by the quantities that derive from the electron coupling to plasmons prepumped into the coherent state $|\mbox{coh}\rangle$. Since the latter is expressed through the eigenvalues $\lambda_{\bf Q}$ which depend on the history of plasmon pumping by external fields, the values of ${\cal W}({\bf r})$, $\beta_{s}(\lambda)$ and $Z_{s}(\lambda)$, and hence of (\ref{eq:TlgSBC}) and (\ref{eq:fullVolkov}), depend on external parameters of the present model.



\subsection{Model calculations of multiplasmon-induced transition rates in the velocity gauge}
\label{sec:Floquetframeworkvg}

In the derivation of state-to-state transition rates in the velocity gauge we again restrict the analysis to vertical transitions from the SS-band, i.e. to the case ${\bf K}_{s}=0$.
We start from (\ref{eq:Tfinvg}) and the same observation that 
the large energy differences between different $n$-contributions to (\ref{eq:Tfinvg}) prevent their constructive or destructive interference in the total state-to-state transition probability $|T_{({\bf K}_{f},f)\leftarrow( {\bf K}_{s}=0,s)}^{(vg)}|^{2}$. Then we follow the same procedure as in Sec. {\ref{sec:Volkovdipole} to obtain in the narrow final state window $[{\bf K}_{f},k_{z,f}]$
\barr
& &\sum_{[{\bf K}_{f},k_{z,f}]}|T^{(vg)}_{({\bf K}_{f},f)\leftarrow( {\bf K}_{s}=0,s)}|^{2}
=
\sum_{n=-\infty}^{\infty}\sum_{[{\bf K}_{f},k_{z,f}]}|T_{({\bf K}_{f},f)\leftarrow( {\bf K}_{s}=0,s)}^{(vg)}(n)|^{2}\nonumber\\
&=&
\sum_{n=-\infty}^{\infty}\sum_{[{\bf K}_{f},k_{z,f}]}W_{({\bf K}_{f},f)\leftarrow( {\bf K}_{s}=0,s)}^{(vg)}(n)2\pi\hbar\rho(E_{k_{f}}^{(n)})\delta_{k_{z,f},k_{z,f}(n)},
\label{eq:Tvgsquared}
\earr
with the $n$-th Floquet component given in the length gauge by   
\begin{widetext}
\bq
 W_{({\bf K}_{f},f)\leftarrow( {\bf K}_{s}=0,s)}^{(vg)}(n)= \frac{2\pi}{\hbar} \left(\tilde{U}_{s}+n\hbar\omega_{sp}\right)^{2}
\left|J_{n}\left(-k_{z,f}Z_{s}(\lambda),-\beta_{s}(\lambda)\right)\right|^{2}\frac{|\tilde{u}_{s}(k_{z,f})|^{2}}{L_{z}}
\delta\left(E_{k_{z,f}}-E_{s}-\tilde{U}_{s}-n\hbar\omega_{sp}\right).
\label{eq:Wvg}
\eq
\end{widetext}
This expression describes the electron state-to-state transition rate per unit time in the $n$-th channel. The associated $2\pi\hbar\rho(E_{k_{z,f}}^{(n)})$ has the meaning of electron traversal time in the $n$-th emission channel, rendering (\ref{eq:Tvgsquared}) dimensionless and free from $L_{z}$. Note also that according to (\ref{eq:Wvg}) all terms in the sum on the RHS of (\ref{eq:Tvgsquared}) represent ascending even order responses of electrons to the action of prepumped plasmonic field. 

Likewise in the case of length gauge described in Sec. {\ref{sec:Volkovdipole}, expression (\ref{eq:Wvg}) is as far as one can go with nonperturbative closed form solution for the plasmoemission probabilities. Expansion of the Bessel functions into a power series then yields perturbative contributions to (\ref{eq:Tvgsquared}) in the various scattering channels. However, in order to comply with (\ref{eq:translatedphi}) we must require that in these processes all intermediate plasmon wavevectors ${\bf Q}$ satisfy momentum conservation in going from ${\bf K}_{s}$ to ${\bf K}_{f}$, i.e. in the present case starting from  ${\bf K}_{s}=0$, to sum to zero. However, instead of adopting this essentially perturbative and impractical procedure we shall follow the line implemented in the length gauge that exploits the analogy with cumulant approach and leads to representation (\ref{eq:ansatzJn}). Thus in (\ref{eq:Wvg}) we also make an "equivalent BA substitution" 
\bq
\left|J_{n}\left(-k_{z,f}Z_{s}(\lambda),-\beta_{s}(\lambda)\right)\right|^{2}
\rightarrow
\left|J_{n}\left(-2k_{z,f}\zeta_{s}^{(off)}(\lambda,z_{s}),-\beta_{s}(\lambda)\right)\right|^{2}
=
w_{n}^{(vg)}(\lambda),
\label{eq:substzeta}
\eq
where for $Z_{s}(\lambda)$ given by (\ref{eq:Zslambda}) the form of $\zeta_{s}^{(off)}(\lambda,z_{s})$ is determined to satisfy 
\bq
2k_{z,f}\zeta_{s}^{(off)}(\lambda,z_{s})= k_{z,f}\sqrt{\sum_{\bf Q} \lambda^{2}\left(\frac{2QV_{Q}}{m\omega_{sp}^{2}}\right)^{2}e^{-2Qz_{s}}}
=
\lambda\frac{(k_{z,f}a_{B})}{\sqrt{2}}\left(\frac{z_{s}}{a_{B}}\right)^{-3/2}\left(\frac{\hbar\omega_{sp}}{1H}\right)^{-3/2}.
\label{eq:zetas}
\eq 
This is also an off-the-energy-shell quantity likewise analogous expression (\ref{eq:xioff}) in the length gauge.  
With the ansatz (\ref{eq:substzeta}) we correctly reproduce the BA results (\ref{eq:Gammavg2}) and (\ref{eq:Gammavg/omegasperp}) obtained from second order cumulant expansion in terms of $k_{z,s}Z_{s}(\lambda)$. The merit of replacing $k_{z,f}Z_{s}(\lambda)$ by $2k_{z,f}\zeta_{s}^{(off)}(\lambda,z_{s})$ in (\ref{eq:Wvg}) is that thereby  both dimensionless variables of generalized Bessel function on the RHS of (\ref{eq:substzeta}) are now free from the quantization lengths $L$ and $L_{z}$, and the exact BA result $|k_{z,f}^{(1)}\zeta_{s}^{(off)}(\lambda,z_{s})|^2$ is retrieved by retaining only the first expansion term of (\ref{eq:substzeta}) for $n=1$. Hence the powers of (\ref{eq:zetas}) and functions thereof are amenable to graphical representation. Figure \ref{w1vglambda} illustrates the $\lambda$-dependence of  the RHS of (\ref{eq:substzeta}) for $n=1$ and $n=2$. 
 These results should be compared with the ones calculated in the length gauge and presented in Fig. \ref{wnlglambda}.

\section{Plasmoemission currents from surface Floquet bands}
\label{sec:emissionFloquet}

\subsection{Gauge specifities of plasmoemission from $\rm{Ag}$(111) surface bands}
\label{sec:discussion}

Signatures of Floquet sidebands in plasmoemission rates predicted by the RHS of $T$-matrix expressions (\ref{eq:Wlg}) and (\ref{eq:Wvg}) in the length and velocity gauge, respectively, largely depend on two factors. The first is related to the values of Bessel functions with parametric variables ${\cal W}_{dip}(\lambda)$, $U_{s}(\lambda)$, $\beta_{s}(\lambda)$ and $Z_{s}(\lambda)$ in relation to $\hbar\omega_{sp}$ of the interacting e-SP system. The second one pertains to the overall magnitude of (\ref{eq:Wlg}) and (\ref{eq:Wvg}) which determine the weight of each $\delta$-function in the sum over the Floquet band index $n$. Here we shall separately examine the constituents of plasmoemission intensities from surface bands on Ag(111) in the length and velocity gauge. In the latter case we shall also inspect the high plasmon frequency regime $\hbar\omega_{sp}\geq U_{s}$ and the quasistatic strong field limit $U_{s}\gg\hbar\omega_{sp}$. 

The structures of closed form solutions for the transition probabilities (\ref{eq:Tlgsquared}) and (\ref{eq:Tvgsquared}) clearly reveal the plasmon field driven surface electronic Floquet bands shifted from the parent one by multiples of $\hbar\omega_{sp}$[\onlinecite{Giovannini}] and in the velocity gauge additionally by the positive ponderomotive energy $\tilde{U}_{s}(\lambda)$. The latter case situation is schematically illustrated in Fig. \ref{SPFloquet}. 
   The weight of each $n$-th Floquet sideband is determined by a multiple  sequence of plasmon-assisted processes generated by the matrix elements of  electron-plasmon  interaction in the corresponding gauge.
Hence, the primary pumping of plasmonic coherent states may give rise to non-Einsteinian electron emission signal at multiples of $\hbar\omega_{sp}$, and this process will be appropriately described by the transition rates  (\ref{eq:Wlg}) and (\ref{eq:Wvg}) provided the final state $|\phi_{{\bf K},f}\rangle$ is  outgoing wave solution for the potential (\ref{eq:Vscal}). 
For definiteness of the calculations we have approximated the wavefunction of this state by the simple form (\ref{eq:phifin}). 
 Therefore, expressions (\ref{eq:Wlg}) and (\ref{eq:Wvg}) constitute the dimensionless probabilities (\ref{eq:Tlgsquared}) and (\ref{eq:Tvgsquared})  which describe electron emission current from the $n$-th Floquet emission channel in the length and velocity gauge, respectively. This is analogous to the photoemission current sought in Refs. [\onlinecite{Adawi,Bunkin,Mahan,Ashcroft}].

Then, for vertical transitions from the zone center ${\bf K}_{s}=0$ the prefactors of $\delta$-functions on the RHS of (\ref{eq:Wlg}) and (\ref{eq:Wvg}) can be factorized to produce gauge-specific state-to-state transition rates, viz.  
\bq 
 W_{f\leftarrow s}^{(lg)}(k_{z,f}^{(n)},n)=
\frac{2\pi}{\hbar L_{z}}(n\hbar\omega_{sp})^2|\tilde{u}_{s}(k_{z,f}^{(n)})|^{2}w_{n}^{(lg)}(\lambda),
\label{eq:Wlgfinal}
\eq
where we use a convenient ansatz form of $w_{n}^{(lg)}(\lambda)$ introduced in (\ref{eq:ansatzJn}) and plotted in Fig. \ref{Gammaslgvg}.
Analogously, in the velocity gauge we have
\bq 
 W_{f\leftarrow s}^{(vg)}(k_{z,f}^{(n)},n)=
\frac{2\pi}{\hbar L_{z}}(n\hbar\omega_{sp}+\tilde{U}_{s})^2|\tilde{u}_{s}(k_{z,f}^{(n)})|^{2}w_{n}^{(vg)}(\lambda),
\label{eq:Wvgfinal}
\eq
with the ansatz for $w_{n}^{(vg)}(\lambda)$ introduced in relation (\ref{eq:substzeta}) and plotted in Fig. \ref{w1vglambda}.
Note at this point that we have justified the forms of these ans\"{a}tze only by the requirement of circumventing the problem of momentum summations in the  intermediate propagations between the initial and final states in the matrix elements (\ref{eq:Wlg}) and (\ref{eq:Wvg}). Hence, their use is problem-specific and may be unfounded outside this context. Next, in both expressions (\ref{eq:Wlg}) and (\ref{eq:Wvg}), and consequently in (\ref{eq:Wlgfinal}) and (\ref{eq:Wvgfinal}), the extra factor $1/L_{z}$ arises from using (\ref{eq:phifin}) in the calculations of the transition matrix elements, and $k_{f}^{(n)}$ and $E_{k_{f}}^{(n)}$ are the values of $k_{f}$ and $E_{k_{f}}$ constrained to the energy shell by the $\delta$-functions on the RHS of (\ref{eq:Wlg}) and (\ref{eq:Wvg}), respectively.  
Finally, the Fourier transform $\tilde{u}_{s}(k_{f}^{(n)})$ in (\ref{eq:Wlgfinal}) and (\ref{eq:Wvgfinal}) plays the role of static form factor for inelastic transitions whereas the only trully gauge-specific constituents are
$w_{n}^{(lg)}(\lambda)$ and $w_{n}^{(vg)}(\lambda)$ which depend on the dynamics of plasmon field.

 The linear coupling limit of (\ref{eq:Wvg}), in which the quadratic coupling ${\bf A}^{2}({\bf r})/2m$ is neglected, is obtained by setting $\beta_{s}=0$ in (\ref{eq:substzeta}) which leaves only $J_{k=0}(0)=1$ in the expansion             (\ref{eq:genJ}). In the first order perturbation with linear coupling (Born approximation) only the terms $n=\mp 1$ survive in (\ref{eq:Wvg}) and (\ref{eq:substzeta}), yielding $w_{1}^{(lg)}({\cal W}(\lambda)/\hbar\omega_{sp})\rightarrow\left|\xi_{s}^{(off)}(\lambda)\right|^{2}$ and $w_{1}^{(vg)}(k_{z,f}^{(n)}Z_{s},\beta_{s})\rightarrow|k_{z,f}^{(1)}\zeta_{s}^{(off)}(\lambda)|^{2}$, respectively.

The static form factor $\tilde{u}_{s}(k_{f}^{(n)})$ in (\ref{eq:Wlgfinal}) and (\ref{eq:Wvgfinal}) may act either as a muffler or an amplifier  for the  transition amplitudes and rates.  Moreover, since the Bessel functions  have zeros on the real axis the variation of their arguments can give rise to resonant and antiresonant behaviour in the transition rates. In the velocity gauge this modulating effect is enhanced by the interference among the various intermittent $k$-fold plasmon absorption and emission processes (cf. Eq. (\ref{eq:genJ})) that lead to the same $n$-th final Floquet state. Provided the preference of electron excitation from the Fermi level of metallic  systems[\onlinecite{plasPE,Hopfield1965}] holds also in hot electron dynamics at surfaces,[\onlinecite{encyclopedia}] such yields would manifest as discernible peaks in the electron emission spectra, as exemplified in Fig. \ref{AllAgFloquet}b.

The transition rates (\ref{eq:Wlgfinal}) and (\ref{eq:Wvgfinal}) have been derived by using various simplifications and approximations described in detail in the preceding sections. Their final operative nonperturbative forms incorporate an interplay of vertex and selfenergy corrections for electron propagation but their separate roles in final transition amplitudes can not be straightforwardly disentangled. However, following the analyses in Secs. \ref{sec:EspectrumLG} and \ref{sec:EPdynvg} it is envisaged that self-energy corrections would give rise to the plasmonically-induced Floquet band structure whereas the vertex corrections should induce electron transitions from such bands into the outgoing wave states supporting the plasmoemission current.  
%

\begin{figure}[tb]
\rotatebox{0}{\epsfxsize=8.5cm \epsffile{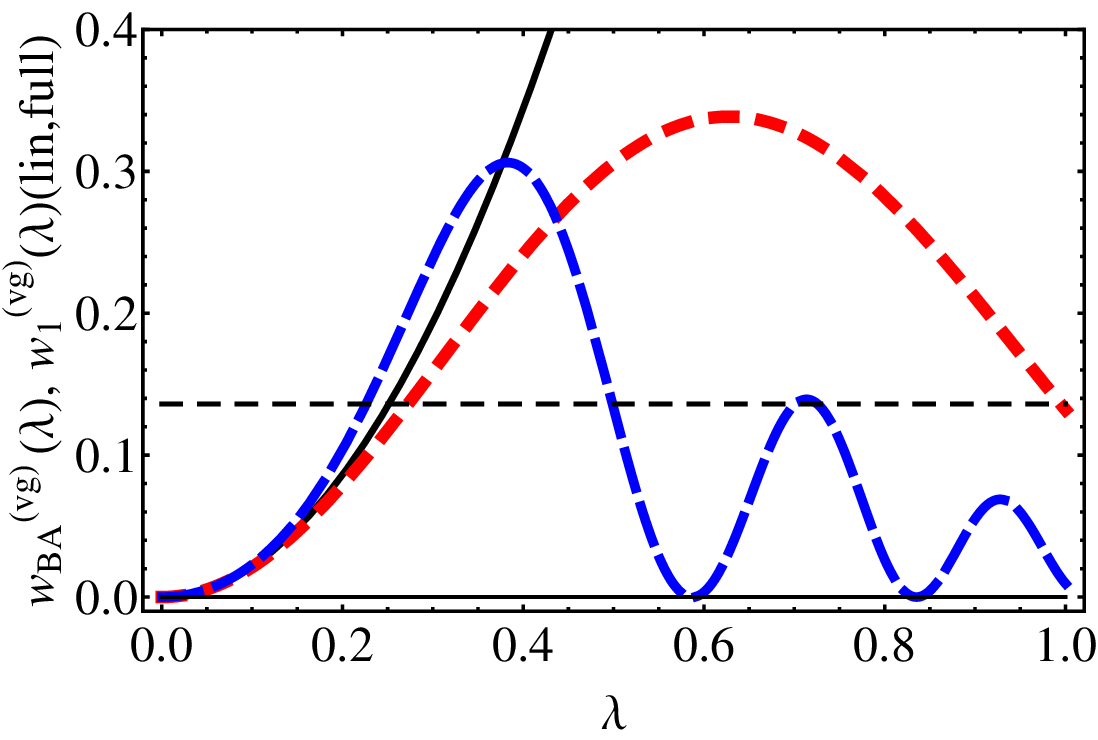}} 
\caption{$\lambda$-dependence of $w_{1}^{(vg)}(\lambda)$ (Eq. (\ref{eq:substzeta})) constituting the transition probability (\ref{eq:Wvg}) appropriate to one-plasmon-induced electron emission from SS-band bottom on Ag(111) surface with reduced work function as calculated in the velocity gauge. Thick black curve: Born approximation (BA) result $w_{BA}^{(vg)}(\lambda)=|\beta_{s}(\lambda)/2|^{2}$. Dashed red  curve: $w_{1}^{(vg)}(\lambda)$ calculated only with linear coupling ${\bf p}\bcalA/m$. Long dashed blue curve: $w_{1}^{(vg)}(\lambda)$ calculated with full linear and quadratic coupling ${\bf p}\bcalA/m+\bcalA^2/2m$. Horizontal thin dashed black line denotes the surface plasmon energy on Ag(111) surface measured in a.u. Notable is the effect of renormalization of one-plasmon-induced emission by the quadratic coupling.
 The BA result makes a good approximation for $\lambda<0.2$ . }
\label{w1vglambda}
\end{figure}

\begin{figure}[tb]
\rotatebox{0}{\epsfxsize=8.5cm \epsffile{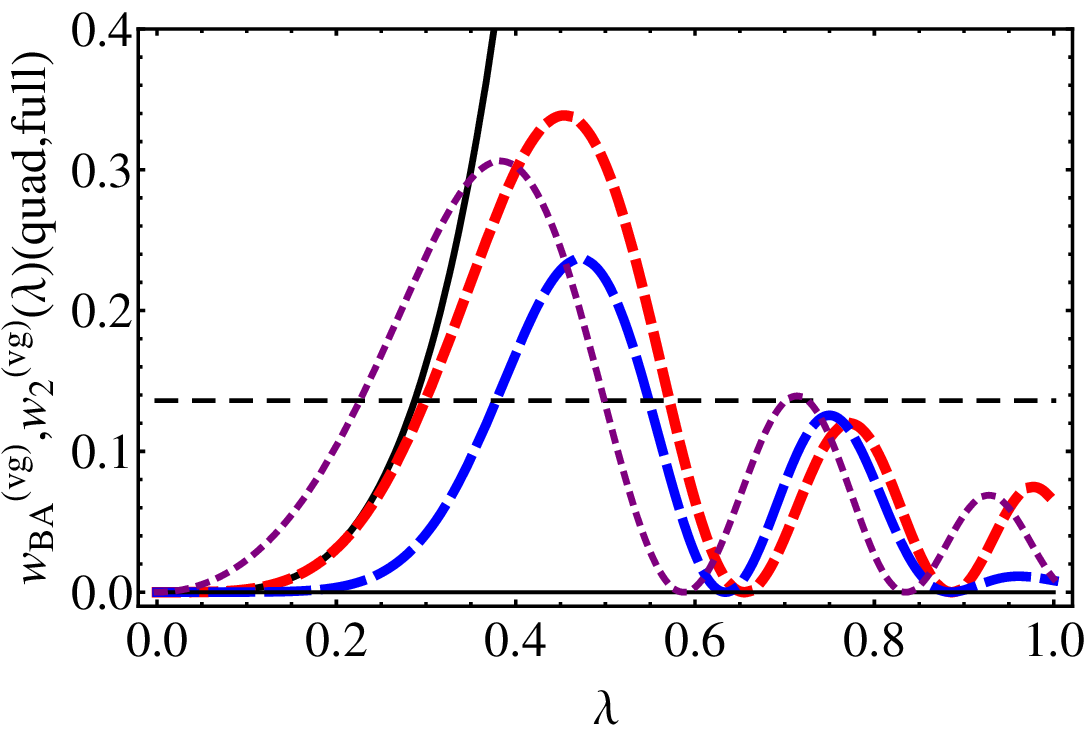}} 
\caption{$\lambda$-dependence of $w_{2}^{(vg)}(\lambda)$ (Eq. (\ref{eq:substzeta})) constituting the transition probability (\ref{eq:Wvg}) appropriate to two-plasmon-induced electron emission from SS-band bottom on Ag(111) surface with reduced work function as calculated in the velocity gauge. Thick black curve: Born approximation (BA) result $w_{BA}^{(vg)}(\lambda)=|\beta_{s}(\lambda)/2|^{4}$ for quadratic coupling $\bcalA^2/2m$. Dashed red  curve: $w_{2}^{(vg)}(\lambda)$ calculated only with quadratic coupling $\bcalA^2/2m$. Long dashed blue curve: $w_{1}^{(vg)}(\lambda)$ calculated with full linear and quadratic coupling ${\bf p}\bcalA/m+\bcalA^2/2m$. Horizontal thin dashed line denotes the surface plasmon energy on Ag(111) surface measured in a.u. The effect of renormalization of two-plasmon-induced emission by the linear coupling is less drastic than in the reverse case illustated in Fig. \ref{w1vglambda}. The two consecutive actions of linear couplings inhibit the effect of quadratic coupling despite their same asymptotic behaviour $\propto\lambda^{4}$ for $\lambda\rightarrow 0$. Purple dotted line shows full $w_{1}^{(vg)}(\lambda)$ from Fig. \ref{w1vglambda} that has been added for comparison. Note its asymptotic $\propto\lambda^2$ behaviour for $\lambda\rightarrow 0$.}
\label{w2vglambda}
\end{figure}

The derived expressions for the transition amplitudes and rates provide a proof of concept for non-Einsteinian electron emission from surface Floquet bands. These bands can be generated by sufficiently populated plasmonic coherent state clouds prepumped in interactions of external EM fields with electrons in metals.[\onlinecite{plasPE}] These expressions were derived within the framework of the length and velocity gauge and the assumption of strong plasmon field localization at the surface that allowed a rather straightforward navigation among the various intermediate steps of the derivation. In this limit numerical assessments of the gauge invariance have also been made for electron interactions with homogeneous EM fields in solids and the results for photodressing and optical conductivities showed close values of the length and velocity gauge results.[\onlinecite{SchuelerPRB,SchuelerJESRP}]


\subsection{Plasmoemission electron currents as indicators of plasmonic coherent states}
\label{sec:vgdynamical}

The gauge-specific quantities $\xi_{s}^{(off)}(\lambda,z_{s})$, $\zeta_{s}^{(off)}(\lambda,z_{s})$, $U_{s}(\lambda)$, $\beta_{s}(\lambda)$ and $\Phi_{s}$ can be readily calculated once $z_s$ is known. We first consider the high plasmon frequency regime $\hbar\omega_{sp}>U_{s}(\lambda)$.

The tunable $\lambda$ introduced in (\ref{eq:lambdalin}) 
serves as a measure of the efficiency of pumping the SP coherent state[\onlinecite{plasPE}] and enables a straightforward evaluation of $\xi_{s}^{(off)}(\lambda,z_{s})$, $\zeta_{s}^{(off)}(\lambda,z_{s})$, $U_{s}(\lambda,z_{s})$ and $\beta_{s}(\lambda,z_{s})$ defined in the dipole approximation in the previous sections. This  gives in atomic units of length $a_{B}$, energy $e^{2}/a_{B}=1\mbox{H}$, momentum $\hbar/a_{B}$ the $\lambda$-scaling functions [\onlinecite{scaling}] which we reiterate for convenience  
\barr
& &\xi_{s}^{(off)}(\lambda,z_{s})=\frac{\lambda}{2}\left(\frac{z_{s}}{a_{B}}\right)^{-\frac{1}{2}}\left(\frac{\hbar\omega_{sp}}{1\mbox{H}}\right)^{-\frac{1}{2}}
\label{eq:xioff2}\\
& &\zeta_{s}^{(off)}(\lambda,z_{s})=\lambda\frac{(k_{z,f}a_{B})}{2\sqrt{2}}\left(\frac{z_{s}}{a_{B}}\right)^{-3/2}\left(\frac{\hbar\omega_{sp}}{1H}\right)^{-3/2}
\label{eq:zeta_s}\\
& &\beta_{s}(\lambda,z_{s})=\frac{\lambda^{2}}{16}\frac{m(m+m^{*})}{mm^{*}}\left(\frac{z_{s}}{a_{B}}\right)^{-3}\left(\frac{\hbar\omega_{sp}}{1 H}\right)^{-2},
\label{eq:betalambda}\\      
& &U_{s}(\lambda)=\frac{\lambda^{2}}{8}\frac{m(m+m^{*})}{mm^{*}}\left(\frac{z_{s}}{a_{B}}\right)^{-3}\left(\frac{\hbar\omega_{sp}}{1 H}\right)^{-1}\times 1H,     
\label{eq:Ulambda}\\
& &\Phi_{s}(z_{s})=\frac{1}{2}U_{s}(\lambda=1,z_{s}),
\label{eq:Phislambda1}\\
& &w_{n}^{(lg)}(\lambda)=|J_{n}(2\xi_{s}^{(off)}(\lambda,z_{s}))|^{2}
\label{eq:wnlglambda}\\
& &w_{n}^{(vg)}(\lambda)=\left|J_{n}\left(-2k_{z,f}\zeta_{s}^{(off)}(\lambda,z_{s}),-\beta_{s}(\lambda)\right)\right|^{2}.
\label{eq:wnvglambda}
\earr  
Thus, the magnitude of $\lambda$ controls the high and low frequency regimes defined as $\beta_{s}(\lambda)<1$ and $\beta_{s}(\lambda)>1$, respectively, as well as the relative contributions of (\ref{eq:Ulambda}) and (\ref{eq:Phislambda1}) to the sum $\tilde{U}_{s}(\lambda)=U_{s}(\lambda)+\Phi_{s}$ appearing in the transition rates (see Fig. \ref{ponderomotives}).

Of particular importance is the $\lambda$-dependence of the main constituents of expressions (\ref{eq:Wlgfinal}) and (\ref{eq:Wvgfinal}), viz.  
\barr
w_{n}^{(lg)}(\lambda)&=&w_{n}^{(lg)}(\lambda,\xi_{s}^{(off)}), 
\label{eq:wnlglambda2}\\
w_{n}^{(vg)}(\lambda)&=&w_{n}^{(vg)}(\lambda,k_{z,f}^{(n)}\zeta_{s}^{(off)}), 
\label{eq:wnvglambda2}
\earr
where in (\ref{eq:wnvglambda2}) the value of $k_{z,f}^{(n)}$ is determined from $E_{k_{z,f}}^{(n)}$ constrained to the energy shell of $n$-plasmon assisted electron emission. Thus, for one SP-assisted emission $E_{z,f}^{(1)}$ and $k_{z,f}^{(1)}$ should correspond to emission from Ag(111) surface with sufficiently reduced work function, e.g. by alkali submonolayer adsorption.[\onlinecite{Horn,Petek2008,PetekJPC2011,Raseev}] For two SP-assisted emission from a clean Ag(111) surface $E_{z,f}^{(2)}$ and $k_{z,f}^{(2)}$ should be representative also of the situation depicted in Fig. \ref{AllAgFloquet}.

 The $\lambda$-dependence of the full and various limiting forms of $w_{1,2}^{(lg)}(\lambda)$ and $w_{1,2}^{(vg)}(\lambda)$ that also illustrate strong interplay of the one- and multi-plasmon-induced emission processes are illustrated in Figs. \ref{wnlglambda}, \ref{w1vglambda} and \ref{w2vglambda}. 
 In particular, the various contributions to $w_{n}^{(vg)}(\lambda)$ for $n=1$ and $n=2$, as  calculated separately for the full linear and quadratic coupling, only linear coupling, and in the first order Born approximation, are shown  in Figs. \ref{w1vglambda} and \ref{w2vglambda} as functions of $\lambda$. Here for the sake of comparing our model results we relate $k_{z,f}^{(n)}$ to Ag(111) surface with work function reduced by $\Delta\phi=1.3$ eV [\onlinecite{Horn}]. In doing so we assert that the initial SS-states remain robust with respect to $\Delta\phi$ (cf. Figs. 1 in [\onlinecite{Petek2008,PetekJPC2011,Raseev}]) whereas the IP-states downshift from the energy interval of plasmonically driven $|SS\rangle\rightarrow|f\rangle$ transition resonances.

The results for $w_{n}^{(lg,vg)}(\lambda)$ presented in Figs. \ref{wnlglambda}, \ref{w1vglambda} and \ref{w2vglambda} convey several important messages: 
\\
{\it (i)}  $w_{n}^{(lg)}(\lambda)$ and $w_{n}^{(vg)}(\lambda)$ are extremely sensitive to the effective number of plasmons (as measured by $\lambda$) that have been pumped into the coherent state for driving plasmoemission.
\\
{\it (i)} The magnitudes of ubiquotus BA limits of $w_{n}^{(lg)}(\lambda)$ and $w_{n}^{(vg)}(\lambda)$ ($\propto\lambda^{2}$) are nearly the same for realistic values of the parameters $z_{s}$ and $\hbar\omega_{sp}$.
\\ 
{\it (iii)} For the velocity gauge form of SP vector potential the interplay between the linear and quadratic coupling is very strong so that they must be treated on an equivalent footing.[\onlinecite{eqfooting,Despoja2016}] Figures \ref{w1vglambda} and \ref{w2vglambda} specifically illustrate this for one- and two-plasmon-driven electron emissions. 
\\
{\it (iv)} Quadratic coupling in the velocity gauge strongly renormalizes also one-plasmon induced transitions (cf. the difference between the red and blue dashed red curves in Fig. \ref{w1vglambda}), in a fashion analogous to the DW factor.[\onlinecite{nonlinscat}] The linear coupling result for one plasmon-induced transition starts to deviate from the first order Born approximation for $\lambda>0.25$.   
\\
{\it (v)}  Linear and quadratic couplings in the velocity gauge give nearly equal contributions to  two-plasmon-assisted electron emission for $\lambda>0.5$ (cf. dashed red and blue lines in Fig. \ref{w2vglambda}). Two successive one-plasmon assisted processes (dashed red curve) are dominant for $\lambda<0.3$. This is in accord with earlier findings for coupled quasiparticle-boson systems.[\onlinecite{nonlinscatt}] However, these trends may be strongly modified by the prefactors of $w_{n}^{(lg,vg)}(\lambda)$ present in Eqs. (\ref{eq:Wlgfinal}) and (\ref{eq:Wvgfinal}).
\\
{\it (vi)} Fully renormalized one- and two-plasmon driven electron emissions obtained from the length gauge are out of phase for $\lambda>0.5$ (cf. Fig \ref{wnlglambda}). By contrast, in the velocity gauge they exhibit pronounced resonant-like behaviour in the same interval of $\lambda$ (cf. Fig. \ref{w2vglambda}).

 It is gratifying that irrespective of the gauge the present model predicts considerable plasmoemission effects for relatively small $\lambda$ (cf. Figs. \ref{wnlglambda}, \ref{w1vglambda} and \ref{w2vglambda}) and consistent with the dynamical limit $\hbar\omega_{sp}\sim U_{s}$. This is important because the pumping of coherent plasmonic states with optimal properties[\onlinecite{plasPE}] represents one of the emission limiting factors. Another limitation affecting expressions (\ref{eq:Wlgfinal}) and (\ref{eq:Wvgfinal}), and thereby the emission currents, may come from the form factor for higher order processes that result in large $k_{z,f}^{(n)}$ and correspondingly small $|\tilde{u}_{s}(k_{z,f}^{(n)})|^{2}$ (cf. inset in Fig. \ref{ponderomotives}). 

 Next we formulate expressions (\ref{eq:Wlgfinal}) and (\ref{eq:Wvgfinal}) for description of experimental situation of $n$-plasmon assisted electron emission with ${\bf K}_{s}=0$ to obtain normal electron emission current ${\cal J}_{s}(n,\lambda)$ from a narrow interval around the final state energy $E_{z,f}^{(n)}$. This gives  
\bq
{\cal J}_{s}^{(lg,vg)}(n,\lambda)= W_{f\leftarrow s}^{(lg,vg)}(k_{z,f}^{(n)},n)\rho(E_{z,f}^{(n)}).
\label{eq:Pn}
\eq
These gauge-dependent expressions are free from the quantization length $L_{z}$ and have the dimension of inverse time which within the normalizations of (\ref{eq:phi_s}) and  (\ref{eq:u(ks)})  signifies the CIS emission current in the $z$-direction (cf.  (\ref{eq:jz}) and Ref. [\onlinecite{LL}]). The same result can be  obtained directly from the generic expression $\sum_{f}j_{z}^{(n)}|T_{{\bf K}_{s}=0,f\leftarrow s}^{(n)}|^2$ by using definitions for the scattering matrix and the current in accord with (\ref{eq:Tlgsquared}), (\ref{eq:Tvgsquared}) and (\ref{eq:current}), respectively. 

Since expressions (\ref{eq:Pn}) display the formal structure of electron transition rates per unit time in the respective gauges we can use them to define nonperturbative analogs of (\ref{eq:Gammalg/omegas}) and (\ref{eq:Gammavg/omegasperp}) that describe the efficiency of multiplasmonic electron excitation from the SS-band. In the two gauges these quantities read
\bq
\Theta_{s}^{(lg)}(n,\lambda)=\frac{{\cal J}_{s}^{(lg)}(n,\lambda)}{\omega_{sp}}
=
\frac{n^2}{(k_{z,f}a_{B})}\left(\frac{\hbar\omega_{sp}}{1H}\right)w_{n}^{(lg)}(\lambda),
\label{eq:nJefficiencylg} 
\eq
\bq
\Theta_{s}^{(vg)}(n,\lambda)=\frac{{\cal J}_{s}^{(vg)}(n,\lambda)}{\omega_{sp}}
=
\frac{n^2}{(k_{z,f}a_{B})}\left(\frac{\hbar\omega_{sp}}{1H}\right)\left(1+\frac{\tilde{U}}{n\hbar\omega_{sp}}\right)^{2}w_{n}^{(vg)}(\lambda).
\label{eq:nJefficiencyvg} 
\eq
In the Born approximation limit $\Theta_{s}^{(lg,vg)}(1,\lambda)$ reduce to expressions (\ref{eq:Gammalg/omegas}) and (\ref{eq:Gammavg/omegasperp}), respectively.
These gauge-specific dimensionless quantities are compared in Fig. \ref{Thetaslgvg} for experimentally relevant $n=2$. It is seen that the length gauge result follows the expected variation dictated by the renormalization of linear coupling. By contrast, for the present set of system parameters quadratic coupling in the velocity gauge introduces dramatically strong  enhancement over the multiplasmon results produced by the corresponding linear coupling. 
%

\begin{figure}[tb]
\rotatebox{0}{\epsfxsize=8.5cm \epsffile{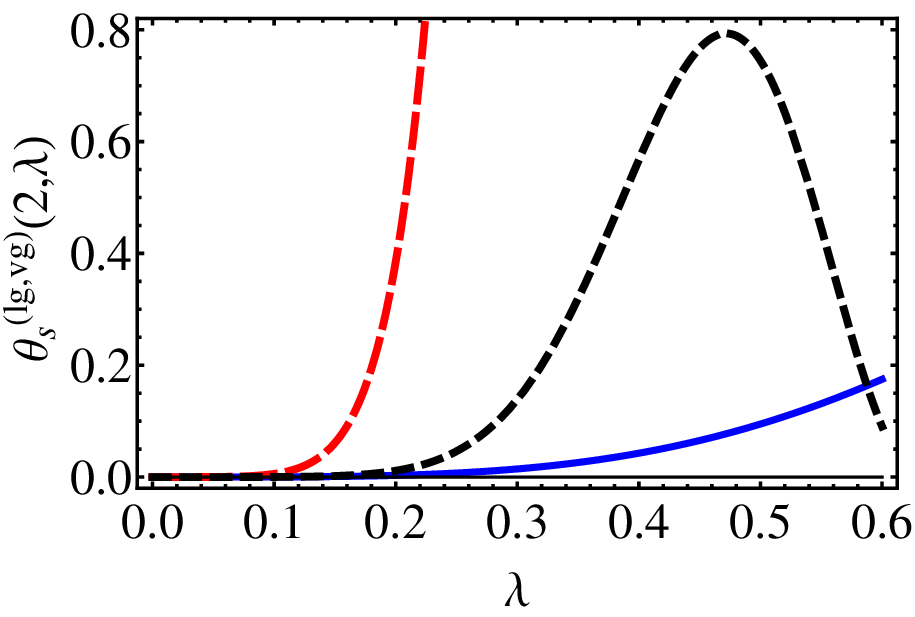}} 
\caption{ Plots of two-plasmon induced electron transition efficiencies $\Theta_{s}^{(lg,vg)}(n=2,\lambda)$ defined in Eq. (\ref{eq:nJefficiencyvg}). Full blue line: $\Theta_{s}^{(lg)}(n=2,\lambda)$ that is a fully renormalized two-plasmon counterpart of the full blue line in Fig. \ref{Gammaslgvg}. Long dashed red line:  $\Theta_{s}^{(vg)}(n=2,\lambda)$ fully renormalized through linear ${\bf p}\cdot\bcalA/m$ and and quadratic couplings $\bcalA^2/2m$. Short dashed black line: same as the latter but with $\tilde{U}$ reduced to zero in (\ref{eq:nJefficiencyvg}), as would occur in the case of initial and final state renormalizations by the $\bcalA^2/2m$ term. The difference between two dashed curves illustrates the strong effect of quadratic coupling in multiplasmon induced transition rates described in the velocity gauge.}
\label{Thetaslgvg}
\end{figure}

\subsection{Calibration of plasmonic coherent states}
\label{sec:assesscohstate}

\begin{figure}[tb]
\rotatebox{0}{\epsfxsize=8.5cm \epsffile{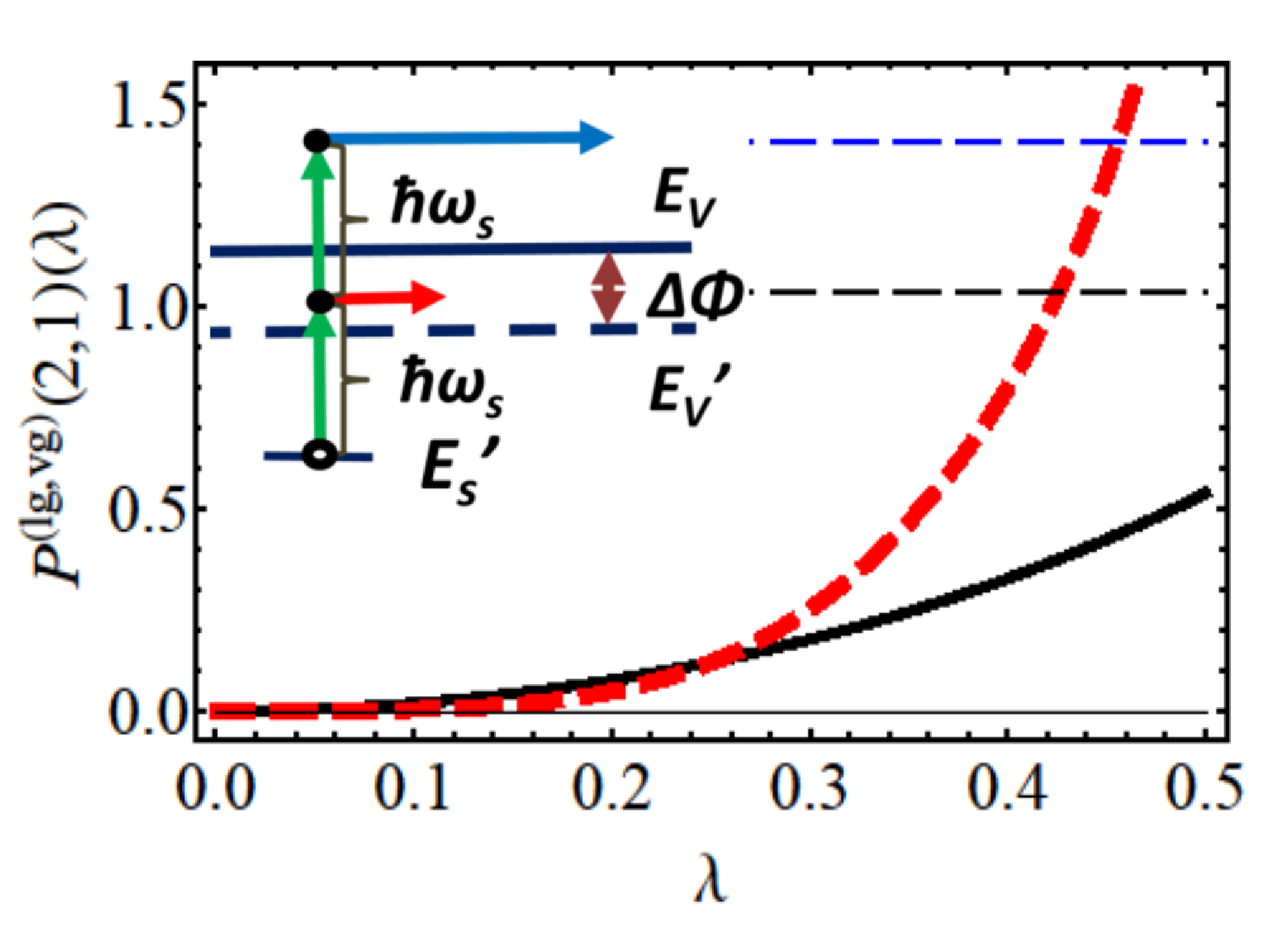}} 
\caption{ Plots of relative intensities of the two- and one-plasmon driven electron emissions $P^{(lg,vg)}(2,1)(\lambda)={\cal J}_{s}^{(lg,vg)}(2,\lambda)/{\cal J}_{s}^{(lg,vg)}(1,\lambda)$, defined in the length and velocity gauge in (\ref{eq:P21lg}) and (\ref{eq:P21vg}), respectively. from the same initial SS-state on Ag(111) surface with unperturbed energy $E_{s}$ and ${\bf K}_{s}=0$. The plots extend over the low plasmonic coherent state population $\lambda<0.5$.  
Upper blue and lower black dashed horizontal lines give the ratio of the lowest order Born approximation results $w^{(vg)}_{BA}(1,\lambda,z_{s})/w^{(lg)}_{BA}(1,\lambda,z_{s})$ for the values $z_{s}=1.2$ (used throughout) and $z_{s}=1.4$ (yielding coincidence), respectively.
Inset: Blue and red horizonatal arrows illustrate two- and one-plasmon driven electron emissions from  the occupied $E_{s}$-level to states above the vacuum level $E_{V}'$ of the Ag(111) surface with work function reduced by $\Delta\phi\approx 1.3$ eV due to alkali submonolayer adsorption[\onlinecite{Horn}] (dashed black line). Scaling of arrow lengths corresponds to $k_{f}^{(2)}/k_{f}^{(1)}=2.14$ which provides complementary information required for full characterization of the SP field-driven electron emission. }
\label{P21}
\end{figure}

One may attempt to assess $\lambda$ and thereby calibrate the pre-excited plasmonic coherent states by  measuring one- and two-plasmon driven electron emissions from one and the same surface with sufficiently reduced workfunction $\phi_{\rm red}<\hbar\omega_{sp}$ (see inset in Fig. \ref{P21}). Here the intensities of one- and two-plasmon induced peaks may enable the estimate of $\lambda$ through the comparison of relative experimental emission intensities with the corresponding theoretical predictions for $f\leftarrow s$ transition probabilities. In the length gauge they are defined by  
\bq
\frac{{\cal J}_{s}^{(lg)}(2,\lambda)}{{\cal J}_{s}^{(lg)}(1,\lambda)}=
4\left(\frac{|\tilde{u}_{s}(k_{z,f}^{(2)})|^{2}/k_{z,f}^{(2)}}{|\tilde{u}_{s}(k_{z,f}^{(1)})|^{2}/k_{z,f}^{(1)}}\right)\frac{w_{2}^{(lg)}(\lambda)}{w_{1}^{(lg)}(\lambda)},
\label{eq:P21lg}
\eq
and in the velocity gauge by
\bq
\frac{{\cal J}_{s}^{(vg)}(2,\lambda)}{{\cal J}_{s}^{(vg)}(1,\lambda)}=
\left(\frac{2\hbar\omega_{sp}+\tilde{U}_{s}}{\hbar\omega_{sp}+\tilde{U}_{s}}\right)^2\left(\frac{|\tilde{u}_{s}(k_{z,f}^{(2)})|^{2}/k_{z,f}^{(2)}}{|\tilde{u}_{s}(k_{z,f}^{(1)})|^{2}/k_{z,f}^{(1)}}\right)\frac{w_{2}^{(vg)}(\lambda)}{w_{1}^{(vg)}(\lambda)}.
\label{eq:P21vg}
\eq
These quantities are illustrated in the main body of Fig. \ref{P21}.
In the present situation where $w(2,\lambda)$ and $w(1,\lambda)$ are of nearly the same magnitude (cf. Figs. \ref{wnlglambda} and \ref{w2vglambda}) the enhancement of ${\cal J}_{s}(2,\lambda)$ over ${\cal J}_{s}(1,\lambda)$ for $\lambda>1/3$ arises mainly from the first factor on the RHS of (\ref{eq:P21lg}) and (\ref{eq:P21vg}) and the-gauge independent channel kinetics ratio 
\bq
\eta_{f\leftarrow s}(2,1) =\frac{|\tilde{u}_{s}(k_{f}^{(2)})|^{2}/k_{f}^{(2)}}{|\tilde{u}_{s}(k_{f}^{(1)})|^{2}/k_{f}^{(1)}},
\label{eq:etaF}
\eq
with the entries retrivable from the insets in Figs. \ref{Gammaslgvg} and \ref{P21}. The overall data from Fig. \ref{P21} convey the message that irrespective of the gauge the described multiplasmon-induced electron emission may be an efficient mechanism already for subsingle mode occupation of prepumped plasmonic coherent states. Hence, with the above developed prerequisites the estimates of $\lambda$ may provide insightful and much desired information on the calibration of plasmonic coherent states generated upon irradiation of surfaces by strong trans-resonant EM fields.[\onlinecite{plasPE}]

\subsection{Quasistatic strong field limit $\hbar\omega_{sp}\ll U_{s} $}
\label{sec:qstatic}

Although not easily realizable with surface plasmons the quasistatic strong field limit (SFL) in which $U_{s}/2\hbar\omega_{sp}=\beta_{s}\gg 1$ is worth exploring in its own right because of the possible analogy with Keldysh's tunneling regime of photoionization.[\onlinecite{Keldysh1965}] Specifically, with the set of parameters employed in Sec. \ref{sec:vgdynamical} the requirement $\beta_{s}\gg 1$ would imply $\lambda\gg 1$, as is also evident from inspection of Fig. \ref{ponderomotives}. 
However, the differences between the present and Keldysh's case exist and arise from the gauges used to represent the form of electron-field interactions (velocity vs. length gauge), the spatial characteristics of the driving fields (surface localization of the plasmon field vs. homogeneous character of the EM field), as well as from the stage (initial vs. final state) at which Volkov ansatz has been implemented. In the present case the point of departure for analysing the limit $U_{s}\gg 2\hbar\omega_{sp}$ is expression (\ref{eq:Tfinvg}) for vertical transitions where in the $\delta$-function on the RHS we neglect the contributions $n\hbar\omega_{sp}$ relative to $U_{s}$. This allows carrying out $n$-summations using addition formulas for generalized Bessel functions 
\bq
\sum_{n=-\infty}^{\infty}(\pm 1)^{n}J_{n}(x,y)=e^{\pm ix},
\label{eq:sumJxy}
\eq
which directly follow from the generating function for generalized Bessel functions (cf. Eq. (B11) in [\onlinecite{Reiss1980}]) 
\bq
e^{[x\sin\phi+y\sin(2\phi)]}=\sum_{n=-\infty}^{\infty}e^{i n\phi}J_{n}(x,y).
\label{eq:genJxy}
\eq
This procedure yields the SFL transition amplitude
\bq
T_{{\bf K},f\leftarrow s}^{SFL}= -2\pi i \tilde{U}_{s}(\lambda)
\left\langle \phi_{f}\left|\exp{\left(Z_{s}(\lambda)\frac{\partial}{\partial z}\right)}\right|u_{s}\right\rangle
\delta\left(\frac{\hbar^{2}{\bf K}^2}{2}\left(\frac{m^{*}-m}{m m^{*}}\right)+E_{f}-E_{s}-\tilde{U}_{s}\right).
\label{eq:TfinvgSFL}
\eq
Here we have expectedly retrieved the static limit $\omega_{sp}\tau\rightarrow0$ of the operator (\ref{eq:expZscos})-(\ref{eq:expZs+}) in the matrix element $\langle \phi_{f}|\dots|u_{s}\rangle$. 
The SFL transition amplitudes (\ref{eq:TfinvgSFL}) are, besides the energy conserving $\delta$-function, controlled by the overlap of the strongly energetically and spatially shifted electron initial state with the final outgoing state wavefunctions.
Here the large energy displacement $\tilde{U}_{s}=\tilde{U}_{s}(\lambda)$ enters through the energy conserving $\delta$-function whereas the spatial shift enters via the displacement operator $\exp[Z_{s}(\lambda)\partial/\partial z]$ acting on $u_{s}(z)$.  
 Taking $\phi_{f}(z)$ as either a localized state, spatial resonance or itinerant state, we arrive at the results describing very different physical situations. However, in all cases the SFL transition rates derived from (\ref{eq:TfinvgSFL}) are governed by the prefactor $\tilde{U}_{s}^{2}$ and read
\bq
 W_{{\bf K},f\leftarrow s}^{SFL}= \frac{2\pi}{\hbar} \tilde{U}_{s}^{2}(\lambda)
\left|\left\langle \phi_{f}\left|\exp\left[3\pi\left(\frac{\ell_{s}}{z_{s}}\right) \sqrt{\beta_{\perp}(\lambda)}\frac{\partial}{\partial z}\right]\right|u_{s}\right\rangle\right|^{2}
\delta\left(\frac{\hbar^{2}{\bf K}^{2}}{2}\left(\frac{m^{*}-m}{m m^{*}}\right)+E_{k_{f}}-E_{s}-\tilde{U}_{s}\right),
\label{eq:wSFL}
\eq 
where $Z_{s}$ has been expressed in terms of $\lambda$-independent 
\bq
\ell_{s}=\left(\frac{4\hbar}{m\omega_{sp}}\right)^{1/2}=2a_{B}\left(\frac{m}{m}\right)^{1/2}\left(\frac{\hbar\omega_{sp}}{1\rm{H}}\right)^{-1/2},
\label{eq:ls}
\eq
and the enhancement factor 
\bq
\beta_{\perp}(\lambda)=\frac{U_{\perp}(\lambda)}{2\hbar\omega_{sp}}.
\label{eq:betaperp}
\eq
 Therefore, since expression (\ref{eq:wSFL}) derives from (\ref{eq:TfinvgSFL}) using (\ref{eq:substzeta}) and is normalized to the current in the $n$-th channel (\ref{eq:jz}), its overal $\lambda$-scaling, as well as of the associated emission current, are expressible through the quadratic $\lambda$-dependence of $U_{s}(\lambda)$ explicated in  Eq. (\ref{eq:Ulambda}). This formulation is reminiscent of the Landau-Zener problem of tunneling between two degenerate states,[\onlinecite{LandauLifshitz,Zener}] viz. the strongly $\tilde{U}_{s}(\lambda)$-upshifted and displaced initial state and the final state. Hence its magnitude strongly depends on the structure and (de)localization of the final state wavefunction $\phi_{{\bf K},f}({\bf r})$.  For the electronic states described by the wavefunctions (\ref{eq:phi_s}) and (\ref{eq:phifin}) we obtain 
\bq
 W_{{\bf K},f\leftarrow s}^{SFL}= \frac{2\pi}{\hbar L_{z}} \tilde{U}_{s}^{2}(\lambda)
|\tilde{u}_{s}(k_{z})|^{2}
\delta\left(\frac{\hbar^{2}{\bf K}^{2}}{2}\left(\frac{m^{*}-m}{m m^{*}}\right)+E_{k_{f}}-E_{s}-U_{s}\right),
\label{eq:WSFL}
\eq 
where $|\tilde{u}_{s}(k_{z})|^{2}$ plays the role of the form-factor associated with the transition $s\rightarrow f$. This simple and intuitive result incorporates the dynamic parameter $\omega_{sp}$ only through the scaling behaviour of the ponderomotive shift $U_{s}$ defined in (\ref{eq:Ulambda}).


\section{Discussion and summary}
\label{sec:Conclusion}

High resolution multiphoton photoemission spectra from flat (111), (100) and (110) surfaces of Ag have been found to exhibit distinct electron emission peaks in the energy interval where in the one-electron picture one would expect only a diffuse, featureless background of "secondary electrons". Since the energies of these peaks do not scale with the frequency of applied radiation field but rather with the multiples of either bulk or surface plasmon frequency of Ag (cf. Fig. \ref{AllAgFloquet}) they were attributed to secondary emission by plasmons.[\onlinecite{plasPE,MarcelPRL,ACSPhotonics}] To interpret and make assignments of these novel discrete features we have proposed a plasmoemission model in which the plasmons prepumped by the primary electron-radiation field interaction act in a secondary process as an electron emitting field.  Since bulk and surface plasmon frequencies in Ag are very close, i.e. $\sim 3.8$ and $\sim 3.7$ eV, respectively, we have proposed in Ref. [\onlinecite{plasPE}] the bulk plasmon-based, and in Ref. [\onlinecite{plasFloquet}] the surface plasmon-based mechanism as complementary models for the interpretation of plasmoemission. The point of departure was the theoretical finding that fixed experimental conditions of surface illumination by EM field generate a plasmon coherent state with $\lambda_{\bf Q}$-specific mode populations (\ref{eq:cohstate}) whose field can support electron emission from the metal (cf. Sec. III in Ref. [\onlinecite{plasPE}]). 

In the present work we have focused on developing several complementary nonperturbative approaches for exploring electron emission induced by multiple quanta of the surface plasmon field. To this end we have coupled only the surface-localized electron states to the plasmon field. This corresponds to the treatment of "intrinsic effects" in photoemission.[\onlinecite{Langreth}] Taking this as a point of departure we have formulated the expressions for and extracted information from the electron excitation spectra and the scattering matrix in the length and velocity gauge representations of electron-plasmon interaction. The obtained expressions can be treated nonperturbatively by employing  cumulant expansion and  $T$-matrix formalism to address electron dynamics and averaging over the prepumped plasmonic coherent state. In both exact approaches approximations are introduced at the same level of neglecting electron momentum correlations between the successive elementary interactions with plasmons. This amounts to the use of second order cumulant expansion for electron propagators and Volkov operator ansatz representation of the evolution operator. 
The thus obtained results smoothly interpolate between the transition amplitudes and energy renormalization shifts derived in fully quantal first order Born approximation and semiclassical multiplasmon electron emission. Using Eliashberg's ansatz[\onlinecite{Eliashberg}] to parameterize the set $\{\lambda_{\bf Q}\}$ determining the plasmonic coherent state by a common scaling parameter $\lambda$ we have derived in Secs. \ref{sec:EPdynlg}-\ref{sec:emissionFloquet} the $\lambda$-scalable expressions for plasmoemitted electron spectra and currents in the length and velocity gauge. 

In our analyses we have established that in both gauges the $\lambda^2$-scaling Born approximation results for plasmoemission currents are close to each other in the experimentally relevant interval of variation of $\lambda$. By contrast, the energy renormalization shifts of plasmoemission-involved electron levels seem to be much more gauge-sensitive, particularly to the ponderomotive interaction brought about by the field-squared term in the velocity gauge representation of electron-plasmon interaction. Likewise in the studies of electron interactions with EM fields this issue still remains to be further explored.   

 In both gauges the main difference between the nonperturbative cumulant and Volkov ansatz approaches appears in the reverse order of taking the matrix elements of the interactions and their exponentiation in transition amplitudes. In cumulant approach the desired plasmonically induced electron (de)excitation amplitudes and probabilities are accessed indirectly via the spectrum or the imaginary part of  Fourier transform of diagonal one-electron propagators. The latter are defined in the length gauge by (\ref{eq:Gcumlength}). Here by construction the cumulants (\ref{eq:Ccumlength}) that appear in the exponent of the propagator (\ref{eq:GKslengthC}) are diagonal spans over the sum of excited eigenstates of $H_{0}^{el}$ and plasmonic coherent state. Then, the approximate form of (\ref{eq:GKslengthC}) is obtained by neglecting higher order cumulants $C_{n>2}(t)$, i.e. complex functions containing the sums of interaction matrix elements correlated through the higher order momentum and energy transfer processes. Analogous procedure is followed in the velocity gauge except that in the latter also the first order cumulant $C_{1}(t)$ yields ponderomotive energy shifts that are quadratic both in $\lambda$ and the coupling strength. This feature cannot be accessed through the finite order cumulant expansion in the length gauge. This is consistent with the common knowledge that the equivalence of two gauges cannot be straightforwardly established by inspecting the convergence of finite order perturbative terms[\onlinecite{{Boyd2004,Reiss2013}}] except in numerical treatments.[\onlinecite{SchuelerPRB,SchuelerJESRP}] 

Conversely, the Volkov evolution operator ansatz offers direct access to excitation amplitudes through the offdiagonal matrix elements of the $T$-matrix. To arrive at operative form of this ansatz as given by (\ref{eq:Uuncorr}) we have neglected the operators arising from a nested commutator expansion on the RHS of (\ref{eq:calGcorr}) that describe higher order correlated electron momentum and energy transfer processes. In the consecutive step we have taken the matrix elements of such reduced evolution operator or Volkov operator ansatz in the expression for the scattering matrix (\ref{eq:TlgVolkovSBC}). This is just the reverse order of steps relative to the cumulant-based calculations which circumvents explicit and mostly formidable summations over the set of intermediate excited states appearing in expressions for cumulants. Completely analogous procedures apply in the velocity gauge.  These subtleties are most clearly manifested in comparison of transition amplitudes obtained from the diagonal propagators (\ref{eq:GKslengthC}) and (\ref{eq:GKsVolkov}), with the offdiagonal ones obtained directly from (\ref{eq:TlgVolkovPsi}) and (\ref{eq:TVolkovPsi}), respectively. In the employed approximations this leads, at least seemingly, to inequivalent  roles of intermediate excited states in the respective fundamental exponentiated contributions whose differences are hard to estimate without inspecting increasingly cumbersome higher order corrections. On the level of first order Born approximation the required entries in the quasiparticle spectrum and the $T$-matrix are found to be identical (cf. comments after Eqs. (\ref{eq:TlgSBC}) and (\ref{eq:Tfinvg})). Thus, although the variants of cumulant approach discussed in Secs. \ref{sec:EPdynlg} and \ref{sec:EPdynvg} do not provide direct insight into plasmoemission yield, they nevertheless enable  better understanding of the physics underlying experimentally accessible plasmoemission currents discussed in Secs.  \ref{sec:Tlength} and \ref{sec:emissionFloquet}.

In the concrete example of plasmoemission from surface bands on Ag(111) the results of calculations displayed in Figs. \ref{wnlglambda}-\ref{P21} indicate that even subsingle mode occupation amplitudes $\lambda<1$ of plasmonic coherent states can give rise to multiplasmon electron emission. 
The specific emission maxima as described by (\ref{eq:Wlg}) and (\ref{eq:Wvg}) are rather gauge insensitive although appearing at somewhat different values of $\lambda$. 
The gauge-specific first order Born approximation results (cf. Figs. \ref{wnlglambda} and \ref{w1vglambda}) start to strongly deviate from the nonperturbative ones already for $\lambda>0.3$. This indicates that even relatively low occupation amplitudes of plasmonic coherent states $\lambda \sim 0.3$ can support discernible multiplasmon electron emission from surface bands. 
At higher values of $\lambda$ the electron emission maxima are followed by the minima that occur due to strong interference between plasmon absorption and emission processes.
The proposed mechanism can be responsible also for the Fermi level plasmoemission spectra from bulk band states overlaping the region of surface plasmon localization at Ag(100) and Ag(110) surfaces shown in Fig. \ref{AllAgFloquet}. Based on these ideas we have also proposed an experimental method of calibration of plasmonic coherent states. The presented theoretical description may be extended to other plasmonic materials and system geometries [\onlinecite{TangNanoLett24}] and calls for more extensive use of nonperturbative methods for treating plasmoemission from surfaces and plasmonics in general.   

The modification of the presented formalism also lends itself to nonperturbative treatment of multiphoton photoemission  induced by strong electromagnetic fields.

\acknowledgments

The author wishes to acknowledge illuminating discussions and fruitfull collaborations on the various aspects of quasiparticle dynamics at surfaces with D.M. Newns, D.C. Langreth, N.W. Ashcroft, K. Wandelt, H. Ueba, A. \v{S}iber, P. Lazi\'{c}, E.V. Chulkov, V.M. Silkin, H. Petek, and D. Novko, that have all stimulated the formulation of the presented theory. Thanks are also due to the referee for helpful suggestions towards improving the manuscript.



\section{Appendices}

\appendix

\section{Heuristic estimate of expression (\ref{eq:corrVI})}
\label{sec:deviationV} 

Model systems in which the commutator (\ref{eq:corrVI}) 
\bq
\frac{1}{\hbar}[H_{0}^{el},V_{S}']= \frac{1}{m\hbar}[v_{ps}({\bf r}),{\bf p}\cdot{\bf A(r)}]=\frac{i}{m}{\bf A(r)}\cdot\left(\nabla v_{ps}({\bf r})\right)
\label{eq:corrVIApp}
\eq
vanishes exactly are those for which $\nabla\cdot v_{ps}({\bf r})=0$, e.g. for free electrons. Another simple limit occurs in the case of homogeneous external fields ${\bf A(r)}(t)={\bf A}(t)$ to which the Volkov ansatz was originally applied.[\onlinecite{Volkov,Madsen}] 

Besides these simple cases there exist a class of systems for which the expression on the RHS of (\ref{eq:corrVI}) when calculated as a transition matrix element may give negligible contribution. This occurs if the position of the maximum of vector potential ${\bf A(r)}$ overlaps the position of the minimum of the binding potential $v_{ps}({\bf r})$. Spatial integration over (\ref{eq:corrVI}) in this situation may produce[ much smaller result than the integral over $V_{S}'$. 
A typical example of such a configuration is provided  by localization of electrons by $v_{ps}({\bf r})\approx v_{ps}(z)$ in Q2D SS-bands on Ag(111) surfaces  (cf. Fig. \ref{SSband}). Here the electron is localized in the effective image potential well where $\nabla\cdot v_{ps}(z)=0$, and this coincides with the position of the dynamical image plane[\onlinecite{Liebsch}] to which ${\bf A(r)}$ is referenced. On the other hand, here $V_{S}'$ reaches maximum. This allows to neglect the contribution of (\ref{eq:corrVIApp}) in further studies of plasmoemission from SS-bands on Ag(111) surfaces.

\section{Evolution operator in the exponential form}
\label{sec:Gross}
\subsection{Evolution operator as exponentiated nested commutator expansion}
\label{sec:nestedU}

We start from a general formulation of the system wavefuction in the interaction representation $\psi_{I}(t)$ which is obtained from the wavefunction in the Schr\"{o}dinger representation $\psi_{S}(t)$ according to (see Ch. 2.5 in Ref. [\onlinecite{GW}])
\bq
\psi_{I}(t)=e^{iH_{0}(t-\tau)/\hbar}\psi_{S}(t)=e^{iH_{0}(t-\tau)/\hbar}U_{S}(t,t_{0})\psi_{S}(t_{0}),
\label{eq:psiIpsiS}
\eq
where in the evolution operator in the Schr\"{o}dinger picture satisfies the equation
\bq
i\frac{\partial}{\partial t}U_{S}(t,t_{0})=H(t)U_{S}(t,t_{0}), \hspace{1cm} U_{S}(t_{0},t_{0})=1,
\label{eq:U_S}
\eq
with $H(t)=H_{0}+V(t)$. The wavefunction $\psi_{I}(t)$ satisfies the Schr\"{o}dinger equation in the interaction respresentation 
\barr
i\hbar\frac{\partial}{\partial t}\psi_{I}(t)&=&V_{I}(t)\psi_{I}(t), \nonumber\\
 V_{I}(t)&=& e^{iH_{0}(t-\tau)}V e^{-iH_{0}(t-\tau)}.
\label{SchrInt}
\earr
where $\tau$ is the coincidence time of the two representations, viz. 
\bq
\psi_{I}(\tau)=\psi_{S}(\tau).
\label{eq:psi_coinc}
\eq
This yields for the evolution operator in the interaction picture[\onlinecite{GW}]
\bq
U_{I}(t,t_{0})=e^{iH_{0}(t-\tau)/\hbar}U_{S}(t,t_{0})e^{-iH_{0}(t_{0}-\tau)/\hbar}.
\label{eq:U_coinc}
\eq
Hence, with the choice for the initial electron wavefunction $\psi_{S}(t_{0})=\phi_{i}e^{-iE_{i}t_{0}}$ in (\ref{eq:Tlength}) and (\ref{eq:Tvel}) we have for the general coincidence time $\tau$
\bq
\psi_{I}(t_{0})=\phi_{i}e^{-iE_{i}\tau}.
\label{eq:phi_i_tau}
\eq

Thus, for the most commonly considered coincidence times we have
\barr
\tau=0, \hspace{0.5cm} & &U_{I}(t,t_{0})=e^{iH_{0}t/\hbar}U_{S}(t,t_{0})e^{-iH_{0}t_{0}/\hbar},\nonumber\\
& &\psi_{I}(0)=\psi_{S}(0), \hspace{1cm} \psi_{I}(t_{0})=\phi_{i};
\label{eq:coinc_0}
\earr
\barr
\tau=t_{0}, \hspace{0.5cm} & &U_{I}(t,t_{0})= e^{iH_{0}(t-t_{0})/\hbar}U_{S}(t,t_{0}),\nonumber\\
& &\psi_{I}(t_{0})=\psi_{S}(t_{0}), \hspace{1cm} \psi_{I}(t_{0})=\phi_{i}e^{-iE_{i}t_{0}}.
\label{eq:coinc_t0}
\earr

Within the SBC  it is most convenient to invoke the choice (\ref{eq:coinc_0}) since the interaction $V$ is assumed to reach the full strength at $t=0$ and is turned off adiabatically at $t\rightarrow\pm\infty$.[\onlinecite{GW}] To proceed with the calculation of (\ref{eq:TvelInt}) and its sequels for $\tau=0$ we exploit the following exponential representation of the evolution operator[{\onlinecite{Gross,PhysRep}]
\bq
U_{I}(H,\tau,t_{0})=e^{-iG^{I}(\tau,t_{0})}=e^{-i[G_{1}^{I}(\tau,t_{0})+G_{corr}^{I}(\tau,t_{0})]},
\label{eq:UGross}
\eq
where 
\barr
G_{1}^{I}(\tau,t_{0})&=&\frac{1}{\hbar}\int_{t_{0}}^{\tau}dt_{1}V_{I}(t_{1}),
\label{eq:G1}\\
G_{corr}^{I}(\tau,t_{0})&=&\sum_{n=2}^{\infty}G_{n}^{I}(\tau,t_{0}),
\label{eq:Gcorr}
\earr
and (\ref{eq:Gcorr}) a nested commutator expansion in powers $n\geq 2$ of perturbation $V_{I}$ [\onlinecite{Gross,PhysRep}] that describes all higher order correlated processes in sequential  interactions of electron and plasmon fields. In the next step we transform a single exponential operator (\ref{eq:UGross}) to a product of two exponential operators using a modifid version of the Baker-Campbell-Hausdorff formula to obtain
\bq
e^{-iG^{I}(\tau,t_{0})}= e^{-iG_{1}^{I}(\tau,t_{0})}e^{-i{\cal G}_{corr}^{I}(\tau,t_{0})}
=
U_{I}^{UCS}(\tau,t_{0})e^{-i{\cal G}_{corr}^{I}(\tau,t_{0})},
\label{eq:GcorrI}
\eq
where 
\bq
U_{I}^{UCS}(\tau,t_{0})=\exp[-iG_{1}^{I}(\tau,t_{0})]
\label{eq:Uuncorr}
\eq
describes uncorrelated scattering processes, and the infinite commutator series 
\bq
{\cal G}_{corr}^{I}(\tau,t_{0})= G_{corr}(\tau,t_{0})
+\frac{i}{2}[G_{1}(\tau,t_{0}),G_{corr}(\tau,t_{0})]
+\dots 
\label{eq:calGcorr}
\eq
 arises due to the noncommutativity of $G_{1}^{I}(\tau,t_{0})$ with $G_{corr}^{I}(\tau,t_{0})$ and describes yet more correlated processes from the coupling of two quantized fields. 
We shall refer to the replacement of the full evolution operator (\ref{eq:UGross}) by the form (\ref{eq:Uuncorr}) as the Volkov operator ansatz.

\subsection{Application to the scattering matrix}
\label{sec:nestedT}

Substitution of (\ref{eq:GcorrI}) back in (\ref{eq:Tlength}) or (\ref{eq:Tvel}) yields
\begin{widetext}
\bq
\lim_{SBC}T_{f,i}(t,t_{0})=
-\frac{i}{\hbar}\int_{t_{0}}^{t} d\tau\langle\langle\mbox{coh}, \phi_{f}|| V_{I}(\tau)e^{[-\frac{i}{\hbar}\int_{t_{0}}^{\tau}dt_{1}V_{I}(t_{1})]} e^{-i{\cal G}_{corr}^{I}(\tau,t_{0})}||\phi_{i},\mbox{coh})\rangle\rangle.
\label{eq:TvelSBCcorr}
\eq
\end{widetext}
This expression is exact insofar (\ref{eq:Tlength}) and (\ref{eq:Tvel}) are exact. 
Further evaluations of the transition probabilities (\ref{eq:TvelSBCcorr}) require  examination and specification of the interaction $V_{I}(t)$. Neglecting the correlated scattering processes described by ${\cal G}_{corr}^{I}(\tau,t_{0})$ expression (\ref{eq:TvelSBCcorr}) reduces in the  UCS regime to 
\begin{widetext}
\bq
\lim_{SBC}T_{f,i}(t,t_{0})|_{UCS}=
-\frac{i}{\hbar}\int_{t_{0}}^{t} d\tau\langle\langle\mbox{coh}, \phi_{f}|| V_{I}(\tau)U_{I}^{UCS}(\tau,t_{0})||\phi_{i},\mbox{coh}\rangle\rangle,
\label{eq:TvelSBC_UCS}
\eq
\end{widetext}
with $U_{I}^{UCS}(\tau,t_{0})$ defined in (\ref{eq:Uuncorr}). This expression was used in the derivation of  (\ref{eq:TVolkov}) via the combined use of the Coulomb gauge (\ref{eq:nablaA}) and the smallness of (\ref{eq:corrVI}).

\section{Quasiparticle spectrum in the limit of electron transitions between iso-energetic levels}
\label{sec:Isoenergetic}

An instructive limit of (\ref{eq:expC2}) arises in the regime of iso-energetic electron transitions for which $E_{{\bf K},s}=E_{{\bf K-Q},f}$. This may typically occur between degenerate localized states or forward scattering processes $|i\rangle\rightarrow |i\rangle$. Applying this in (\ref{eq:C1omegas}) and (\ref{eq:C2omegas}) and assuming that the corresponding scattering matrix elements are nonvanishing, we have in an obvious shorthand notation ($t_{0}=0$) 
%
\bq
C_{s}^{(2)}(t)=
\mu_{s}\left(e^{-i\omega_{sp}t}-1\right)-i\nu_{s}t 
+
\sum_{{\bf Q},f}\lambda_{\bf Q}^{2}\frac{|V_{f,s}^{\dag}|^{2}}{(\hbar\omega_{sp})^{2}}\left[\cos(2\omega_{sp}t)-1\right],
\label{eq:isoC2}
\eq
%
where
\bq
\mu_{s}=\sum_{{\bf Q},f}\frac{|V_{f,s}^{\dag}|^{2}}{(\hbar\omega_{sp})^{2}},
\hspace{1cm}
\nu_{s}=-\omega_{sp}\mu_{s}.
\label{eq:nu_s}
\eq
The first line on the RHS of (\ref{eq:isoC2}) describes only direct one-plasmon emission and subsequent reabsorption processes and hence is $\lambda_{\bf Q}$-free.  The remaining $\lambda_{\bf Q}$-weighted one-plasmon processes proceeding via the coherent state and contributing the terms $\propto[\cos(\omega_{sp}t)-1]$ mutually cancel out in the sum of (\ref{eq:C1omegas}) and (\ref{eq:C2omegas}) leading to (\ref{eq:isoC2}). Such cancellations do not occur in the case of nonzero momentum transfer. This leaves only the second line which describes the $2\omega_{sp}$ processes proceeding via the coherent state cloud and hence weighted by $\lambda_{\bf Q}^{2}$.  The quasiparticle spectral density generated by (\ref{eq:isoC2}) appears as a convolution of a Poisson distribution of one-plasmon processes[\onlinecite{GW+C}] and the distribution of two-plasmon processes obtained by application of (\ref{eq:BesselI}) to the exponent of the second line from the RHS of (\ref{eq:isoC2}). 
We also observe that the short time limit of (\ref{eq:isoC2}) subject to Eliashberg's ansatz exhibits the ballistic or Zeno-like propagation
\bq
\lim_{t\rightarrow 0}C_{s}^{(2)}(t)=-\mu_{s}\omega_{sp}^{2}\left(\frac{1}{2}+2\lambda^{2}\right)t^{2}.
\label{eq:C2Zeno}
\eq
This Zeno-like behaviour is universal to all coupled electron-boson systems subject to sudden switching on of the interaction.[\onlinecite{Zeno}] In the forward scattering limit $V_{f,i}\rightarrow V_{i,i}\delta_{f,i}$ expression (\ref{eq:isoC2}) provides exact solution to the XPS core level problem[\onlinecite{Langreth}] in the environment of coherent plasmonic states.


\section{Alternative representations of E\lowercase{q}. (\ref{eq:phipsi})}
\label{sec:alternativeT}

In this appendix we present several equivalent representations of the scattering amplitude (\ref{eq:phipsi}) that can be conveniently used in the derivations of the corresponding scattering matrix.
Using the generating functions for Bessel functions (\ref{eq:BesselJ}) and (\ref{eq:BesselI})  we can write expression (\ref{eq:expZs+}) in alternative forms
\begin{widetext}
\barr
\exp\left[Z_{s}(\lambda)\cos(\omega_{sp}\tau)\frac{\partial}{\partial z}\right]
&=&
\sum_{n=-\infty}^{\infty}e^{in\omega_{sp}\tau}I_{n}\left(Z_{s}(\lambda)\frac{\partial}{\partial z}\right),
\label{eq:Ilin}\\
\exp\left[iZ_{s}(\lambda)\sin\left(\frac{\pi}{2}-\omega_{sp}\tau\right)\frac{-i\partial}{\partial z}\right]
&=&
\sum_{n=-\infty}^{\infty}(i)^{n}e^{-in\omega_{sp}\tau}J_{n}\left(Z_{s}(\lambda)\frac{-i\partial}{\partial z}\right),
\label{eq:Jlin-}\\
\exp\left[iZ_{s}(\lambda)\sin\left(\omega_{sp}\tau-\frac{\pi}{2}\right)\frac{i\partial}{\partial z}\right]
&=&
\sum_{n=-\infty}^{\infty}(-i)^{n}e^{in\omega_{sp}\tau}J_{n}\left(Z_{s}(\lambda)\frac{i\partial}{\partial z}\right).
\label{eq:Jlin+}
\earr
\end{widetext}
Similarly, convenient alternatives for $2\omega_{sp}$-component in the exponent of (\ref{eq:phipsi}) are
\bq
e^{i\beta_{s}\sin(\pi-2\omega_{sp}\tau)}=\sum_{m=-\infty}^{\infty}(-1)^{m}e^{-2im\omega_{sp}\tau}J_{m}(\beta_{s}),
\label{eq:Jqu+}
\eq
\bq
e^{-i\beta_{s}\sin(2\omega_{sp}\tau-\pi)}=\sum_{m=-\infty}^{\infty}(-1)^{m}e^{2im\omega_{sp}\tau}J_{m}(-\beta_{s}).
\label{eq:Jqu-}
\eq
Using the definition (\ref{eq:genJ}) 
we can represent the oscillating phase component of (\ref{eq:phipsi}) either as
\begin{widetext}
\barr
& &
\exp\left[i\beta_{s}(\lambda)\sin(2\omega_{sp}\tau)+Z_{s}(\lambda)\frac{\partial}{\partial z}\cos(\omega_{sp}\tau)\right]
\rightarrow
\exp\left[i\beta_{s}(\lambda)\sin\left(2(\pi/2-\omega_{sp}\tau)\right)+i\left(Z_{s}(\lambda)\frac{-i\partial}{\partial z}\right)\sin(\pi/2-\omega_{sp}\tau)\right]\nonumber\\
&=&
\sum_{n=-\infty}^{\infty}(i)^{n} e^{-i n\omega_{sp}\tau}J_{n}\left(Z_{s}(\lambda)\frac{-i\partial}{\partial z},\beta_{s}(\lambda)\right).
\label{eq:oscphase-}
\earr
or equivalently
\barr
& &\exp\left[i\beta_{s}(\lambda)\sin(2\omega_{sp}\tau)+\left(Z_{s}(\lambda)\frac{\partial}{\partial z}\right)\cos(\omega_{sp}\tau)\right]
\rightarrow
\exp\left[i(-\beta_{s}(\lambda))\sin\left(2(\omega_{sp}\tau-\pi/2)\right)-i\left(Z_{s}(\lambda)\frac{-i\partial}{\partial z}\right)\sin(\omega_{sp}\tau-\pi/2)\right]\nonumber\\
&=&
\sum_{m=-\infty}^{\infty}(-i)^{m}e^{im\omega_{sp}\tau}J_{m}\left(-Z_{s}(\lambda)\frac{-i\partial}{\partial z},-\beta_{s}(\lambda)\right),
\label{eq:oscphase+app}
\earr
\end{widetext}
where we have made use of the notation of generalized Bessel functions (\ref{eq:genJ}) [\onlinecite{Reiss1980}].
For obtaining the RHS of (\ref{eq:oscphase-}) we have combined (\ref{eq:Jlin-}) and (\ref{eq:Jqu+}), and for  (\ref{eq:oscphase+app}) the alternatives (\ref{eq:Jlin+}) and (\ref{eq:Jqu-}). In expression (\ref{eq:oscphase+app}) the positive multiple $m\omega_{sp}$  in the exponent of $e^{im\omega_{sp}\tau}$ is conventionally associated with plasmon absorption by the electron.   
Substituiting (\ref{eq:oscphase+}) back in (\ref{eq:phipsi}) we can bring the latter to the Floquet form (\ref{eq:phipsi})
\begin{widetext}
\bq
\langle\phi_{f}|\psi_{{\bf K},s}(\tau)\rangle=\exp\left[-i\left(\frac{(\hbar{\bf K})^2}{2m^{*}\hbar}+\frac{E_{s}+U_{s}(\lambda}{\hbar}\right)\tau\right]
\sum_{n=-\infty}^{\infty}(-i)^{n} e^{in \omega_{sp}\tau}\int dz \phi_{f}^{*}(z)
J_{n}\left(-Z_{s}(\lambda)\frac{-i\partial}{\partial z},-\beta_{s}(\lambda)\right)u_{s}(z).
\label{eq:fullVolkovapp}
\eq
\end{widetext}
%




\begin{thebibliography}{11}
\bibitem{Glauber} These states are completely analogous to photonic coherent states excited by electronic motion (current) as described by Eqs. (9.16)-(9.23) in Ref. [\onlinecite{GlauberPR}]. Quite generally, their occupations  are determined by the plasmon excitation dynamics and various distributions of excited plasmonic states can be constructed once their values or generating functions are known.[\onlinecite{SunkoBG}]
\bibitem{Truscott1} The validity of Volkov ansatz for a general class of surface potentials was later demonstrated in Ref. [\onlinecite{Truscott2}]. For the special case of Coulomb potential it was demonstrated in Ref. [\onlinecite{Faisal}], see Eqs. (4) and (6) therein.
\bibitem{scaling} The displayed $\lambda$-scaling of expressions (\ref{eq:xioff2})-(\ref{eq:Ulambda}) arises from and signifies the excitation dynamics of the coupled e-SP system. This is in contrast to the system ground state energy and related quantities which scale with the interelectron spacing parameter $r_{s}$.
\end{thebibliography}
\end{document}